\def\Msun{M$\sb{\odot}$}
\def\Msuny{M$\sb{\odot}$~y$^{-1}$}
\def\Lsun{L$\sb{\odot}$}

\newcommand{\kms}{~km~s$\sp{-1}$}
\newcommand{\Teff}{$T\sb{\rm eff}$}

\newcommand{\cc}{[12] $-$ [25]}
\newcommand{\ccc}{[25] $-$ [60]}

\newcommand{\f}{12~$\mu$m}
\newcommand{\ff}{25~$\mu$m}
\newcommand{\fff}{60~$\mu$m}
\newcommand{\Fp}{$F_{\rm p}$}
\newcommand{\Fz}{$F_{\rm z}$}

\documentstyle[lscape,epsf,epsfig]{l-aa}

\begin{document}

\title{Circumstellar shells and mass loss rates: Clues to the
evolution of S stars\thanks{Based on
observations carried out at the Caltech Submillimeter Observatory (Mauna Kea,
Hawaii)}}

\author{A. Jorissen \inst{1,2}\thanks{Research Associate, F.N.R.S., Belgium}
\and    
        G.R. Knapp\inst{2}
}
\offprints{A. Jorissen (at the address in Belgium)}
\institute{
Institut d'Astronomie et d'Astrophysique,
                  Universit\'e Libre de Bruxelles, C.P.226,
                  Boulevard du Triomphe,
                  B-1050 Bruxelles,
                  Belgium
\and 
Department of Astrophysical Sciences,
Princeton University,
Princeton, NJ 08544, U.S.A.}

\date{Received date; accepted date}

\thesaurus{08.16.4, 08.12.1, 08.13.2, 13.09.6, 13.19.5}

\maketitle
\markboth{A. Jorissen \& G. Knapp: Circumstellar shells and mass loss rates of 
S stars}
{A. Jorissen \& G. Knapp: Circumstellar shells and mass loss rates of S stars}

\begin{abstract}

It is the purpose of this paper to rediscuss the circumstellar properties of S
stars and to put these properties in perspective with our current understanding
of the evolutionary status of S stars, in particular the intrinsic/extrinsic
dichotomy. This dichotomy states that only Tc-rich (`intrinsic') S stars are
genuine thermally-pulsing asymptotic giant branch stars, possibly involved in the
M--S--C evolutionary sequence.  
Tc-poor S stars are referred to as `extrinsic' S stars, because they are the
cooler analogs of barium stars, and like them, owe their chemical peculiarities
to mass transfer across their binary system.

Accordingly, an extensive data set probing the circumstellar environment of S stars
(IRAS flux densities,
maser emission, CO rotational lines) has been collected and critically
evaluated. This data set combines new observations (9 stars have been observed in the
CO $J=2-1$ line and 3 in the CO $J=3-2$ line, with four new detections) with
existing material (all CO and maser observations of S stars published in the
literature). The IRAS flux densities of S stars have been re-evaluated by 
co-adding the
individual scans, in order to better handle the intrinsic variability of these
stars in the IRAS bands, and possible contamination by Galactic cirrus.

In the ($K - [12], [25] - [60]$) color-color diagram, 
S stars segregate into five distinct regions according
to their Tc content and ZrO/TiO, C/O and IR spectral indices. 
Stars with photospheric colors (populating `Region A') may be
identified with extrinsic S stars. For the other regions characterized
by different excess levels in the 12, 25 and 60 $\mu$m bands,
several diagnostics (like the IRAS spectral class, maser emission, and
shape of CO rotational lines) have been collected to infer the physical
properties of the dust shell. 
A simple radiative-transfer code has also been used to infer the chemical nature
(carbonaceous or silicate) of the dust grains from the observed IR
colors.
S stars with large $K - [12]$ excesses and moderate
\ccc\ excesses (populating Regions B and C) exhibit the signatures of
oxygen-rich shells (9.7 $\mu$m silicate emission and SiO maser
emission). The situation is less clear for S stars with small $K - [12]$
and moderate  $[25]-[60]$ indices (populating Regions D
and E). Their IR colors are consistent with carbonaceous grains (as is
their featureless IRAS spectrum, and absence of silicate or SiO maser
emission), but these features may equally well be explained by a detached shell. 
For many of these stars with a large 60~$\mu$m excess,
the shell is indeed resolved by the IRAS beam at 60 $\mu$m.
The prototypical SC star FU Mon is among these. Since SC stars are
believed to be in a very short-lived evolutionary phase where C/O = 1 within 1\%,
FU~Mon may be a good candidate for the `interrupted mass-loss' scenario advocated
by Willems \& de Jong (1988). The CO line profile of FU Mon is
also peculiar in being quite narrow ($V_e = 2.8$~\kms), suggesting that the mass
loss has just resumed in this star.     

Mass loss rates or upper limits have been derived for all S stars observed in the
CO rotational
lines, and range from $< 2\;10^{-8}$~\Msun~y$^{-1}$ for extrinsic S stars to 
$1\;10^{-5}$~\Msun~y$^{-1}$ (the Mira S star W~Aql). These mass-loss rates
correlate well with the $K - [12]$ color
index, which probes the dust loss rate, provided that $\mathaccent 95 M \ga 10^{-
8}$~\Msun~y$^{-1}$. Small mass-loss rates are found for extrinsic S stars,
consistent with their not being so evolved (RGB or Early-AGB) as the Tc-rich S
stars. This result does not support the claim often made in relation with
symbiotic stars that binarity strongly enhances the mass-loss rate.


\end{abstract}

\section{Introduction}
\label{Sect:Intro}

The S stars are late-type giants whose spectra resemble those of M
giants, with the addition of distinctive molecular bands of ZrO
(Merrill 1922). The presence of 
ZrO bands is often considered as a direct consequence of
molecular equilibrium in the special circumstances where the atmospheric C/O
ratio is within 10\% of unity (e.g. Scalo \& Ross 1976). 
However, Piccirillo (1980) has shown
that the above statement is only valid for stars  with $T < 3000$~K. At
higher temperatures, an enhanced Zr abundance, rather than a C/O ratio close to
unity, is the dominant factor in the development of strong ZrO bands. Detailed
abundance analyses (Smith \& Lambert 1990) have shown that the envelopes of S stars
are enriched in heavy elements like Zr, and so bear the signature of the 
s-process of nucleosynthesis (K\"appeler et al. 1989). Although larger than in M
giants,  the C/O ratio of S stars is not necessarily close to unity (Smith \&
Lambert 1990), except in the so-called SC stars (Dominy et al. 1986).  

When S stars were still believed to be objects with C/O close
to unity, they were naturally considered as transition objects 
between M giants (C/O $< 1$) and carbon stars (C/O $> 1$) on the asymptotic giant
branch (AGB) (Iben \& Renzini 1983). Support to this scenario is
provided by observations of S stars on the upper AGB of globular
clusters in the Magellanic Clouds (Bessell et al. 1983; Lloyd Evans
1983, 1984, 1985).
In this evolutionary phase, low- and intermediate-mass
stars are characterized by a double (H, He) burning-shell
structure which is thermally unstable. The thermal instabilities (`thermal
pulses') developing in the He-burning shell are the site of a rich nucleosynthesis
(Frost \& Lattanzio 1995), probably including the s-process, although
its detailed mode of operation remains poorly understood (e.g. Sackmann
\& Boothroyd 1991; Herwig et al. 1997). In the receding phase of the
thermal instability, the
convective outer envelope may plunge (`third dredge-up') into the intershell zone
containing the He-burning ashes, and bring fresh
carbon and s-process elements to the surface.

However, several observations have challenged this traditional M--S--C evolution
sequence. The first set of observations relates to Tc, an element with no stable
isotopes, discovered in the spectra of some S stars by Merrill (1952). If the
s-process indeed occurred during recent thermal pulses in S stars, Tc should 
be observed at the surface along with the other s-process elements (Mathews et al.
1986). Little et al. (1987) found however that only long-period Mira or
semiregular S stars (i.e., intrinsically bright S stars) exhibit Tc lines.      
Second, the broad range of IRAS colors exhibited by S stars (e.g., Jorissen et
al. 1993) is difficult to reconcile with the idea that they represent a brief
transition phase as the star evolves from an oxygen-rich M
giant with C/O $< 1$ into a C star with C/O $> 1$. M and C stars occupy 
well-defined regions in the IRAS color-color diagram, and it is not clear how S stars
fit into the (much debated) evolutionary sequence joining M stars to C stars
in that diagram 
(Willems \& de Jong 1986, 1988; Chan \& Kwok 1988; Zuckerman \& Maddalena 1989;
de Jong 1989).
  
These problems received a new impetus with the discovery that the barium stars,
a family of peculiar red giant (PRG) stars of spectral type G and K, are all
members of binary systems (McClure et al. 1980; McClure 1983). Iben \& Renzini
(1983) were the first to propose that Tc-poor S stars could perhaps be the cooler
analogs of the barium stars. Long-term radial-velocity
monitorings confirmed this suggestion, and
it is now clear that Tc-poor S stars are binary stars
(Smith \& Lambert 1988; Brown et al. 1990; Jorissen \& Mayor 1992; Jorissen et al.
1993; Johnson et al. 1993) with orbital elements identical to those of barium
stars (Jorissen et al. 1997). Tc-poor
S stars are now referred to as `extrinsic S stars', because, like barium stars,
they owe their chemical peculiarities to mass transfer across the binary system. 
On the contrary, Tc-rich, `intrinsic S stars' are genuine
thermally-pulsing (TP) stars on the TP-AGB. Since the C/O ratio of
extrinsic S stars depends on the details of the
mass accretion process, it is not necessarily close to unity, but as discussed
above, neither abundance analyses nor predictions of molecular chemical
equilibrium really require C/O to be close to unity in S stars.

In Paper I (Jorissen et al. 1993; see also Groenewegen 1993), it
was shown that the correlation Tc-poor/binary found for S stars could be extended
to their IRAS colors, since another distinctive property of binary, Tc-poor S
stars is the absence of IR excesses. Their IRAS colors simply reflect 
the photospheric colors, contrary to Tc-rich S stars
which usually exhibit IR excesses. These IR excesses are caused by circumstellar
dust, and are indicative of substantial mass loss, thus suggesting that Tc-rich S
stars are more massive and/or more evolved than Tc-poor S stars. 

The possibility offered by the circumstellar properties of S stars to probe their
evolutionary status has been used in several recent studies (Jura 1988; Chen \&
Kwok 1993; Bieging \& Latter 1994; Sahai \& Liechti 1995). However, these studies
still rely on hypotheses not fully consistent with the dichotomy recently found
among S stars, as discussed above, and their conclusions may therefore be somewhat
biased. For example, it makes no sense to test the AGB evolutionary sequence 
M--S--C using a sample of S stars not properly cleaned from its extrinsic content. 
In addition,
models inferring the chemical nature (carbonaceous or silicate) of the dust
grains from models assuming that C/O is close to unity in the photosphere do not
sample the whole parameter space occupied by these stars.
A similarly incorrect corollary consists of concluding that dust production, and
thus mass loss, is not very efficient in S stars because dust-seed molecules no
longer form in the absence of any free C or O atoms, 
since these are all tied up in CO when C/O $\sim 1$.

It is the purpose of this paper to rediscuss the circumstellar properties of S
stars and to put these properties in perspective with our current understanding
of the evolutionary status of S stars, in particular the intrinsic/extrinsic
dichotomy. An extensive data set probing the circumstellar environment of S stars
(IRAS colors, maser emission, CO rotational lines) has therefore been collected
and critically evaluated. This data set combines new material with existing results
collected from the literature.

The IRAS colors of S stars provide a first way to probe their circumstellar
environment, and more specifically, to evaluate the amount and nature of the dust
surrounding these objects. A re-evaluation of the flux
densities listed in the IRAS {\it Point Source Catalogue} (IRAS
Science Team 1988; PSC) was necessary to
circumvent the problems of interpretation
related to the intrinsic IR variability of these sources and
to their possible contamination by Galactic cirrus emission. These effects are
not always properly handled in the PSC. This re-evaluation was performed by
co-adding the raw scans (Sect.~2.2). The clean flux densities were then used to  
define five regions in the $(K-[12], [25]-[60])$ color-color diagram
(Sect.~2.3) which contain stars of similar extrinsic or intrinsic nature
and of similar ZrO/TiO, C/O and IR spectral indices.
Several S stars with shells resolved at 60 (and sometimes 100)~$\mu$m have been
found by comparing the source profile with the IRAS point source response
function (Sect.~2.4).
Resolved shells at 60~$\mu$m appear to be correlated with large 60~$\mu$m
excesses, suggesting that these shells have detached from their parent star. 
A simple model of the dust shell has been used to predict its IR colors and to
infer the chemical nature of the dust grains (Sect.~4), using constraints provided
by the detection or non-detection of maser emission (Sect.~3).       
Finally, mass loss rates have been derived in an homogeneous way (Sect.~5.3) from
the intensities of the CO millimeter-wave
lines, derived from new observations with the
Caltech Submillimeter Observatory (Sect.~5.1) or collected from the literature
(Sect.~5.2). The mass loss rates, wind expansion velocities, IR colors and
extrinsic/intrinsic nature are then discussed together in Sect.~6.

\section{Infrared colors of S stars}
\subsection{The sample}
\label{Sect:IRAS}

The sample of S stars considered in this paper was selected from
the list of Chen et al. (1995), which provides associations between
IRAS sources from the PSC and S stars from the 
{\it General Catalogue of Galactic S
Stars} (GCGSS, Stephenson 1984) and from an additional later list
(Stephenson 1990).
Among these, only stars having flux densities of good quality  
(i.e. with a quality flag of
3 in the PSC) in the 12, 25 and \fff\ bands have been retained.

Several stars that have probably been misclassified as S stars were
removed from the final sample. For example, stars from the original
Westerlund \& Olander (1978) sample were later recognized by
Lloyd Evans \& Catchpole (1989) as
actually being heavily-reddened M giants or supergiants.  
A few other cases of M giants or supergiants misclassified as S stars 
were identified by Keenan \& McNeil (1989; HR
3296 = GCGSS 500) and Winfrey et al. (1994; GCGSS 1314 and
star 41 in Stephenson 1990). A detailed heavy-element abundance
analysis (Lambert et al. 1995) has shown that the stars DE~Leo (HR~4088) and 
HR~7442, although often considered as S stars, have normal abundances.   
According to Meadows et al. (1987), GCGSS 886 (IRAS 15194-5115) is
now classified as a carbon star and thus has not been retained in the
final sample of S stars.
Our final sample may still be somewhat
contaminated by M supergiants misclassified as S stars and by a few carbon stars.
The star T Cet
for example was classified as M5-6Se in the original paper by Keenan (1954)
defining the S class, but it was reclassified as M5/M6Ib/II in the
Michigan Spectral Survey (Houk \& Cowley 1975). Since at the low plate
dispersions  used in classification work, these two types of spectra
look similar (e.g. Lloyd Evans \& Catchpole 1989), T Cet has been  kept
in our final list until higher resolution spectra resolve these
conflicting classifications. The same holds true for TV Dra and OP
Her (see Table~2 of GCGSS). Note,
however, that SC stars like RZ Peg, FU Mon and GP Ori are retained in
the sample, since they may provide important
clues to the evolutionary status of S stars as a whole. 
Similarly interesting are the two CS stars TT Cen and BH Cru that
are known to exhibit ZrO bands at some times and C$_2$ bands at others
(Stephenson 1973; Lloyd Evans 1985). 

\subsection{The IRAS flux densities}
\label{Sect:IRASflux}

The final sample consists of 124 S stars having flux densities at
12, 25 and \fff\ flagged as being of good quality in the PSC. 
These stars are listed in Table 1.
However, PSC flux densities suffer from several shortcomings   
that make them inadequate for the present study. First, the \fff\ flux
density may in some cases be seriously contaminated by Galactic cirrus
emission, as shown by Ivezi$\acute{\rm c}$ \& Elitzur (1995). Second, 
the PSC flux densities are not appropriate in the case of slightly extended or
variable sources, as several S stars appear to be. Moreover, hysteresis of the
detectors hampers the search for extended sources and should be properly
identified. Finally, the detectors may saturate on very bright sources, those with
flux densities in excess of 1000 Jy (like $\chi$~Cyg, see Appendix A).

In order to correctly handle these effects, the raw IRAS data 
for all S stars in our sample were reprocessed 
through the {\sc ADDSCAN} procedure provided by the {\it Infrared Processing and
Analysis Center} (IPAC\footnote{IPAC is operated by the {\it Jet Propulsion
Laboratory} and {\it California Institute of Technology} for NASA}). This
procedure
has been used to co-add all scans passing within 1\farcm7 of the
target position. Before being co-added, the raw data are first
interpolated using cubic splines and resampled at 10 data points per
arcminute in all bands. A baseline is then defined for each individual
scan by fitting a parabola to the data in a window extending from 
$r_{\rm in}$ to $r_{\rm out}$ from the target 
($r_{\rm in} = \pm 2', \pm 2', \pm 2\farcm5, \pm 4'$ and 
$r_{\rm out} = \pm 7\farcm5, \pm 7\farcm5, \pm 10', \pm
13\farcm5$ in the 12, 25, 60 and 100~$\mu$m bands, respectively).   
The r.m.s. residual $\sigma_i$ of the data around the baseline fit   
is an indication of the background noise in a given scan $i$
(note that possible nearby
sources with a peak flux density exceeding $2.5 \sigma_i$ are automatically
removed and do not enter the final residual calculation).
The different scans are then co-added with weighting factors equal to
$1/\sigma_i^2$. The noise $\sigma$ in the co-added scan, computed in a
similar way as for the individual scans, is smaller than
that in any individual scan, and the reduction may be substantial 
when scans with very different orientations, sampling
different regions of the sky around the target, are available,
since structure in the Galactic cirrus emission contributes to
the noise.
Finally, a point-source template with adjustable width is fitted to
the co-added scan in the target window $\pm r_{\rm in}$. 
Three different estimates of the flux densities 
in each band are then computed: the
peak flux density $F_{\rm p}$, the template flux density $F_{\rm t}$ and the
`zero-crossing'  
flux density $F_{\rm z}$. The template flux density 
$F_{\rm t}$ is derived from
the  template fit to the data, and generally agrees with $F_{\rm p}$, unless the
source is  extended or hysteresis effects are important. For extended
sources, the `zero-crossing'  flux density $F_{\rm z}$ must be preferred.
It corresponds to the
integrated flux density between the `zero crossings', 
which are defined as
the first locations, moving outwards from the peak location, where
the source profile intersects the baseline.
In Sect.~\ref{Sect:extended}, a criterion based on the
comparison of  $F_{\rm z}$ and $F_{\rm p}$ has been designed to
identify sources with possibly resolved shells. If that criterion is
met, the $F_{\rm z}$ flux density (identified by a `+' in
Table~\ref{Tab:IRAS}) has been adopted instead of $F_{\rm t}$.   
In the case of bright sources, detector hysteresis 
results in a trail extending along the scan direction, thus disturbing
the template fit. In that case, $F_{\rm p}$ has been adopted.    
The template fit to 
sources embedded in strong Galactic cirrus emission
generally resulted in an abnormally narrow profile, because the adopted baseline
is then too high with respect to the base of the source signal.
These contaminated sources were thus readily identified by their
narrow template fit, and were rejected.
Finally, individual scans clearly contaminated by nearby sources have 
been eliminated.

\begin{figure}
  \vspace{9cm}
  \vskip -0cm
  \begin{picture}(8,8.5)
    \epsfysize=9.8cm
    \epsfxsize=8.5cm
    \epsfbox{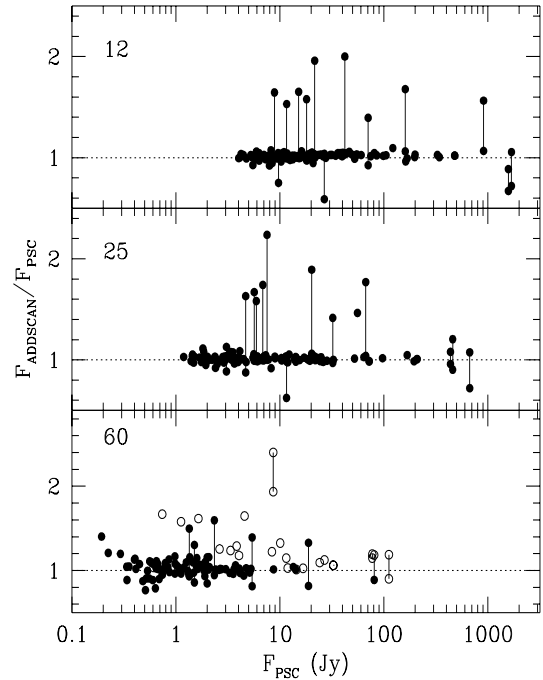}
  \end{picture}
  \vskip -0.5cm
\caption[]{\label{1}
Comparison between the PSC flux densities and those derived from the {\sc
ADDSCAN} procedure (see text) in the 12, 25  and \fff\ bands. Flux densities
of variable sources at two different epochs are connected by a vertical line.  
At \fff, open circles correspond to extended sources (see
Sect.~\ref{Sect:extended})
}
\end{figure}

The adopted IRAS flux densities
are listed in Table~\ref{Tab:IRAS} in columns 5--8. The 2.2~$\mu$m
flux density is listed in column~3. The calibration of Beckwith et al.
(1976; 620 Jy corresponds to $K = 0$) has been used to derive the
2.2~$\mu$m flux density from the $K$ magnitude provided by different authors,
as listed in column~4:  
1: Catchpole et al. (1979); 2: Neugebauer \& Leighton (1969; {\it
Two-Micron Sky Survey}); 3: Mendoza \& Johnson
(1965); 4: Price (1968); 5: Chen et al. (1988); 6: Noguchi et al. (1991);
7: Guglielmo et al. (1993); 8: Epchtein et al. (1990); 9: Epchtein et
al. (1987); 10: Lloyd Evans \& Catchpole (1989). 

The Tc content (from Paper~I, and Lambert et al. 1995) 
is listed in column~9. The
columns labelled `LRS' and `VC' provide the classification of the IRAS
low-resolution spectrum, according to the IRAS {\it Low Resolution
Spectrometer Catalogue} (Olnon et al. 1986)
or to the Volk \& Cohen (1989) schemes, respectively. The optical spectral
type is from the GCGSS. The variability type, the period and the
range of variation of the magnitude (lowest minimum - highest maximum) are 
from the {\it General Catalogue of
Variable Stars} (Kholopov et al. 1985; GCVS). The variable name and/or the HD/HR
designations, when available, are given in the last column. If maser emission has
been detected (see Sect.~\ref{Sect:maser} and
Table~\ref{Tab:maser}), the maser type (SiO or OH) is given between parentheses
after the stellar designation.  

Several sources observed by IRAS a few months apart turn out to be strongly
variable in the IRAS bands. For these, the {\sc ADDSCAN} procedure has
been run separately on the two groups of data, and the flux densities are
listed on two separate lines in Table~\ref{Tab:IRAS}. The
corresponding approximate Julian dates have been derived from
the `Satellite Operation Plan' number attached to the scans and from the
mission chronology provided in Table III.C.1 of the IRAS {\it Explanatory
Supplement} (1988).

The flux densities derived from the {\sc ADDSCAN} procedure generally agree
with the PSC flux densities
within 20\%, as shown in Fig.~\ref{1}. In all bands, the
scatter is larger at low flux
densities. Several stars show much larger deviations because
of intrinsic variability. For these, it turns out that the
PSC flux densities correspond to one epoch, whereas the flux densities
at the other epoch  
may differ by as much as a factor of 2. An example of a strongly variable source,
$\chi$~Cyg, is discussed in Appendix A.   
In the \fff\ band, large discrepancies are also found for extended sources, as
expected (see Sect.~\ref{Sect:extended}).

\subsection{The infrared color-color diagrams}
\label{Sect:IRcolor}

\begin{figure}
  \vspace{9cm}
  \vskip -0cm
  \begin{picture}(8,8.5)
    \epsfysize=9.8cm
    \epsfxsize=8.5cm
    \epsfbox{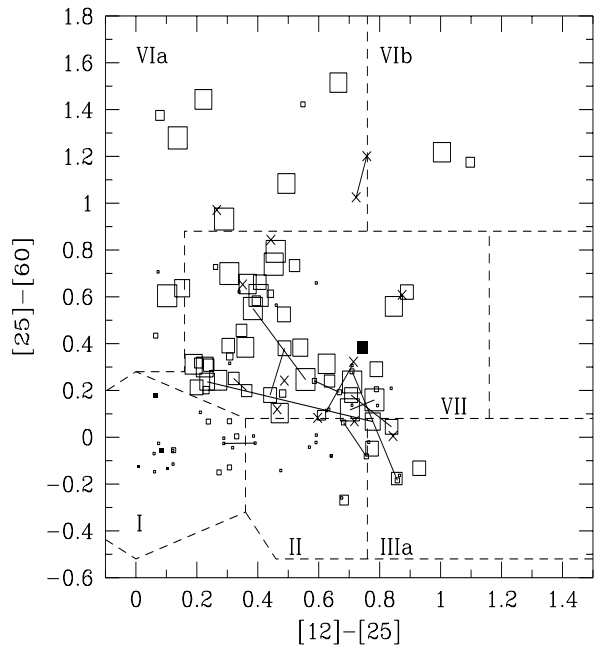}
  \end{picture}
  \vskip -0.5cm
\caption[]{\label{2}
The (\cc, \ccc) diagram for S stars. The size of the square is
proportional to the ZrO/TiO or C/O spectral index of the S star
(whichever is available), in the 
spectral classification scheme of Keenan (1954) or Keenan \&
Boeshaar (1980), respectively. MS and SC stars were assigned classes 1
and 9, respectively. Crosses correspond to stars with no spectral
type available in the GCGSS. Tc-poor S stars are represented by
filled squares, and colors of variable stars measured at different epochs are
connected by a line segment
}
\end{figure}

The use of the (\cc, \ccc) diagram to probe the
circumstellar shells surrounding late-type stars was first
demonstrated by Hacking et al. (1985) and by van der Veen \& Habing (1988,
VH).   
VH defined defined various regions (labelled I to VII)
corresponding to circumstellar shells with relatively homogeneous properties. 
Figure~\ref{2} presents the (\cc, \ccc) diagram for the
sample of S stars defined in Sect.~\ref{Sect:IRAS}. 
In this paper, the color index $[i]-[j]$ is  defined as 
$[i]-[j] = -2.5 \log [(F(i)/F_0(i))/(F(j)/F_0(j))]$ (where $F(i)$
refers to the non-color corrected flux density in the IRAS band $i$, and
$F_0(i)$ is a normalization flux density as given in the IRAS {\it Explanatory
Supplement} 1988). With this normalization, black bodies in the
Rayleigh-Jeans limit have a color index of 0. The colors for stars whose IRAS flux
densities show large variations are plotted as two symbols joined by a line
segment.    


There is a strong correlation between the position of an S star
in the color-color diagram and the intensity of the
spectral features distinctive of the S star class (as measured by the
ZrO/TiO or C/O spectral indices in the spectral
classification schemes devised by Keenan 1954 and Keenan \& Boeshaar
1980, respectively; both will be called `ZrO index' for simplicity in
the following). In particular, S stars with ZrO indices
larger than 3 are mainly found in Regions VII and VIa with a few more
in Region III, while S stars with ZrO indices smaller
than or equal to 2 are found mainly in Regions I and II. 
This segregation of S stars in the (\cc, \ccc) diagram  
is a clear indication of the inhomogeneous nature of this family of
peculiar red giants. In the center of Region I lie the S stars with
photospheric colors. These are in fact the binary and Tc-poor S stars
(`extrinsic' S stars) that  owe their chemical
peculiarities to mass transfer in a binary system (Sect.~1). 
Another group of weak S stars is found at the boundary between Regions I
and II, a zone generally devoid of stars in the (\cc, \ccc) diagram
(see e.g. VH,  and Lewis 1989). This group
includes the prototypical Tc-rich Mira S variable $\chi$ Cyg.
The few S stars located in Region II, defined by VH as
comprising O-rich stars with `young' circumstellar shells, indeed have small
ZrO indices.
Most S stars populate Region~VII, 
where C-rich circumstellar shells are found according to VH, and
it is therefore not surprising that those S stars have large ZrO indices.
Finally, several stars (among which are many of type SC) are located in
Regions VIa and b, many of them having conspicuously resolved
circumstellar shells (see Sect.~\ref{Sect:extended}).


\begin{figure}
  \vspace{9cm}
  \vskip -0cm
  \begin{picture}(8,8.5)
    \epsfysize=9.8cm
    \epsfxsize=8.5cm
    \epsfbox{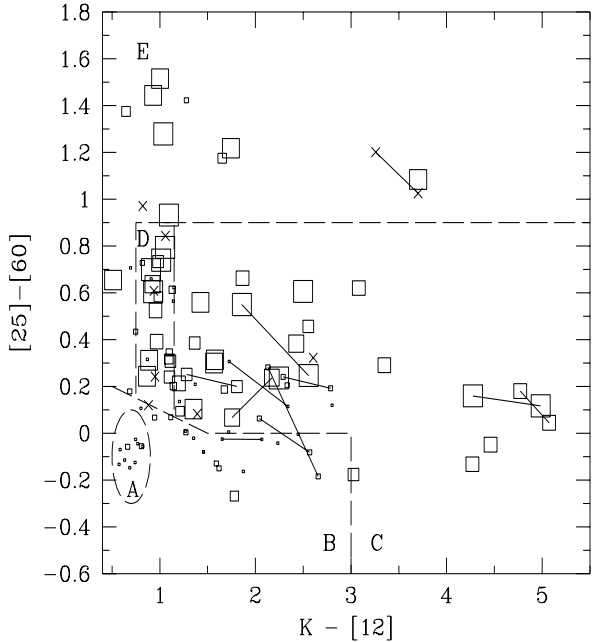}
  \end{picture}
  \vskip -0.5cm
\caption[]{\label{3}
Same as Fig.~\protect\ref{2} for the ($K - [12]$, \ccc)
diagram. Note that the $K - [12]$ index may be inaccurate for variable
stars, since their $K$ and [12]
magnitudes were not measured simultaneously
}
\end{figure}

Several authors (see e.g. Habing 1996) have argued that the  $K -
[12]$ index is superior to the \cc\ index for tracing the mass loss
rate, the focus of the present study, because of its greater wavelength
range and the fact that the 
photosphere emits more strongly at $K$ while the  shell emits more
strongly at 12~$\mu$m.
Therefore, we felt that it was more meaningful to use the 
($K - [12]$, \ccc) diagram
to define groups of S stars with homogeneous IR properties, as
follows (Fig.~\ref{3}):\\
Region A: S stars with photospheric IR colors (`extrinsic S
stars');\\
Region B: S stars with no excesses at \fff, and small ZrO indices;\\
Region C: S stars with excesses in all three \f, \ff\ and \fff\ bands,
and large ZrO indices;\\
Region D: S stars with large ZrO indices and \fff\ excesses, but small \f\
excesses;\\
Region E: mainly SC stars with large \fff\ excesses and often
resolved shells.


\begin{figure}
  \vspace{9cm}
  \vskip -0cm
  \begin{picture}(8,8.5)
    \epsfysize=9.8cm
    \epsfxsize=8.5cm
    \epsfbox{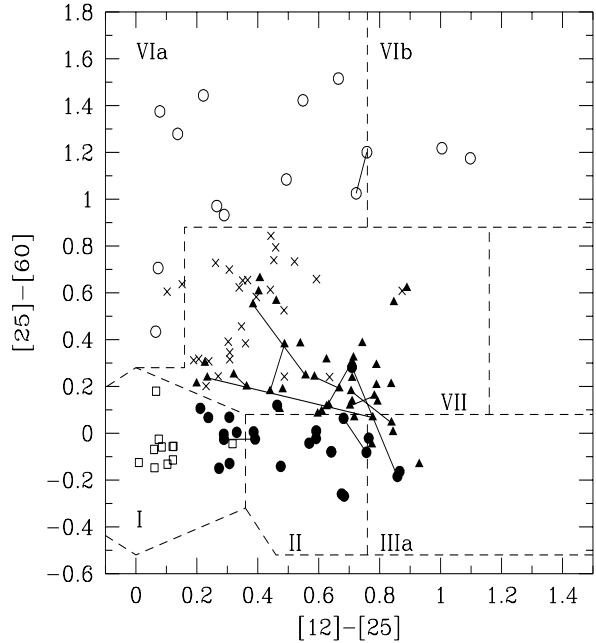}
  \end{picture}
  \vskip -0.5cm
\caption[]{\label{4}
Same as Fig.~\protect\ref{2}, where the stars have been drawn
with a symbol corresponding to the region they belong to in the ($K -
[12]$, \ccc) diagram (Fig.\protect\ref{3}), as follows:
open squares: Region A; filled circles: Region B; filled triangles:
Region C; crosses: Region D; open circles: Region E
}
\end{figure}

\begin{figure}
  \vspace{9cm}
  \vskip -0cm
  \begin{picture}(8,8.5)
    \epsfysize=9.8cm
    \epsfxsize=8.5cm
    \epsfbox{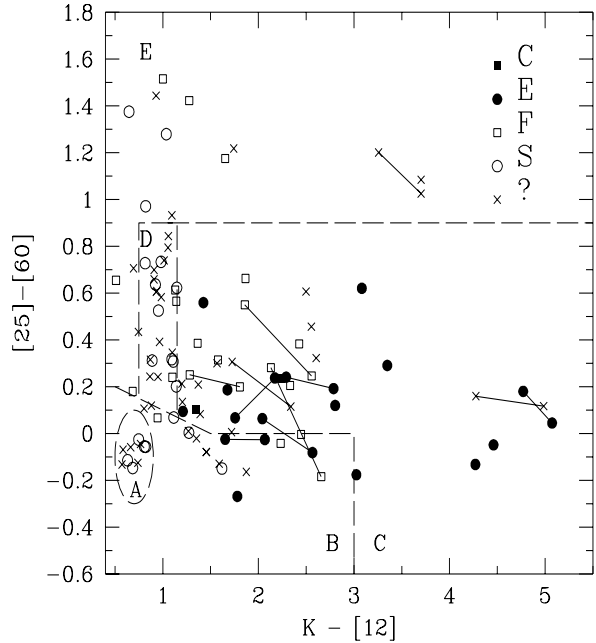}
  \end{picture}
  \vskip -0.5cm
\caption[]{\label{5}
Same as Fig.~\protect\ref{3}, with symbols referring to the
spectral class of the IRAS low-resolution spectrum  
as defined by Volk \& Cohen (1989): S: stellar
continuum; F: featureless spectrum; E: silicate emission; C: SiC
emission; ?: not available. Spectral type assignments for  
S stars are from Chen et al. (1995)  
}
\end{figure}

\begin{figure}
  \vspace{9cm}
  \vskip -0cm
  \begin{picture}(8,8.5)
    \epsfysize=9.8cm
    \epsfxsize=8.5cm
    \epsfbox{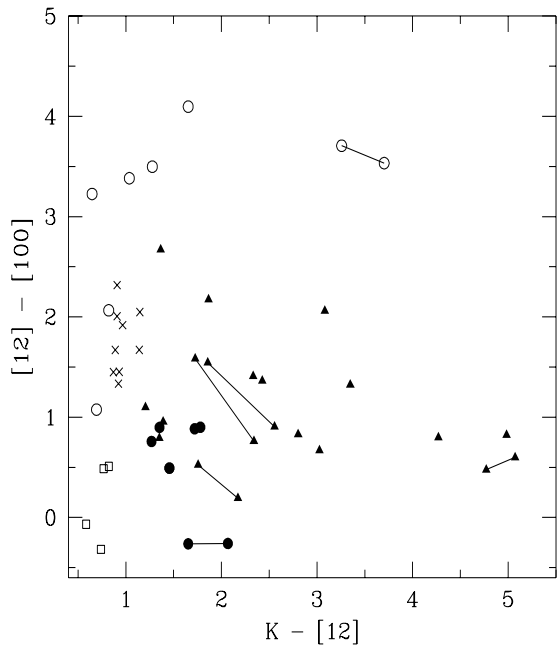}
  \end{picture}
  \vskip -0.5cm
\caption[]{\label{6}
The ($K - [12], [12] - [100]$) diagram. 
Symbols are as in Fig.~\protect\ref{4}
}
\end{figure}

The stars classified this way are plotted in the (\cc, \ccc) color-color diagram
in Fig.~\ref{4}, and it is seen that the above 
regions almost exactly correspond to those of VH, with the
exception of Region B which encompasses both Regions I and II, and
Region VII which is a blend of Regions C and D (though stars from
Region D are concentrated in the upper left of Region VII). 
The motivation for creating Region D is apparent in Fig.~\ref{5},
which presents the distribution of the various IR spectral types (as
defined by Volk \& Cohen 1989) across
the ($K - [12]$, \ccc) diagram.
Region D differs from Region C in having many stars exhibiting a stellar
continuum in the IR (class S) and none with silicate emission (class E).
As it will be shown in Sect.~4, these features suggest
that the circumstellar shells in Region D contain 
C-rich grains, whereas silicate grains are found in the circumstellar
shells of Region C. Note that Regions C and D are also quite distinct 
in the ($K - [12]$, $[12] - [100]$) diagram (Fig.~\ref{6}).


Finally, let us remark that 
the different regions defined above have been denoted by
uppercase letters to avoid confusion with the Regions a to d
defined in Paper I  from the lower-accuracy PSC flux densities.   
The new classification into Regions A--E is actually not very
different from that in Paper I (matching uppercase with lowercase letters,
and with Region d splitting into D and E), but it is clearly superior since
several stars that appeared exceptional in Paper~I fit well into the present
classification.
For instance, $\chi$ Cyg, NQ Pup and $o^1$ Ori were exceptional 
as being Tc-rich stars in Region a. With the more accurate {\sc
ADDSCAN} flux densities, $\chi$ Cyg moves to Region B (see also Appendix A)
while  NQ Pup and $o^1$ Ori move to Region E. 
The only remaining outliers in this respect are HR Peg,
the only Tc-rich star in Region A, and DY Gem, the only Tc-poor
star not in Region A (but in C).  DY Gem is however exceptional in
many other respects: it is a very cool star (S8,5 corresponding to
\Teff\ = 3000~K; Smith \& Lambert 1990) and a SRa variable with a very
long period (1145~d). Moreover, it has the largest $[12]-[100]$ index
among stars in Region~C (see Fig.~\ref{6}). 

\subsection{S stars with envelopes resolved at 60 $\mu$m}
\label{Sect:extended}

\begin{figure}[t]
\vspace {0cm}
  \vspace{9cm}
  \vskip -0cm
  \begin{picture}(8,8.5)
    \epsfysize=9.8cm
    \epsfxsize=8.5cm
    \epsfbox{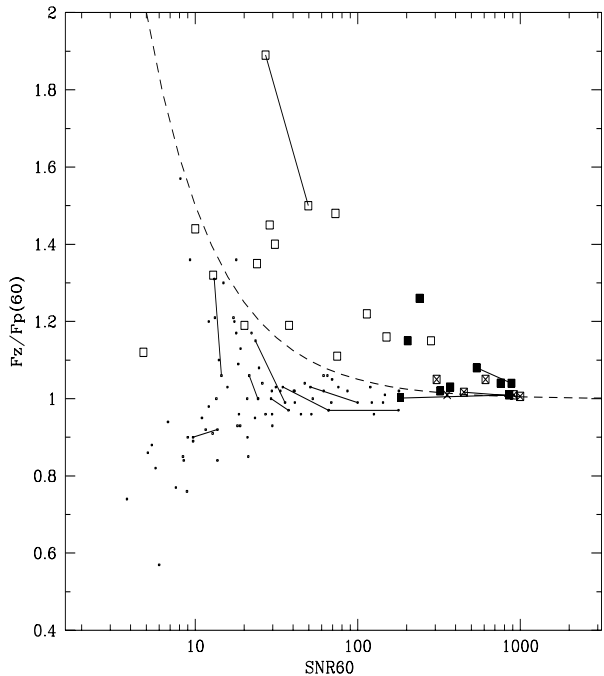}
  \end{picture}
  \vskip -0.5cm
\caption[]{\label{7}
The ratio \Fz/\Fp\ between the `zero-crossing flux density' and the `peak
flux density' in the 60~$\mu$m band,  
vs the S/N ratio along the baseline (see text for details).
Stars lying above the dashed line have a flux density excess in the \fff\ band with a
quality factor larger than 5. 
Stars with a flux density excess above the 0.3~Jy 
threshold are represented by squares. 
Squares lying above the dashed line thus satisfy both criteria 
defined in the text and
are probably truly resolved sources.
Filled squares and crosses correspond to 
sources flagged by YPK as extended or non-extended, respectively,
at 60~$\mu$m. Observations of the same star at two different epochs
are connected by a line segment.
Sources confused with a nearby source are not plotted  
}
\end{figure}

The possibility that some S stars may have circumstellar
shells resolved by IRAS is now examined. 
As argued by Young et al. (1993a; YPK),
the \fff\ band is best suited for that purpose, because it is not
contaminated by Galactic cirrus emission as severely as is
the 100~$\mu$m band.
The \fff\ co-added scans were examined, and characterized as follows.
First, the width of the template profile fitted to the source is
compared to that expected for point sources, namely 2\farcm05 and
1\farcm44 full widths at the 25\% and 50\% levels, respectively 
(Levine et al.  1993). 
However, this criterion is not sensitive to resolved shells showing
up as a weak plateau at the base of the profile.
A simple criterion has therefore been devised, based on the comparison
of the `peak flux density' $F_{\rm p}$  with the `zero-crossing flux
density' $F_{\rm z}$ (see
Sect.~\ref{Sect:IRASflux}).
These quantities are standard outputs from the {\sc ADDSCAN}
procedure. In the case of a
point source, \Fz\ and \Fp\ are identical within a few times the noise, measured
as the rms $\sigma$ of the residuals along the baseline 
outside the signal
range (i.e. between 2\farcm5 and $10'$ from the target position in
either directions) after baseline subtraction (see Sect.~\ref{Sect:IRASflux}). In
the case of an
extended source, the fraction $\epsilon$ of flux density in excess of that 
of a point source is expressed by 
$\epsilon = ($\Fz\ $-$ \Fp)/\Fp. 
For estimating the significance of this excess, one has to be aware of
the following effects. First, very bright ($> 500$ Jy) sources may
have a characteristic six-pointed star shape due to reflection from
the telescope secondary mirror struts. Since approximately 5\% of the peak
flux density may be contained in the star pattern, \Fz/\Fp\ values
of the order of 1.05 may be of instrumental origin (Levine
et al. 1993).   
Bright sources are also affected by  hysteresis in the
detector which causes a trail in the signal along the scan direction in the
outgoing part of the scan.
This trail is easily recognized on individual scans as it causes an
asymmetry in the template fit. However, if the
co-added scan results from individual scans made in opposite
directions, as is often the case, trails will be present in both
directions and will mimick an extended plateau at the base of the profile.
For bright sources ($> 100$ Jy), this effect may thus also cause spurious
excesses of the order of a few percent. Finally, for fainter
sources, an inhomogeneous background may also cause spurious
detections. In this case, a way to estimate the significance of the
excess $\epsilon$
is to compare it with the inverse signal-to-noise ratio $1/SNR = \sigma$/\Fp, 
which is nothing more than the relative flux density excess 
expected from the background noise. The significance, or `quality
factor' $QF$, of the observed
flux density excess can then be expressed as 
$QF \equiv \epsilon$ \Fp$/\sigma = \epsilon\; SNR$, 
so that $QF > 5$ for a detection at the $5\sigma$ level. Figure~\ref{7}
presents the ratio \Fz/\Fp\ vs $SNR$ for S stars not confused by nearby sources,
and several stars with $QF > 5$  are found. 
One has to be aware, however, that the co-addition process will
be mostly effective in lowering the baseline noise far away from
the target, where scans with different orientations sample 
different regions of the sky.
But since they all 
intersect approximately on the target, the noise-averaging process will not be
as effective on the target. A criterion based on the absolute noise level
present on-target has therefore been considered as well, by requiring
that, to be considered significant, an
excess \Fz\ $-$ \Fp\ should  
 not be smaller than some given threshold value of the order of the
background fluctuations in the \fff\ band.
It was found that meaningful results are obtained
by adopting 0.3~Jy as the typical background fluctuation on-target, 
combined with a quality factor of 5. 
In Fig.~\ref{7}, sources satisfying the \Fz\ $-$ \Fp\ $ > 0.3$~Jy criterion 
have been represented by open squares.
Since very few square symbols are located below
the dashed line in Fig.~\ref{7}, both criteria are fulfilled simultaneously
for most of the stars, and thus provide consistent conclusions regarding the
resolved nature of these sources.


Sources flagged as extended at \fff\ by the above criteria are
listed in Table~\ref{Tab:extended}.
Columns~1, 2 and 3 give the IRAS name, the variable star name
and variability type when available, respectively. The zero-crossing
flux density \Fz\ is in column~4. Column 5 lists the flux density ratio 
\Fz/\Fp, and columns 6 and 7 the full widths $W25$ and $W50$ 
at the 25\% and 50\%
levels, respectively, to be compared with 2\farcm05 and
1\farcm44 for a point source (Levine et al. 1993). 
The quality factor $QF$ is listed in column 8. Mass loss rates and
wind terminal velocities are given in columns 9 and   10 (see
Sect.~\ref{Sect:massloss}).

\begin{figure}
  \vspace{9cm}
  \vskip -0cm
  \begin{picture}(8,8.5)
    \epsfysize=9.8cm
    \epsfxsize=8.5cm
    \epsfbox{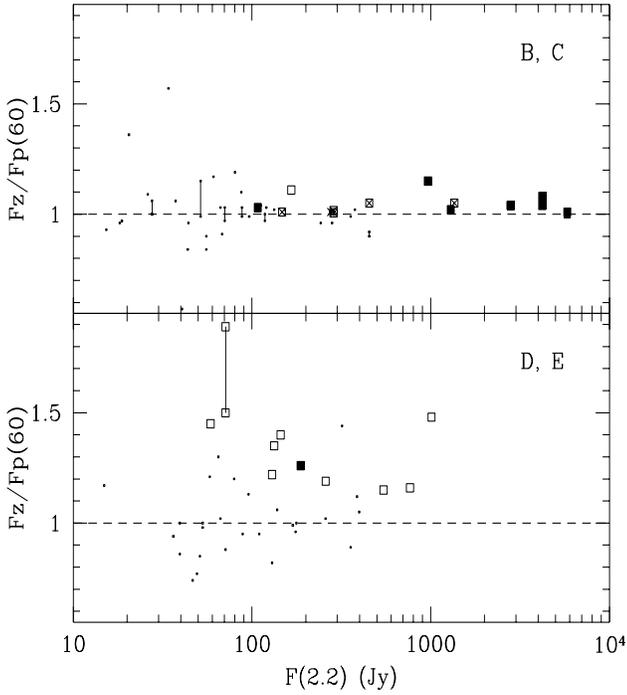}
  \end{picture}
  \vskip -0.5cm
\caption[]{\label{Fig:9}
Flux ratio \Fz/\Fp\ vs $F(2.2)$. Symbols are as in
Fig.~\protect\ref{7}, except that large squares now denote stars
fulfilling simultaneously our two criteria for extended envelopes.
Stars have been separated according to Regions B, C (upper panel)
and D, E (lower panel)
}
\end{figure}

A similar search for late-type giants with resolved shells was
performed by Young et al. (1993a,b). These authors used a more 
sophisticated method based on the possibility of successfully fitting the    
signal by a point source surrounded by a circumstellar shell having
`reasonable' properties. Our simpler 
approach has the advantage of being applicable to fainter stars, and
several have indeed been detected, as seen in
Fig.~\ref{Fig:9}. 
However, for bright sources, our method is more vulnerable to
spurious detections due to hysteresis (YPK used only the scan data
taken {\it prior} to passing over the source). 

As can be seen in
Table~\ref{Tab:extended} (in column 11, YPK+ and YPK$-$ denote 
sources flagged by YPK as extended or not at \fff, respectively),  the two methods
give conflicting results for
four stars, R And, S Cas, W Aql and T Cet among the 11 bright objects
common to the two samples. In the first three cases, 
our detection is probably an artefact due to detector hysteresis,
since {\it individual} (as opposed to co-added) 
scans show an extended tail only on the side posterior to the passage over the source
(see Fig.~\ref{11} below). The situation is
less clear for T Cet, as its 100~$\mu$m
profile is wider than the point-source template, suggesting that this
source may be truly extended.

\begin{figure}
  \vspace{9cm}
  \vskip -0cm
  \begin{picture}(8,8.5)
    \epsfysize=9.8cm
    \epsfxsize=8.5cm
    \epsfbox{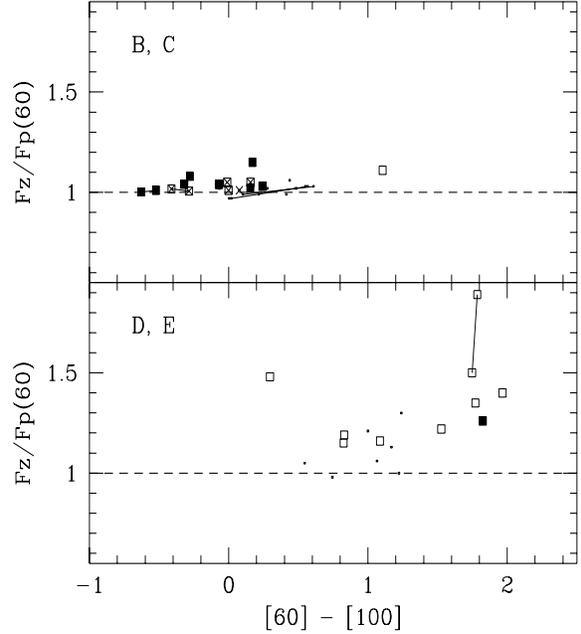}
  \end{picture}
  \vskip -0.5cm
\caption[]{\label{10}
Same as Fig.~\protect\ref{Fig:9} for \Fz/\Fp\ vs $[60] - [100]$
}
\end{figure}

Figure~\ref{Fig:9} presents the flux density excess \Fz/\Fp\ vs $F(2.2)$, the flux
density at 2.2 $\mu$m, and reveals
that the properties of the resolved
shells in Regions~B and C are very different from those of Regions~D
and E. In Regions~B and C, the flux density
excess is of the order of a few percent, with a maximum of 15\% for Y~Lyn.
Because the flux density excess is so small, the envelopes around stars in
Regions B and C can be resolved only for the stars closest to the sun
[i.e. with large $F(2.2)$], as shown by
Fig.~\ref{Fig:9}. By contrast, stars in Regions D and E have
much larger flux density excesses at \fff, which make them detectable at much 
lower total flux density levels.
Stars in Regions B and C also differ markedly from those in Regions D
and E as far as the $[60] - [100]$ index is concerned
(Fig.~\ref{10}; see also Fig.~\ref{6}): the
resolved envelopes in Regions D and E go along with large 100 $\mu$m
excesses, suggestive of cool dust in detached envelopes, contrary to
the situation prevailing in Regions B and C (the only exception being
T Sgr, which appears as a border case between C and D).  
In the case of RZ Sgr, which is extended in both the  
\fff\ (YPK) and 100~$\mu$m bands (according to the IRAS {\it Small Scale Structure
Catalogue} 1985), 
an optical nebula has even been reported by Whitelock (1994).

In Region E, $o^1$ Ori and S929 do not follow the general trend.  
In those cases, there may be a confusing background
source responsible for the strongly asymmetric \fff\ profile (see
Fig.~\ref{11}), as the 100~$\mu$m profile is offset by about
$1'$ in the direction of the \fff\ asymmetry. The extension observed for
S929 may be real; its $[60] - [100]$ color
index changed by only 2\% between the two IRAS 
observations, while the \fff\ flux density changed by 20\%. This color stability
is observed in other
variable IRAS sources like $\chi$ Cyg (see Appendix A), but would
probably not be preserved if the excess flux density were due to a background source.

Finally, the flux density excess in Regions D and E is associated with a
widening of the whole \fff\ (and sometimes 100 $\mu$m) source
profile, whereas the excess for stars in Regions B and C is caused by
a weak plateau at the base of the profile, as apparent on Fig.~\ref{11}.  
For the resolved shells in Regions D and E, the full widths at the
25\% and 50\% flux density levels (Table~\ref{Tab:extended}) are indeed much
larger than those expected for a point source. 
HD 191630 and RZ Sgr are listed in the IRAS {\it Small Scale
Structure Catalogue} as extended at \fff\ and 100~$\mu$m, respectively.  
In fact, this distinctive feature of the resolved shells in Regions D
and E has been used to include in
Table~\ref{Tab:extended} two stars (BI And and AA Cam) with wide
profiles, despite  quality factors $QF < 5$
that would normally not qualify them. However, these are distant stars
with small \fff\ flux densities, so that the \Fz/\Fp\ ratio cannot be
determined accurately (and is therefore not listed in
Table~\ref{Tab:extended}). 


\begin{figure}
  \vskip -0cm
  \vspace{9cm}
  \begin{picture}(10,9.5)
    \epsfysize=9.8cm
    \epsfxsize=8.5cm
    \epsfbox{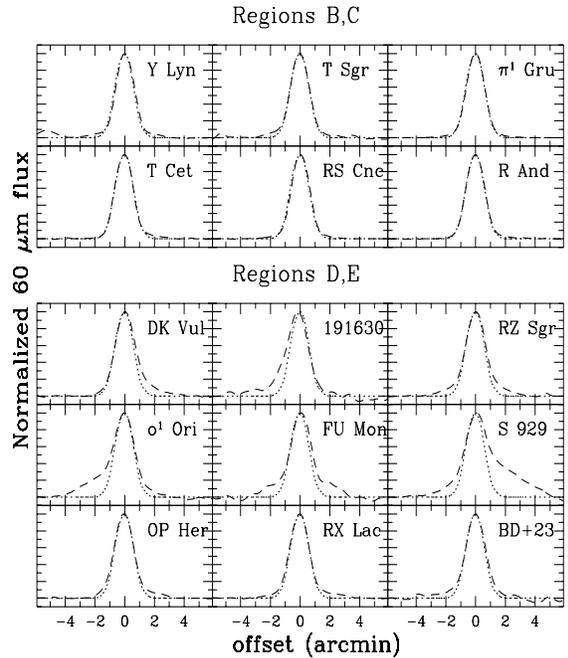}
  \end{picture}
  \vskip -0.5cm
\caption[]{\label{11}
The \fff\ co-added, spline-interpolated scans (dashed lines) for
all stars with a possibly resolved shell in Regions D -- E, and
for selected cases in Regions B -- C (see Table~2).  
The dotted line is the template 60 $\mu$m profile as provided by IPAC.
Note how larger the deviations from the template are in Regions D -- E
as compared to Regions~B -- C. In the latter case, 
the deviation of the observed profile from the template profile
is due to a weak extended tail. Such detections are
therefore vulnerable to detector hysteresis (see text)
}
\end{figure}


\begin{figure}
  \vskip -0cm
  \vspace{9cm}
  \begin{picture}(10,9.5)
    \epsfysize=9.8cm
    \epsfxsize=8.5cm
    \epsfbox{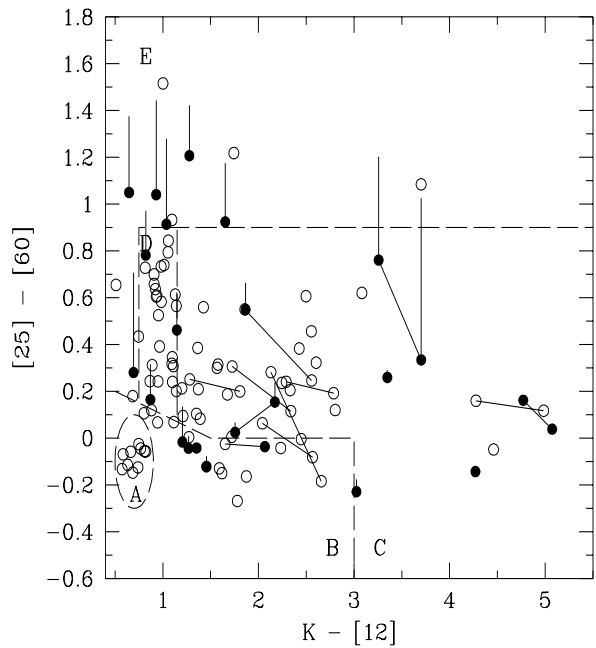}
  \end{picture}
  \vskip -0.5cm
\caption[]{\label{13}
Position of the sources with resolved shells (filled circles) in the
the ($K - [12]$, \ccc) diagram. The filled circles represent the 
color computed from the `template' flux density $F_{\rm t}$ at \fff,
i.e. they roughly correspond to the flux density of the star alone. The upper 
end of the vertical segment is located at the color computed from the
`zero-crossing' flux density \Fz\ at \fff, i.e. the
shell + star color
}
\end{figure}

Stars with resolved shells are represented as black dots in
the ($K - [12]$, \ccc) diagram 
(Fig.~\ref{13}). 
This figure illustrates the relative contribution of the extended shell to the
\ccc\ index: a line segment joins the \ccc\ indices computed
from the `zero-crossing' flux density \Fz\ at \fff\ (measuring the
combined contributions of the star and its resolved shell)
and from the `template' flux density $F_{\rm t}$, which is assumed to
be a rough measure of the stellar contribution alone (represented by a
black dot in Fig.~\ref{13}).
Although that assumption is certainly a very
rough one, it is not totally unreasonable as
stars from Region~E now move down, and  most reach Regions B, C and D
when adopting $F_{\rm t}$ instead of \Fz\ to represent the \fff\ {\it photospheric}
flux density. Some stars, however, do not quite
leave Region~E, presumably because the point source fitting yielding $F_{\rm t}$
does not entirely remove the contribution of the resolved shell in those cases.   

Especially interesting is the fact that Y Lyn and OP Her move from Regions C and
D, respectively, to Region B, which is well in line with their small ZrO index
(see Fig.~\ref{3} where they appear as outliers). With this adjustment,
all three SRc variables in our sample (RS Cnc, Y Lyn and
T Cet)
now belong to Region B, and moreover have resolved shells (tentative in the
case of T Cet).
As noted by Young et al. (1993a,b) and Habing (1996), resolved shells
are a common property of semi-regular variable stars.

Finally, it should be mentioned that several of the stars with a
shell resolved by IRAS turn out to have an extended CO shell as well,
as derived in Sect.~\ref{Sect:masslossresults} from the modelling of the CO data.
The inferred radius of the CO shell (Table~\ref{Tab:COlit}) is larger than $10''$ for
$\chi$~Cyg ($18''$), W~Aql ($23''$), $\pi^1$~Gru ($15''$) and FU~Mon ($60''$).

\section{Masers in S stars}
\label{Sect:maser}

\begin{figure}
  \vskip -0cm
  \vspace{9cm}
  \begin{picture}(10,9.5)
    \epsfysize=9.8cm
    \epsfxsize=8.5cm
    \epsfbox{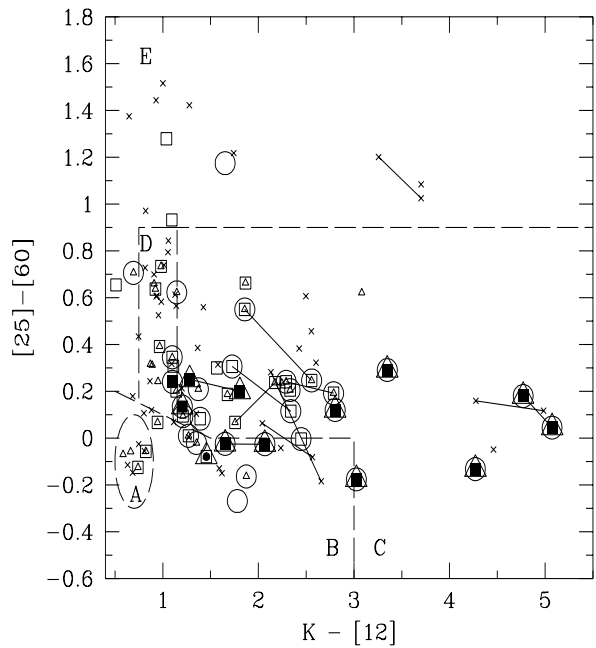}
  \end{picture}
  \vskip -0.5cm
\caption[]{\label{15}
Location of the maser sources in the ($K - [12]$, \ccc) diagram.
Positive  detections
correspond to filled symbols, non-detections to open symbols, and
stars not probed for maser emission to crosses. Squares stand for SiO
masers, triangles for H$_2$O masers and circles for OH masers 
}
\end{figure}

This section presents data collected from the literature on SiO, OH or
H$_2$O maser emission from S stars (relying on the compilation of
Benson et al. 1990 for the earlier literature). These data are summarized
in Table~\ref{Tab:maser}. The detection of SiO, OH or
H$_2$O maser
emission is a clear indication that the circumstellar shell is oxygen-rich (see
e.g. Deguchi et  al. 1989, Lewis 1996), thus providing an important constraint on
the chemical nature of the dust grains (see Sect.~4). In
Table~\ref{Tab:maser}, `Y' or `N' in the SiO, OH or H$_2$O columns means that the
corresponding maser is or is not present, respectively, whereas a dash indicates
that the maser has not been searched for in a given star.  

The masers which are found in S stars are mainly SiO masers 
forming in the densest part of the circumstellar envelope near the
stellar photosphere.
OH maser emission has been detected in RS~Cnc by Rudnitskij (1976) but not
by Kolena \& Pataki (1977), though the latter authors probed a different
transition from that (1667 MHz) detected by Rudnitskij (1976). 

The S stars with SiO maser emission 
lie in the lower part of Region~C, with some overlap with
Regions B and D (Fig.~\ref{15}). 
As expected, the region occupied by the masers exactly matches the
region delineated by the stars showing the 9.7 $\mu$m silicate feature in
emission (IRAS LRS class E;
Fig.~\ref{5}). Indeed, 6 of the 8 SiO masers with an available
LRS class are of class E, the only exceptions being R~Lyn and EP~Vul (class F).
On the other hand, no maser emission is observed for stars 
in Regions D and E (with the exception of EP~Vul, but that star lies
very close to the boundary with Region C). 

\section{A simple model of dusty circumstellar shells}

\subsection{Ingredients of the model}
\label{Sect:model}

In order to relate the diversity of IR color indices observed among  S stars to
the underlying physical parameters, synthetic IR color indices of a star embedded
in a circumstellar shell have been computed for various input parameters. In our
simple model, the star is assumed to radiate as a
black body at a temperature \Teff. The mass-losing star is surrounded by a
spherically-symmetric dust shell extending from $r_{\rm in}$ to $r_{\rm out}$,
with density decreasing as $r^{-2}$, $r$ being the distance from the central
star. This is equivalent to assuming a steady mass-loss rate
$\mathaccent 95 M$ at constant
outflow speed. The inner radius of the dust shell must be larger than or
equal to the radius where grains start condensing
(i.e., to the radius where the shell temperature drops below 1300~K for silicates,
or 1500~K for graphite and amorphous carbon).
The shell outer radius is chosen such that $r_{\rm out}/r_{\rm in} = 10^4$, which
ensures that the color indices of the shell have reached an asymptotic value.
A roughly logarithmic radial mesh is defined in the dust envelope so that each
shell is optically thin. In each shell, the grains are assumed to be in thermal
equilibrium, so that the energy absorbed by the grains exactly balances the energy
re-emitted. At the inner boundary, the radiation field is that
of a black body of temperature \Teff. The model IRAS flux densities are
calculated from the emergent spectrum by convolving it with the IRAS
filter bandpass (IRAS {\it Explanatory Supplement}, 1988).
The main shortcoming of the
code is the neglect of the scattering contribution, since only absorption
is taken into account.

Three types of dust grains have been considered: silicates (with a specific mass
of 3.5~g~cm$^{-3}$), graphite (with a specific mass of 2.25~g~cm$^{-3}$) and
amorphous carbon (with a specific mass of 1.85~g~cm$^{-3}$).
The grain radius is 0.2~$\mu$m in all cases. The absorption
coefficients as a function of wavelength are taken from Draine \&
Lee (1984) and Draine (1985) for silicates and graphite. For amorphous
carbon, the absorption coefficients have been generated  with the
usual Mie formulae using the
optical constants provided by Rouleau \& Martin
(1991).
At wavelengths $\lambda > 50\;\mu$m, the spectral index of the  emissivity 
coefficient has been taken equal to $-2$ for graphite grains, and to $-1.5$ for
amorphous-carbon and silicate grains (Ivezi$\acute{\rm c}$ \& Elitzur 1995).

\subsection{Synthetic color indices}
\label{Sect:results}

\begin{figure*}
  \begin{picture}(15,9.5)
    \epsfig{file=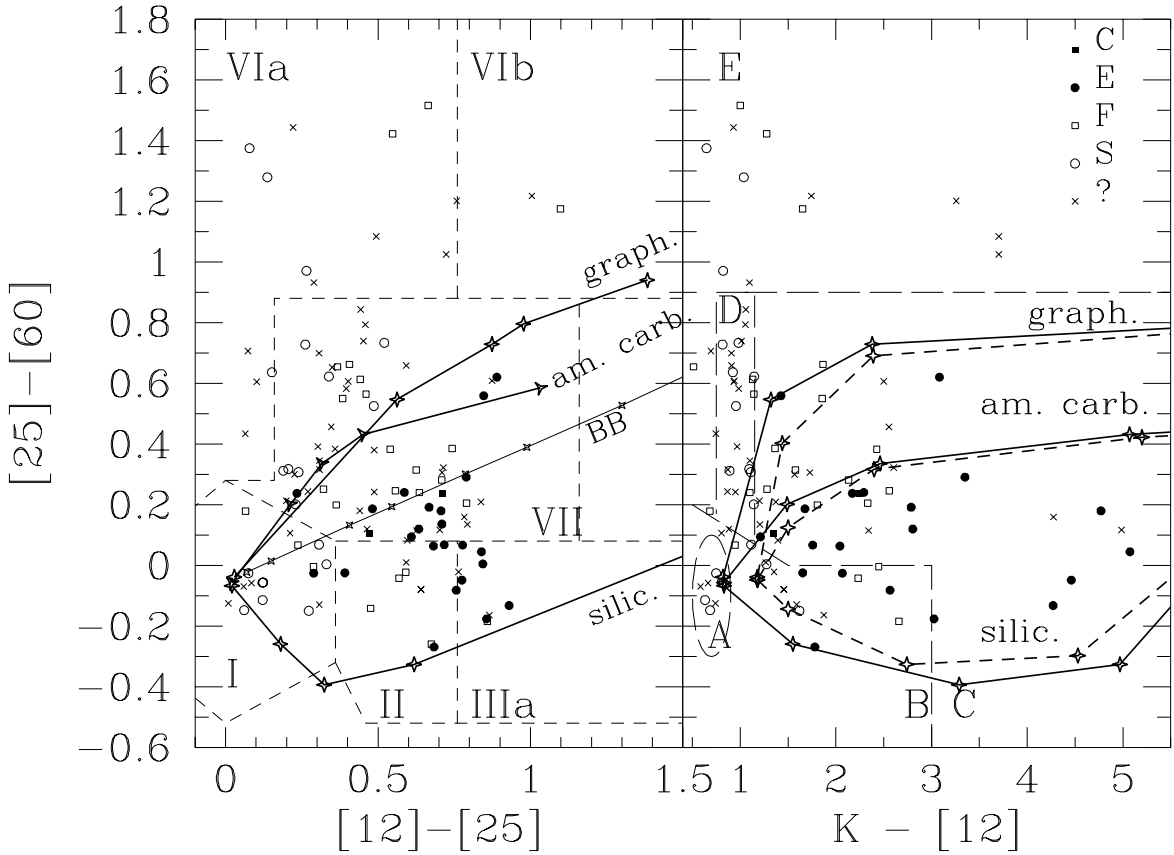,angle=270,width=17cm}
  \end{picture}
\vspace{12cm}
\caption[]{\label{16}
Left panel: The (\cc, \ccc) colors predicted for  
circumstellar shells (with $r_{\rm out}/r_{\rm in} = 10^4$) made
of silicate grains, graphite grains or amorphous-carbon grains,
surrounding a star with $L = 5000$~\Lsun\ and  \Teff $= 4000$~K (solid lines) or
3000~K (dashed lines). 
The diamonds along the curves correspond to dust mass loss rates of
$10^{-12}$, $10^{-10}$, $10^{-9}$, $10^{-8}$ and
$10^{-7}$~\Msun~y$^{-1}$ (from left to right), with a wind velocity of 14 \kms.
Black bodies fall along the central straight line labelled BB.
Right panel: same as left for ($K - [12], [25] - [60]$)
}
\end{figure*}

The results of this model 
are presented in Fig.~\ref{16} for dust shells made of either
silicate grains, graphite grains or amorphous carbon grains. 
Constant {\it dust} mass loss rates
of $10^{-12}, 10^{-10}, 10^{-9}$, $10^{-8}$ and $10^{-7}$ \Msuny\ with a wind
velocity of 14 \kms\ have been adopted. The shell inner radius is set by the
grain-condensation temperature, so that the dust shells 
computed in Fig.~\ref{16} are
not `detached' (in the sense of Willems \& de Jong 1988). The central
star has been assigned effective
temperatures \Teff\  of  4000~K
(solid line) and 3000~K (dashed line), and a luminosity $L =
5000$~\Lsun; these parameters turn out to have little impact on the
shell colors. 

In the (\cc, \ccc) and $(K - [12]$, \ccc) diagrams, the silicate track is 
quite distinct
from the graphite or amorphous-carbon tracks. 
The graphite and amorphous-carbon tracks go directly from Region I to upper VII 
(i.e., from A to D and upper C),
whereas the silicate track goes from I to II and IIIa (i.e., from A to
B and lower C). 
These differences observed in Fig.~\ref{16} between carbon- and
oxygen-rich shells must be related to the different emissivities of
silicate and carbonaceous grains in
the IRAS bands, as discussed by Ivezi$\acute{\rm c}$ \& Elitzur (1995).

The tracks for carbonaceous and silicate grains predicted by this
simple model outline the segregation observed in the
color-color diagrams between S stars with an oxygen-rich shell, as indicated
by the 9.7~$\mu$m silicate feature (IRAS LRS class E),
and S stars with 
featureless IR spectra (IRAS LRS class S -- `stellar' -- or F
-- `featureless'). It is important to note here that featureless 
spectra are indeed predicted for carbonaceous grains
(Ivezi$\acute{\rm c}$ \& Elitzur 1995). 

The data on maser emission collected in Sect.~\ref{Sect:maser} are compatible with
this segregation. 
Those S stars which are SiO or OH masers (Table~\ref{Tab:maser}) must
have an oxygen-rich circumstellar environment, and lie indeed close to the
silicate track (Fig.~\ref{15}). On the contrary,
no SiO, OH or H$_2$O maser emission has been detected for S stars in Regions D and
upper C, along the tracks corresponding to carbonaceous
grains.  Those S stars might therefore possibly 
be surrounded by C-rich circumstellar shells, especially since they
occupy the same region of the color-color diagram as the optical
carbon stars (Chan \& Kwok 1988).

However, the above picture is not totally satisfactory, since
(i) the S stars with silicate emission are actually located {\it
in between}
the silicate and carbonaceous tracks, (ii) many stars in Regions D
and E have large 60~$\mu$m excesses that cannot be reproduced by the
tracks displayed in Fig.~\ref{16}, (iii) at least one star
(R Gem) moves from the region of silicate dust (lower C) into the
region of carbonaceous dust (upper C) during its variability cycle,
and (iv) the photosphere of S stars is oxygen-rich, so that their
circumstellar shell may be expected to be oxygen-rich as well. 
The first two mismatches are in fact not specific to S stars, and 
possible solutions were already suggested by
Ivezi$\acute{\rm c}$ \& Elitzur (1995). They include either (a) 
invoking a mixture of
silicate and carbonaceous grains (see however the discussion of
Sect.~\ref{Sect:maser} on maser emission), (b) decreasing the
spectral index of the
emissivity at long wavelengths to values smaller than $-1.5$, 
or (c) considering detached envelopes in the sense advocated by
Willems \& de Jong (1988). 

As discussed in Sect.~\ref{Sect:extended}, 
the S stars located in Regions D and E have very specific
properties that may help to identify the origin of their large 60~$\mu$m
excess. First, the
prototypical SC stars GP Ori and FU Mon located in Region E
have a C/O ratio equal to unity to within 1\% (Dominy et al. 1986).  
Similarly, TT Cen is a rare CS star which
exhibits ZrO bands at some times and C$_2$ bands at other times
(Stephenson 1973). According to Stephenson (1973),
these  variations are probably
caused by temperature changes in an atmosphere with a {\it C/O ratio very
close to unity}, or perhaps even to a secular change in the atmospheric C/O
ratio. 
Most of the stars in Regions D and E are actually SC
stars or at least S stars with a large C/O spectral index. 
Finally, several stars in Region~E have very
extended shells that are resolved at 60~$\mu$m and, in one case
(RZ~Sgr), visible in the optical (Whitelock 1994).   
All these properties point towards these stars being in a very 
rare and short-lived evolutionary phase.
Based on this evidence, we suggest that the large 60~$\mu$m excess of
SC stars populating Regions~D and E
finds a natural explanation in the much debated concept of
interrupted mass loss first proposed by Willems \& de Jong (1988).
The cessation of mass loss when C/O gets close
to unity (because all C and O atoms are then locked into the CO molecule
instead of being involved in dust-forming molecules) 
causes the dust shell to detach from its parent star, and to cool down
as it expands into the interstellar medium without being replenished
at its inner side. As shown by Willems \& de Jong (1988) and Chan \&
Kwok (1988), the colors of the dust shell then describe a
counter-clockwise loop in the color-color diagram, starting from the
region of stars with silicate emission and ending close to the photospheric
point (Region A) when the shell has dissolved into the interstellar
medium, after  passing through Region E. The mass loss resumes when
C/O  reaches values above
unity, and the star then enters the region of heavily-obscured
infrared carbon stars (Chan \& Kwok 1988; lower VII in
Fig.~\ref{16}). The SC stars found in that region (RZ Peg, UY
Cen and BH Cru, a sister case of TT Cen uncovered by Lloyd Evans 1985) 
may actually be on the lower part of that loop. 
The application of the Willems \& de Jong scenario (implying that the
loop described in the color-color diagram corresponds to a brief
evolutionary phase) to the
rare, supposedly  short-lived SC phase would thus not face 
the difficulty of inconsistent time scales 
generally used to argue against it (e.g., Zuckerman \& Maddalena 1989).  
Further support for this idea comes from
the peculiar CO line profiles observed for many of the stars
populating Region E (FU Mon: see
Fig.~\ref{18} below; DK Vul, RZ Sgr, TT Cen and UY Cen: Sahai
\& Liechti 1995),  as discussed in
Sect.~\ref{Sect:Mdotcolors}.

\section{Molecular line data and mass loss rates}
\label{Sect:massloss}
\subsection{New CO data}

Observations of the CO(2$-$1) line at 230 GHz and of the CO(3$-$2)
line at 345 GHz were made in January 1996 and January 1997 with the 10.4 m
telescope of the Caltech Submillimeter Observatory (CSO) on Mauna Kea,
Hawaii. The CSO is equipped with SIS
junction receivers cooled to liquid helium temperatures.  The effective 
single-sideband system temperatures for these observations, including the
effects of atmospheric emission and absorption, were about 500 K and 800 K
at 230 and 345 GHz, the telescope half-power beamwidths were respectively
$\rm 30''$ and $\rm 20''$ and the main-beam efficiencies 76\% and 65\%.

The spectra were observed using two 1024 channel acousto-optic spectrographs
(AOS) simultaneously.  The first has a total bandwidth of $\sim$ 500 MHz 
($\rm \sim 800 ~ km~s^{-1}$ at 230 GHz) and a velocity resolution of 
$\rm \sim 1 ~ km~s^{-1}$.  The second has a bandwidth of $\sim$50 MHz 
and a velocity resolution of $\rm 0.1 ~ km~s^{-1}$.  The observations were 
made by chopping between the star position and an adjacent sky position,
offset $\rm 90''$ in azimuth, at a rate of 1 Hz, and consisted
of pairs of chopped observations with the source placed alternately in 
each beam.  The spectral baselines resulting from this procedure are linear
to within the r.m.s. noise.  The telescope pointing errors were measured by 
mapping the spectral line emission from a nearby CO bright star before 
each observation was made, and the pointing accuracy is better than $\rm
\sim 5''$ for all of the observations.

The temperature scale and atmospheric opacity were measured by chopping against
a hot (room temperature) load.  The line temperature was corrected for the 
main-beam efficiency, and the resulting scale is the Rayleigh-Jeans
equivalent main beam brightness temperature $T_{\rm MB}$, i.e. that 
measured by a perfect 10.4 m antenna above the atmosphere.  The spectrometer
frequency was calibrated using an internally generated frequency comb,
and the velocity scale is corrected to the Local Standard of Rest (LSR).

\begin{figure}
  \vskip -0cm
  \begin{picture}(10,9.5)
    \epsfig{file=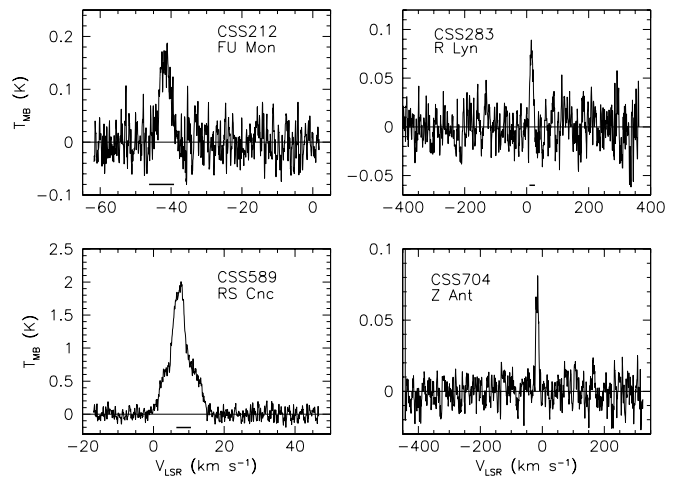,angle=270,width=9cm}
  \end{picture}
  \vspace{7cm}
\caption[]{\label{17}
CO(2-1) line profiles of four S stars observed at CSO.
The abscissa is velocity with respect to the LSR and the ordinate main
beam brightness temperature.  The profiles for GCGSS 212 and 589 are observed
with a velocity resolution of $\rm \sim 0.1 ~ km~s^{-1}$, while GCGSS 283
and GCGSS 704 are observed with 1 $\rm km~s^{-1}$ resolution. The horizontal
bars show the velocity range observed in optical spectra (Table 4).  There
is no optical velocity available for GCGSS 704
}
\end{figure}

Twelve S stars were observed at the CSO (Table~\ref{Tab:COstars}), 
nine in the CO(2--1) line and three in the CO(3--2)
line, and emission was detected from four (Table~\ref{Tab:COresults}).
The line profiles for the detected stars are shown in Fig.~\ref{17}.

Table~\ref{Tab:COstars} lists the stars that were observed:
the GCGSS number 
and the variable star name are in columns 1 and 2, 
respectively.
Next is the observed position: we used positions accurate to 
$\rm \sim 1''$ from the HST Guide Star Catalog and other sources (see 
Chen et al. 1995).  Columns 5--9 list the galactic longitude and latitude,
the spectral type from the GCGSS, the variable type and the period
from the GCVS.  Column 10 gives the stellar radial velocity with
respect to the LSR measured from optical spectra; since the radial velocities
of red giants vary as they pulsate, Table~\ref{Tab:COstars} lists the range of
reported radial velocities. Finally, column 11 gives the channel-to-channel
r.m.s. noise in the 500 MHz AOS.

\begin{figure}
  \begin{picture}(10,9.5)
    \epsfig{file=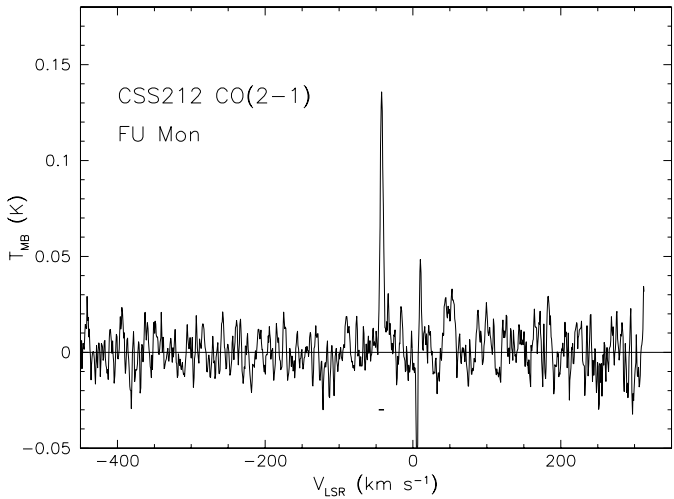,width=7cm,angle=270}
  \end{picture}
  \vspace{7cm}
\caption[]{\label{18}
Broad-band CO(2--1) spectrum in the direction of GCGSS 212 (FU Mon).
The horizontal bar shows the range of the observed optical velocities
}
\end{figure}

Table~\ref{Tab:COresults} gives the line parameters of the four detected
stars: the CO(2--1)
line flux in $\rm K \times km~s^{-1}$, the peak line temperature, the central
velocity $V_{\rm c}$ and the half-width of the line at zero power, $V_{\rm e}$,
which gives the terminal wind outflow speed.  These quantities are determined 
by fitting a parabolic line model to the data.  The agreement between the 
optical and CO radial velocities is good.

Seven of the stars in Table~\ref{Tab:COstars}, 
GCGSS 89, 117 (GP Ori), 422 (NQ Pup) 626 (FM Hya), 704 (Z Ant), 796 (HR 4755)
and 803 (S UMa) have not previously been observed.  We detect one of these,
Z Ant.  GCGSS 816 (UY Cen) is weakly detected by Sahai \& Liechti (1995, SL95)
with good agreement between the CO and optical radial velocities.  This
star was not detected in the present observations, but our sensitivity 
is lower.  The detection of circumstellar CO for this star is of 
particular interest, given its rare SC spectral type
(see Sect.~\ref{Sect:Mdotcolors}). GCGSS 149
(NO Aur) and 796 (HR 4755) were previously observed by SL95 and by Bieging \&
Latter (1994, BL94); like the present observations, these did not detect
CO emission.

RS Cnc has been observed by many authors (see Loup et al. 1993, for example).
Margulis et al. (1990) point out that the CO line profile for this star, as
for several others, more closely resembles a triangle or gaussian in
shape than the parabolic profile typical of circumstellar winds. 
The high velocity resolution observations in Fig.~\ref{17} show that
the line profile actually consists of two parabolic components of different
widths centered at the same velocity.  The parameters for these profiles,
estimated by eye, are given in Table~\ref{Tab:COresults}.  
This line shape may indicate the
presence of two molecular winds. Such line profiles have
been seen for several other stars (e.g. Margulis et al. 1990,  SL95,
Kahane \& Jura 1996, Knapp et al. 1997a) and may be quite common.

GCGSS 283, R Lyn, has previously been detected by BL94, and the data
in Table~\ref{Tab:COresults} 
are in good agreement.  FU Mon (GCGSS 212) was detected by SL95, who observe
two narrow features at $-$44 and +14 $\rm km~s^{-1}$ which they attribute to
the blue- and red-shifted components of an expanding circumstellar shell.
However, the optical radial velocity of the star is about $-43$
$\rm km~s^{-1}$ (Table~\ref{Tab:COstars}), 
and further, the narrow component at +14
$\rm km~s^{-1}$ is likely to be interstellar, as shown by our observations
with the 500 MHz AOS in Fig.~\ref{18}.  We conclude that the emission
at $\rm -44 ~ km~s^{-1}$ (best seen on the high resolution profile
in Fig.~\ref{17}) 
is from the circumstellar envelope.  The outflow speed, 
3 $\rm km~s^{-1}$, is very low, but similar low values are found for
some other Mira and semi-regular variables (Young 1995; Kerschbaum et
al. 1996).

\subsection{CO data from the literature}

Table~\ref{Tab:COlit} 
summarizes (under the header `Observations') 
CO millimeter wavelength observations of S stars 
published since 1990, including those in the present paper.  Data
observed prior to 1990 can be found in Loup et al. (1993).  The stars in
Table~\ref{Tab:COlit} are grouped according to their location in the
($K - [12]$, \ccc) color--color
diagram (see Sect.~\ref{Sect:IRcolor}) 
and arranged in order of right ascension within
these groups.  Several observations are not listed in
Table~\ref{Tab:COlit}  because they
were made at positions which are too discrepant from the optical
position;  the stars are  GCGSS 133, GZ Peg, and T Cam.

Table~\ref{Tab:COlit} gives the IRAS name (an asterisk before the IRAS
name refers to a note at the end of the table),
the star name and the results of CO observations of the star:
the line observed; the telescope half power beamwidth in arcseconds;
the channel-to-channel r.m.s. noise in K; the integrated line brightness
in $\rm K \times km~s^{-1}$; the peak brightness temperature in K;
the central velocity $V_{\rm c}$ with respect to the LSR; the wind outflow
speed $V_{\rm e}$; and the reference.  All temperatures are expressed in
main-beam brightness temperature.  Dashes for any of these quantities 
mean either that no emission was detected from the star or that the quantity
in question was not quoted in the paper.  Table~\ref{Tab:COlit}
contains observations of 56 stars, with 35 detections.

To first order, the peak brightness temperatures and the integrated CO
line intensities should scale inversely with the square of the telescope
beamwidth, since the CO lines are usually fairly optically thick and the
envelopes in general smaller than the beam.  As Table~\ref{Tab:COlit} 
shows, this is 
roughly the case, and the agreement among the observations is in general
good, with no serious discrepancies.

\begin{figure}
  \begin{picture}(10,9.5)
    \epsfig{file=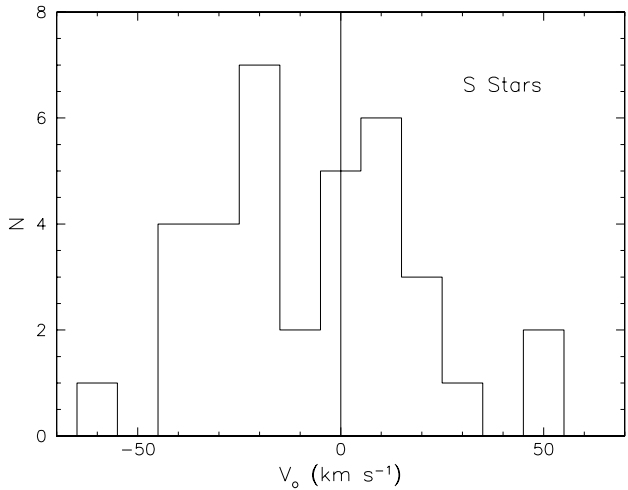,angle=270,width=9cm}
  \end{picture}
  \vspace{7cm}
\caption[]{\label{19}
Distribution of radial velocities (with respect to the LSR)
from CO observations of intrinsic S stars
}
\end{figure}

Figure~\ref{19} shows the histogram of the stellar systemic
radial velocities (with respect to the LSR). 
The values adopted for stars with multiple observations are straight
averages of the individual values.  The mean velocity for 
S stars detected in CO 
is $\rm <V> ~=~ -9.1 \pm 5.0 ~ km~s^{-1}$ and the
radial-velocity dispersion is $\rm \sigma ~ = ~ 25.9 \pm 4.5 ~
km~s^{-1}$. This value refers to a sample of intrinsic S stars,
since no extrinsic S stars have been
detected in CO. This dispersion is typical of a young-disk population 
(Mihalas \& Binney 1981). Such a population
should have a scale-height above the galactic plane of about 200 pc,
in agreement with the results of Van Eck et
al. (1997).

\begin{figure}
  \vskip -0cm
  \vspace{9cm}
  \begin{picture}(10,9.5)
    \epsfysize=9.8cm
    \epsfxsize=8.5cm
    \epsfbox{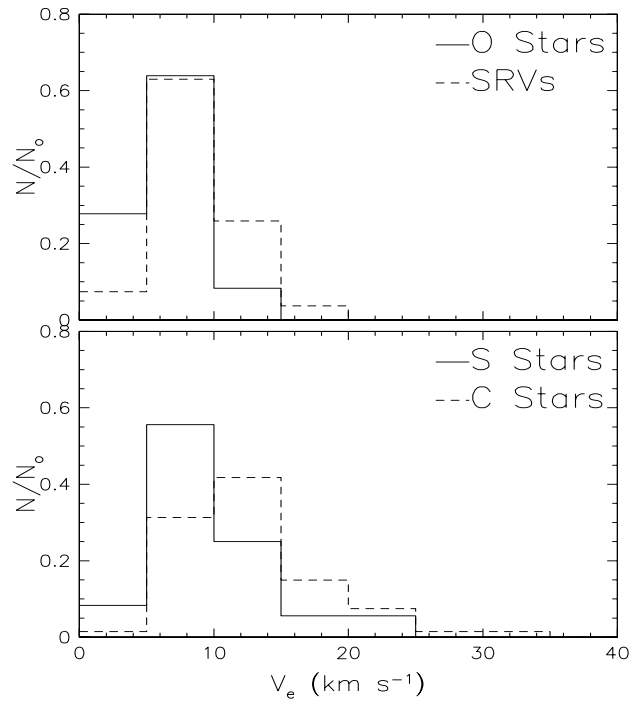}
  \end{picture}
  \vskip -0.5cm
\caption[]{\label{20}
Normalized histogram of terminal wind outflow speeds
of (a) oxygen-rich Miras (Young 1995) and semiregular variables
(Kerschbaum et al. 1996) and (b) S stars (present work) and carbon stars
(Olofsson et al. 1993)
}
\end{figure}

Figure~\ref{20} 
shows the histogram of the wind terminal velocity $V_{\rm e}$
compared with the distributions for three other sets of molecular
line observations; those for nearby oxygen-rich Mira variables
(Young 1995), for semi-regular (SRa and SRb) variables (Kerschbaum et
al. 1996) and for carbon stars (Olofsson et al. 1993).
The outflow speeds $V_{\rm e}$ for the S stars are taken from 
Table~\ref{Tab:COlit}.  We used average values for stars with 
multiple observations except for RZ Sgr (20120$-$4433), for
which CO(1--0) and (2--1) observations give discrepant values
(14 and 8.8 $\rm km~s^{-1}$; SL95).  We use the velocity derived
from the CO(2--1) observation since this has a much higher 
signal-to-noise ratio, but note that the CO(1--0) line may really be 
broader; the larger telescope beam at this wavelength could be
detecting gas at larger distances from the extended envelope of this
star, which could have a larger outflow velocity.

\begin{figure}
  \vskip -0cm
  \vspace{9cm}
  \begin{picture}(10,9.5)
    \epsfysize=9.8cm
    \epsfxsize=8.5cm
    \epsfbox{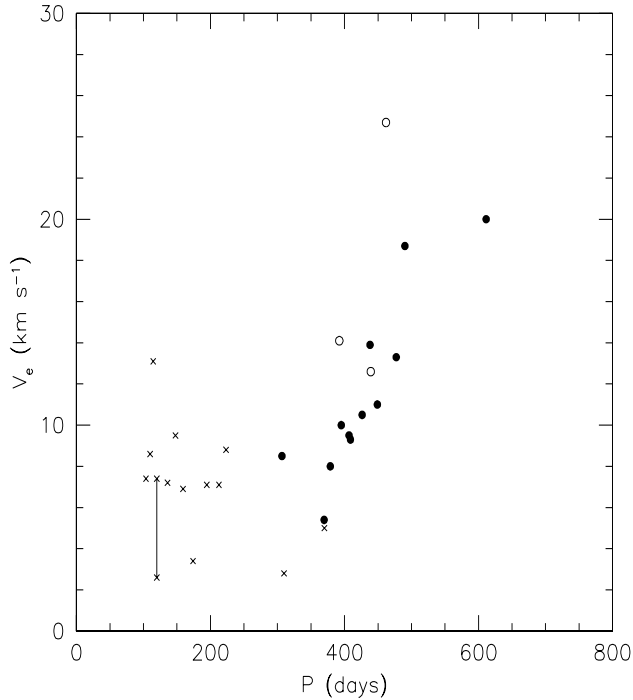}
  \end{picture}
  \vskip -0.5cm
\caption[]{\label{21}
Terminal wind outflow speed of S stars measured from CO millimeter 
wavelength emission
lines versus period of photometric variations (from the GCVS). 
Circles: Mira variables (open circles are for stars for which there 
is only one measurement of $V_{\rm e}$); crosses: semi-regular variables. The
points corresponding to the two outflow speeds for RS Cnc are connected
by a vertical line.  The data for DY Gem, 06331+1414 (SRa; $P = 1145$~d),
are not included in the figure
}
\end{figure}

Figure~\ref{20} shows that oxygen-rich Miras have the smallest outflow
velocities (median 6.5 $\rm km~s^{-1}$, largest value 12.7 $\rm 
km~s^{-1}$), while those of the SRVs cover a similar range (median
8.0 $\rm km~s^{-1}$, largest value 15.6 $\rm km~s^{-1}$).  Carbon
stars have the largest outflow velocities (median 12.0 $\rm km~s^{-1}$,
largest value 33.2 $\rm km~s^{-1}$) while as expected S stars are
intermediate (median 8.5 $\rm km~s^{-1}$, largest value 24.7
$\rm km~s^{-1}$). The largest  $V_{\rm e}$ in our sample of S stars
is observed for the CS star TT Cen, a border case between S and C stars 
(see Sect.~\ref{Sect:IRAS}).
Figure~\ref{21} shows that, among Mira S stars, the outflow velocity
correlates well with the period of the photometric variations, a
result already discussed by Heske (1990) and by Olofsson et al. (1993).
Jura (1988) finds an almost identical dependence on period for the ratio of
flux densities at 25 and 2.2 $\mu$m.

\subsection{Mass loss rates}
\label{Sect:masslossresults}

The CO data from Table~\ref{Tab:COlit} 
were used to calculate mass loss rates for the
detected S stars and upper limits for the non-detected stars.
The CO lines were modeled
using a code based on that of Morris (1980) which assumes 
spherically-symmetric mass loss at a constant rate and constant
outflow 
speed\footnote{The mass loss rates estimated in the present paper
might therefore be inadequate for those stars in Regions D and E with
a resolved -- possibly detached -- shell as observed at 60 $\mu$m}
with excitation by collisions and by infrared photons at
4.6 $\rm \mu m$.  The envelope
outer radius was taken to be that at which the CO is photodestroyed by
the diffuse interstellar radiation field using the calculations of
Mamon et al. (1988).  The details are given by Knapp
et al. (1997b).  The relative abundance of CO to $\rm H_2$ was assumed to
be $\rm 6.5 \times 10^{-4}$ for all stars (Lambert et al. 1986;
Smith \& Lambert 1990).

The infrared radiation field was approximated as that of a black body
of temperature 2500 K and radius $\rm 2.5 \; 10^{13}$ cm. Models show that
the CO line strength is only weakly dependent on the radiation
field, so this simplifying assumption is unlikely to produce an
uncertainty of more than 20\% in the derived mass loss rates.

The distances are derived by adopting absolute magnitudes which
depend on location in the  IR color-color diagram. 
For stars in Region A (photospheric colors) we assume $M_{\rm bol}
 ~ = ~ -2$, i.e. one magnitude below the RGB tip for solar-metallicity
stars with $M = 1$~M$_{\odot}$ (Schaller et al. 1992).  The bolometric
correction $M_{\rm bol} ~ - ~ M_{\rm K}$ is derived from 
the apparent bolometric
flux for BD Cam (HR 1105) 
obtained by integrating the flux densities corresponding to
the $UBVRIJKL$ magnitudes from Lee (1970) plus the IRAS flux densities.
The corresponding absolute $K$ magnitude is $-4.6$.  Other authors, e.g. Jura
(1988) have used $M_{\rm K} = -8.1$ for all S stars; this absolute
magnitude is derived from carbon stars in the solar neighborhood
and AGB stars in the Magellanic Clouds.

For stars in the other regions, which are supposedly more evolved than
are those in Region A, we adopt $M_{\rm K} = -8.1$.
That value of $M_{\rm K}$ yields distance moduli that are consistent  
with direct determinations when available.
For the composite system $\pi^1$ Gru (S5,7e + G0V), 
Ake \& Johnson (1992) derive a distance modulus of
6.0 from a fit to the UV spectrum, corresponding to a distance of 160 pc, 
identical to the value derived from
the $K$ magnitude (Table~\ref{Tab:IRAS}). For T Sgr, a distance of 1000 pc (as
compared to 810 pc from the $K$ distance modulus) is derived by Culver \& Ianna
(1975) from the spectral type F3IV assigned to its companion.
For $\chi$ Cyg, a distance of 136 pc is obtained from the distance
modulus in the $K$ band, consistent with that (106 pc) 
derived from the Hipparcos parallax ($\pi = 9.43 \pm 1.36$~mas; van Leeuwen 
et al. 1997).

The evolutionary status of stars in Regions D and E is unclear, as is
their relationship to stars in other regions of the color-color diagram,
and so their absolute magnitudes are uncertain.  Since these are generally
stars of spectral type SC, the choice $M_{\rm K} = -8.1$ appears
to be a reasonable one.

We modeled the wind from RS Cnc as two separate components, fit only to 
the CO(2--1) observations described in the previous section. 

The upper limits to the mass loss rate of S stars not detected in CO 
were calculated using the median outflow speed of
8.5 $\rm km~s^{-1}$ found for the detected sample and assuming
that we can detect a line of brightness temperature three times the
r.m.s. noise (examination of the data in Table~\ref{Tab:COlit} 
suggests that this is reasonable).

The best fit mass loss rates were found by calculating the model peak
antenna temperature and integrated line intensities for a given input
mass loss rate and comparing it with the observations.  The mass loss rates
were adjusted until reasonable agreement with all of the observations
was found, and were corrected for the mass of helium (see Knapp \& Morris 1985).

The results are listed in the rightmost columns of
Table~\ref{Tab:COlit} (under the header `Model'),  which list 
the distance, the mean outflow speed (the same for all lines; the
value of 8.5 $\rm km~s^{-1}$ used for calculating the upper limits
is given in parentheses), the CO photodissociation radius in cm and in arcseconds,
and the predicted CO peak line temperature and integrated CO
line intensity for each of the observed lines.  
Comparison between the calculations and observations
shows that uncertainties in the observations introduce about a
factor of two uncertainty into the mass loss rate.

\section{Discussion}

\subsection{Comparison with mass-loss rates of Oxygen and Carbon stars}

\begin{figure}
  \vskip -0cm
  \vspace{9cm}
  \begin{picture}(10,9.5)
    \epsfysize=9.8cm
    \epsfxsize=8.5cm
    \epsfbox{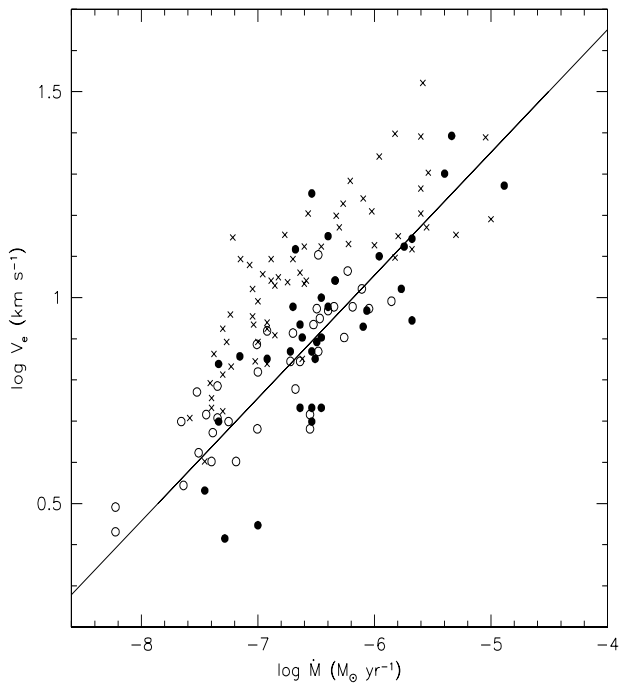}
  \end{picture}
  \vskip -0.5cm
\caption[]{\label{22}
Wind outflow speed $V_{\rm e}$ versus mass loss rate 
for: open circles: oxygen-rich Miras (Young 1995); filled circles:
S stars (current work); crosses: carbon stars (Olofsson et al. 1993).
The relationship for oxygen Miras found by Young (1995) is also shown
}
\end{figure}

Several authors (e.g. Netzer \& Elitzur 1993; Young 1995) have 
shown that AGB stars with higher mass loss rates have higher
outflow velocities.  The corresponding relationship for S stars
is shown in Fig.~\ref{22} and compared with the results for oxygen-rich
Miras (Young 1995) and carbon stars (Olofsson et al. 1993).  The
relationship between $V_{\rm e}$ and $\mathaccent 95 M$
found for oxygen stars by Young (1995) is also shown.

It is not clear at present whether the relationship between
$V_{\rm e}$ and $\mathaccent 95 M$ demonstrated by these data
is real, in the sense that it has a physical origin, or is
due to selection effects.  For the present purpose, we will
simply treat it as a convenient way to display the data and compare
samples.  The mass loss rates in Fig.~\ref{22} have been calculated 
assuming a CO abundance of $f = n({\rm CO})/n({\rm H}_2)  = 3 
\times 10^{-4}$ for oxygen stars (Young 1995), $\rm 6.5
\times 10^{-4}$ for S stars and $\rm 9 \times 10^{-4}$
for carbon stars (Olofsson et al. 1993).  The S and oxygen stars
show essentially the same correspondence between $\mathaccent
95 M$ and $V_{\rm e}$ (though there is a lot more scatter in the
S star data), giving some confidence in the assumed value
of $f$ and the resulting mass loss rates.  The apparently higher
values of $V_{\rm e}$ for a given mass loss rate for carbon stars
(see also Fig.~\ref{20}) will be discussed elsewhere.

\subsection{Comparison of mass loss rates and color excesses}

\begin{figure}
  \vskip -0cm
  \vspace{9cm}
  \begin{picture}(10,9.5)
    \epsfysize=9.8cm
    \epsfxsize=8.5cm
    \epsfbox{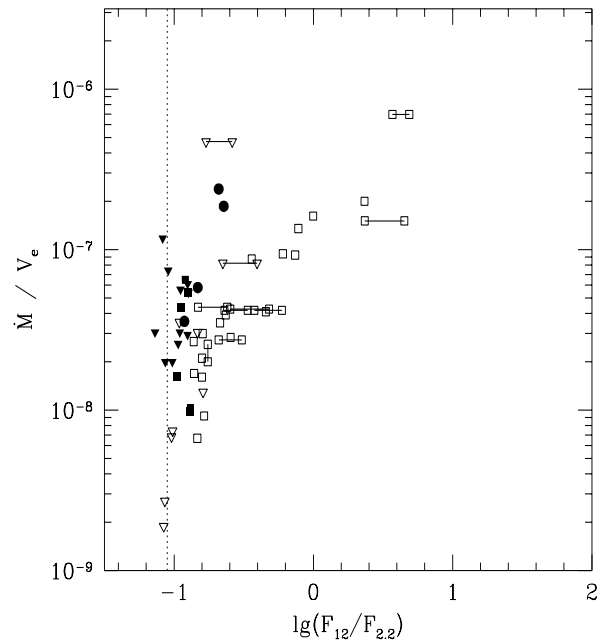}
  \end{picture}
  \vskip -0.5cm
\caption[]{\label{23}
Envelope density $\mathaccent 95 M/V_{\rm e}$ (measured
in $\rm M_{\odot}~y^{-1}$ and $\rm km~s^{-1}$) versus the ratio of 12$\rm
\mu m$ to 2.2$\rm \mu m$ flux densities.  The vertical dotted line shows 
the photospheric color for a star of temperature 3000 K. Open symbols:
stars in Regions A, B and C of the IR color-color diagram
(Fig.~\protect\ref{3});  
filled squares: stars in Region D; filled circles: stars in Region E.
The inverted triangles show upper limits
}
\end{figure}

The color excess can also be used to give the mass loss rate if the
gas-to-dust ratio and the outflow speed are known.  This is
illustrated in Fig.~\ref{23}, which shows the envelope density 
$\mathaccent 95 M/V_e$ versus the 12$\rm \mu m$/2.2$\rm \mu m$
color.  At low mass loss rates ($\mathaccent 95 M ~ \leq ~
10^{-7}$~M$_{\odot}$~y$^{-1}$) the broad-band colors are dominated
by the colors of the photosphere, as expected, while the stars with
higher mass loss rates show a strong correspondence between the
mass loss rate and the 12 $\rm \mu m$ color, showing that the gas
to dust ratio in these envelopes is roughly constant.  The
relationship in Fig.~\ref{23} is in agreement with that
for oxygen stars shown by Habing (1996).

The two stars TT Cen and RZ Sgr in Region E markedly depart from this
relationship, however. These are probably in a transitory  
phase of evolution; TT Cen is a rare CS star where
ZrO bands seem to have disappeared while C$_2$ bands appeared
(see Sect.~\ref{Sect:IRAS} and Stephenson 1973), while RZ Sgr is
surrounded by an optical (Whitelock 1994)
and IR (YPK) nebula. 

Stars with roughly photospheric 12$\rm \mu m$/2.2$\rm \mu m$
colors show a wide range in envelope densities.  This large 
scatter may be due to the imperfect coupling between dust and
gas at these low densities, which sets a lower limit to the 
mass loss rate for a radiation-pressure driven wind (cf. 
Netzer \& Elitzur 1993; SL95).  Empirically, this limit is a few
$\rm 10^{-8}$~M$_{\odot}$~y$^{-1}$ (Fig.~\ref{23} and
Table~\ref{Tab:COlit}).

\subsection{Mass loss rates and the IRAS color-color diagram}
\label{Sect:Mdotcolors}

\begin{figure}
  \vskip -0cm
  \vspace{9cm}
  \begin{picture}(10,9.5)
    \epsfysize=9.8cm
    \epsfxsize=8.5cm
    \epsfbox{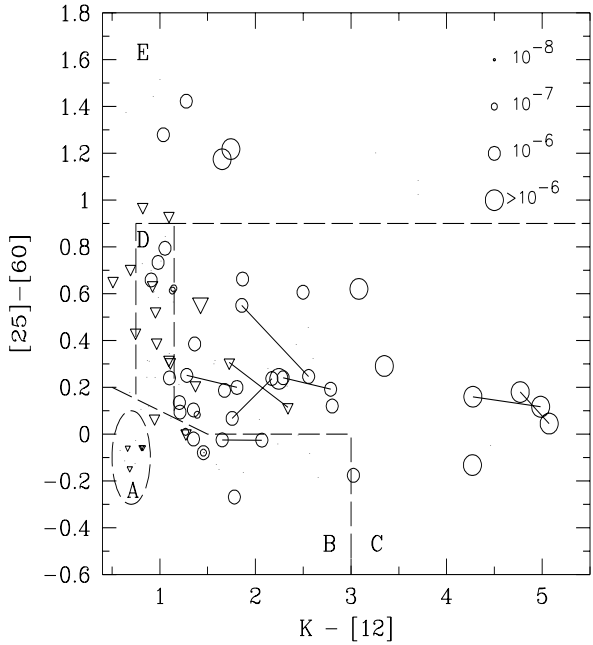}
  \end{picture}
  \vskip -0.5cm
\caption[]{\label{24}
Mass loss rates of S stars (as measured from CO) in the ($K - [12]$, \ccc)
diagram. The diameter of
the circle is proportional to the mass loss rate, as labeled.
Inverted triangles correspond to upper limits
}
\end{figure}

Figure~\ref{24} presents the mass loss rates of S stars
as a function of their location in the ($K - [12]$, \ccc) diagram.
None of the stars in Region A (stars with
photospheric colors) has detectable circumstellar CO emission, with limits
on the mass loss rates of $\rm < 6 \times 10^{-8} ~ M_{\odot}~y^{-1}$
and envelope densities well below those of the detected stars 
(Fig.~\ref{23}).
This confirms the lower limit at which a star can lose mass by a 
radiation-pressure driven wind estimated by Netzer \& Elitzur (1993). The low
mass-loss rates inferred for extrinsic S stars in Region A  
also confirm an earlier suggestion (Paper I) that these stars are
less evolved than the intrinsic S stars populating  the other Regions of
the IR color-color diagram.

Stars in Regions B and D have moderate mass loss rates, in the range
$5\; 10^{-8}$ to  $5\; 10^{-7}$ \Msun~y$^{-1}$. One star from Region~B
(RS Cnc) has been found to exhibit a double wind (Table~\ref{Tab:COlit}).
Most of the observed stars, and most of the detections, lie in Region C,
which contains stars with moderately optically thick circumstellar envelopes,
likely containing silicate dust.  CO emission is detected from 20 of these 23 
stars, and the mass loss rates are typically larger than several $\rm \times 
10^{-7}~ M_{\odot}~ y^{-1}$.  The undetected stars have
upper limits greater than this value, so that the data are consistent with the 
conclusion that all stars in this region lose mass at a rate larger than several
$10^{-7}$ \Msuny.

\begin{figure}
  \vskip -0cm
  \vspace{9cm}
  \begin{picture}(10,9.5)
    \epsfysize=9.8cm
    \epsfxsize=8.5cm
    \epsfbox{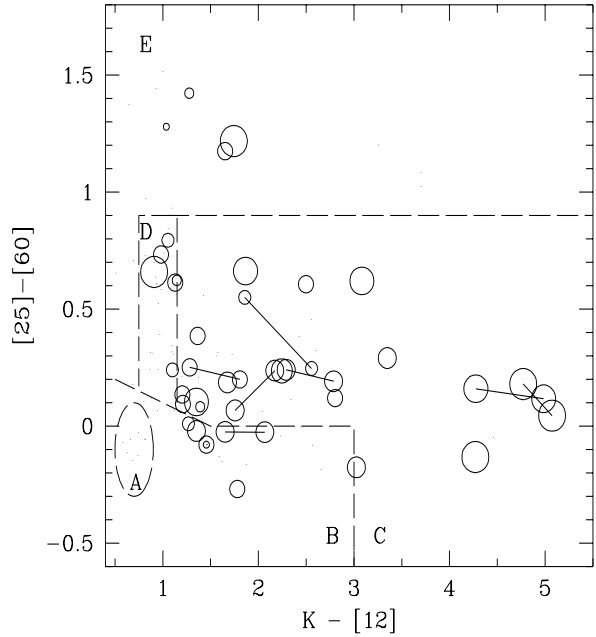}
  \end{picture}
  \vskip -0.5cm
\caption[]{\label{25}
The outflow velocity as measured from CO for S stars in the ($K -
[12]$, \ccc) diagram. 
The diameter of
the circle is proportional to the outflow velocity.
Compare to Fig.~10 of Olofsson et al. (1993) for C stars
}
\end{figure}

The stars in Region D, which show roughly photospheric 12$\rm \mu
m$/2.2$\mu$m colors despite large \ccc\ indices, 
generally have low mass-loss rates (like Region B). 
Stars in Region E have mixed properties. Some, like TT Cen and RZ Sgr,
lose mass at a very large rate (several 10$^{-6}$ \Msun~y$^{-1}$),
while others like FU Mon lose mass at a more moderate rate (a few
10$^{-7}$ \Msun~y$^{-1}$). 
Figure~\ref{25} presents the variation of the wind velocity
across the ($K - [12]$, \ccc) diagram, and shows that stars in Region
E also have a wide range of wind velocities, from very low (FU Mon:
2.8~\kms) to very large (TT Cen: 24.7 \kms).

FU Mon also has a resolved IR envelope (Sect.~\ref{Sect:extended}), 
and this fact taken together with a low mass loss rate and a small
outflow velocity suggest that mass loss has just resumed  in that
star. Olofsson et al. (1990; 1993) detected several carbon stars (S
Sct, U Ant, TT Cyg) with
a detached shell and a double wind that would also fall in our Region
E. The older, detached shell was produced by a massive, fast wind,
whereas the recent shell is caused by a slow (5~\kms), low mass loss rate
wind. Since FU Mon is
an SC star with  a C/O ratio close to unity (Dominy et al. 1986), 
the mass loss may have come to a halt when the  
the C/O ratio approached  unity, as already suggested by      
Willems \& de Jong (1988) and Chan \& Kwok (1988).
This scenario may well hold for all SC
stars, even though some, like UY Cen, are in fact located in Region C.
As argued in Sect.~\ref{Sect:results}, that star may be at the end of
the loop in the IR color-color diagram. The horn-shaped CO line
profile of UY Cen observed by SL95, indicative of a detached shell and
fossil mass loss, supports this idea. Finally, other stars lying in
Region E like TT Cen, RZ Sgr and DK Vul have peculiar CO line profiles
(narrow central feature superimposed on a broader less intense
feature) suggesting that they have multiple winds.

\subsection{Mass loss rates and binaries}
   
Several intrinsic S stars in the sample considered
in this paper are binaries with main sequence
companions, as revealed by the composite nature of their spectrum at
minimum light (WY Cas, W Aql, T Sgr; Herbig 1965; Culver \&
Ianna 1975). A composite spectrum is also suspected for S Lyr from its
shallow lightcurve (Merrill 1956). The star $\pi^1$ Gru has a close
G0V visual companion (Feast 1953). 
W Aql, WY Cas and S Lyr are surely among the S stars with the largest
mass loss rates, while T Sgr and $\pi^1$ Gru have mass loss rates
close to the average for stars in Region C. Clearly, a definite
conclusion as to whether binarity can indeed reinforce the mass loss rate,   
e.g. by the companion-reinforced attrition process (Eggleton 1986),
would require the knowledge of the orbital separation of
these systems. 

It has sometimes been argued (e.g. Eggleton 1986; 
Tout \& Eggleton 1988; Kenyon 1994;  Han et al.
1995) that the mass loss rates of  giant stars  in
binary systems must be larger than those of single red
giants. More precisely, Han et al.
(1995) suggest that the mass loss rate of a giant approaching its
Roche lobe in a binary system
exceeds by more than a factor 10$^3$ the rate predicted by the Reimers
formula (Reimers 1975). 
As far as the binary, extrinsic S stars are
concerned, this effect, if present,
is clearly not large enough to bring their mass loss rates to the
level of the intrinsic S stars.
The difference in the evolutionary stages
of extrinsic and intrinsic S stars, believed to be RGB (or Early-AGB,
according to the terminology of Iben \& Renzini 1983) and TP-AGB
stars, respectively, thus appears to be of greater importance for the mass loss
rate than their binary or non-binary
character.

\section{Summary}

Our extensive discussion of the circumstellar properties of S stars confirms the
dichotomy extrinsic/intrinsic S stars and leads to the following conclusions.  

{\it Extrinsic} S stars have the lowest mass loss rates among S stars ($< 2\;10^{-
8}$~\Msun~y$^{-1}$), and undetectable circumstellar shells. This is consistent
with the hypothesis that these stars are much less evolved than intrinsic S stars,
populating the RGB or Early-AGB rather than the TP-AGB like intrinsic S stars. 
The binary character of these stars does not seem to increase their mass loss
rate, as it has sometimes been suggested, e.g., in discussions relating to
symbiotic or barium stars. Note, however, that the few binary intrinsic S stars
are among the stars with the largest mass loss rates in our sample.

Among {\it intrinsic} S stars, various subclasses must be distinguished:\\
1. A few Tc-rich S stars (e.g., HR Peg) have very low mass-loss rates, similar 
to those of the extrinsic S stars, with undetectable circumstellar shells [Region
A of the $(K - [12], [25]-[60])$ color-color diagram]. These are S stars with weak
chemical peculiarities, barely distinguishable from normal M giants.
\smallskip\\ 
2. S stars with weak chemical peculiarities and tenuous O-rich circumstellar
shells (small $K - [12]$ excess), fed by a small
albeit measurable mass loss rate of a few $10^{-7}$~\Msun~y$^{-1}$ (Region B).
These are short-period ($P \sim 100$ to 150~d) SR variables or short-period ($P <
500$~d) Mira variables. This region may be contaminated by (possibly normal M)
supergiants of variability type SRc (T Cet, RS Cnc, Y Lyn). These SRc supergiants
have resolved shells at 60~$\mu$m, and for RS Cnc, a double wind.  
\smallskip\\ 
3. S stars with strong chemical peculiarities and dense O-rich circumstellar
shells (as indicated by SiO maser emission, the 9.7 $\mu$m silicate feature and large
$K - [12]$ indices), populating the lower part of Region C. Mass loss rates range
from several $10^{-7}$ to $10^{-5}$~\Msun~y$^{-1}$. Almost all of them
are long-period ($P > 300$~d) Mira variables.
\smallskip\\ 
4. S stars with strong chemical peculiarities (often classified as SC in the
optical), with neither SiO maser emission nor 9.7 $\mu$m silicate
emission, with featureless IRAS spectra and  moderate to large $[25] - [60]$ indices.
These stars populate Region
D and the upper part of Region C (like optical carbon stars) and have moderate
mass loss rates (a few $10^{-8}$ to a few $10^{-7}$~\Msun~y$^{-1}$) with small
wind velocities (generally $\la 8$ \kms). They are mainly semi-regular (with
periods ranging from 60 to 360~d) or irregular variables, with a few short-period
($P < 370$~d) Mira variables. These properties are reminiscent of carbon
stars. All their features may equally well be explained by a
carbon-rich circumstellar shell, or by a detached shell.   
The relationship between stars in Regions C and D is far from being clear.
It may be a difference of galactic populations, as suggested by their different
properties as variable stars, or stars might oscillate between Regions C and D at
different phases of their variability cycle, as observed for some variable IRAS
sources in our sample, or they may be on different parts of the loop
associated with the detachment of a dust shell. 
\smallskip\\ 
5. A few S stars have well-resolved (and thus very extended) IR shells (also
visible in the optical in the case of RZ Sgr), probably detached from their parent
star, as indicated by strong 60 and 100~$\mu$m excesses (Region E). The mass loss
rates span a wide range, from a few $10^{-7}$ to several $10^{-6}$~\Msun~y$^{-1}$,
as do the wind velocities (25 \kms\ for TT Cen to
2.8 \kms\ for FU Mon). These stars with extreme properties are often SC or CS
stars, and may be experiencing a loop in the IRAS color-color diagram as first
proposed by Willems \& de Jong (1988), triggered by their C/O ratio being very
close to unity. In TT Cen, the dominant spectral features are ZrO bands at some
times, and C$_2$ bands at other times. This is a rare and short-lived phase, and
thus does not face the difficulties generally opposed to the Willems \& de Jong
scenario. Some SC or CS stars in Region C, like BH Cru (a sister case of TT Cen)
and UY Cen, may be at the end of the counter-clockwise loop in the
IRAS 
color-color diagram, entering the region of heavily-obscured IR carbon stars. 
Region E is probably fed by stars leaving Region C when their C/O ratio reaches
values close to unity. However, there is no need that all S stars experience such
a loop.

\setcounter{table}{0}
\renewcommand{\thetable}{A\arabic{table}}
\begin{table*}
\caption[]{IRAS flux densities for $\chi$ Cyg}
\begin{tabular}{crrrrrrcr}
$JD-2\ts440\ts000$& $\phi$& $F12$ & offset& det& $F25$ & $F60$ 
& [12]$-$[25]& [25]$-$[60]\cr
       &     & (Jy)  & ($'$) &    & (Jy)  & (Jy)  \cr
\hline\cr
5452.8& 0.12& 1860.6& 0.40& 48& 643.5& 99.4& 0.41& $-$0.15\cr
& & & & & 540.2& & 0.22& 0.04\cr
5453.7& 0.12& 1778.6& $-$1.01& 23& 533.2& 97.7& 0.25& 0.04\cr
& & 1880.7& 1.23& 51& & 92.6& 0.19& $-$0.02\cr
5466.1& 0.15& 1157.5& $-$1.79& 28& 626.6& 86.5& 0.89& $-$0.27\cr
& & 1975.0& 0.63& 48& 525.0& 99.8& 0.37& $-$0.11\cr
& & & & & & & 0.70& $-$0.08\cr
& & & & & & & 0.18& 0.08\cr
5466.2& 0.15& 1624.5& $-$0.03& 53& 520.7& 98.7& 0.32& 0.08\cr
& & & & & 583.0& & 0.45& $-$0.05\cr
5639.7& 0.58& 1248.7& $-$0.22& 30& 456.3& 72.6& 0.47& $-$0.11\cr
& & & & & 422.1& & 0.38& $-$0.03\cr
5639.8& 0.58& 1172.5& $-$0.52& 24& 433.5& 69.5& 0.48& $-$0.11\cr
& & 1266.5& 1.72& 49& & & 0.40& \cr
5640.3& 0.58& 1184.6& $-$0.04& 24& 404.9& 70.8& 0.39& $-$0.01\cr
5640.6& 0.58& 1249.0& 0.13& 27& 383.0& 74.7& 0.28& 0.11\cr
& & 1214.4& 1.75& 51& & & 0.31& \cr
\hline\cr
\end{tabular}
\end{table*}

\noindent {\footnotesize {\it Acknowledgments.} 
We thank the staff at CSO, especially Ken Young (Taco), Antony Schinkel,
Maren Purves and Tom Phillips, for the observing time and for
their help with the observations.  Astronomical 
research at the CSO is supported by the National Science Foundation
via grants AST96-15025.  We thank the staff at IPAC for swiftly
processing the  numerous requests for IRAS archive data sent over the 
Internet.  We especially thank Ron Beck for providing us with the IRAS 
template data. We thank Princeton University, the {\it
Fonds National de la Recherche Scientifique} (Belgium) and
the National Science Foundation (USA), via grant AST96-18503,
for partial 
support of this work. The CO line formation modeling was based on
code by Mark Morris, and the figures were drawn using software
by Robert Lupton and Patricia Monger. 
This research has made use of the Simbad data base, operated at CDS,
Strasbourg, France.
A.J. is Research Associate, F.N.R.S. (Belgium). 
}

\section*{Appendix A: Stellar variability and the IRAS flux densities}

As discussed in Sect.~\ref{Sect:IRASflux}, the flux densities obtained
from co-addition of the
IRAS data generally agree well with those in the PSC with, however,
some notable exceptions, as shown in Fig.~\ref{1}. 
Several stars have flux 
densities which are in disagreement at a level well outside that expected
from noise.  In this Appendix we investigate whether these disagreements 
can be explained by stellar variability, using the data for $\chi$ Cyg
as an example.  The ephemerides listed in the GCVS give 
a variation of about $\rm 6$ mag at $V$ between maximum and minimum light,
a period of $\rm 408$~d and the epoch of zero phase which is closest to the 
IRAS launch date of JD~2445404.4.

The colors of $\chi$ Cyg, calculated from the IRAS PSC 
flux densities (see Table~A2), are photospheric, 
despite the strong circumstellar CO emission (discussed in 
Sect.~\ref{Sect:massloss}). On the other hand, 
model IRAS colors calculated from
the mass loss rate given by the CO lines
(several 10$^{-7}$~M$_{\odot}$~y$^{-1}$), assuming a normal gas to dust 
ratio and silicate
grains (the star is an SiO maser;  e.g. Patel et al. 1992), are well
displaced from photospheric values. This discrepancy  does {\it not} mean that the 
circumstellar envelope is dust-free, however; the ratio of the
12$\rm \mu m$ to 2$\rm \mu m$ flux densities is well in excess
of the photospheric value for a temperature of 2400 K (Haniff et
al. 1995), demonstrating the presence of an appropriate amount
of circumstellar dust.

We therefore re-examined the raw IRAS data for this star.  
The individual IRAS 12$\rm \mu m$, 25$\rm \mu m$, and 60$\rm \mu m$ 
observations for $\chi$ Cyg are listed in Table A1.  Column 1 gives the
Julian date of the observations
computed from the `Satellite Operation Plan' number attached to each
scan and from the mission
chronology provided by the {\it Explanatory Supplement}. Column 2
lists the phase.  Next are the flux 
density, the offset between the scan center and the stellar position, and the
detector number for the 12~$\mu$m data.  The final columns contain
the 25~$\mu$m and 60~$\mu$m flux densities and the $[12] - [25]$
and $[25] - [60]$ colors calculated from combinations of the observed
flux densities.  The colors calculated from the flux densities
derived from individual scans
locate $\chi$ Cyg in Region B of the color-color diagram.

Table A1 shows that there are occasionally large discrepancies between the
flux densities measured by different detectors at the same time; see,
for example, the 12~$\mu$m flux densities observed at JD2445466.1.
The low flux density is probably due to the large offset between the
stellar and detector center positions; such discrepant observations are
usually filtered out in the IRAS data processing. Real variability
is also apparent.  The data were obtained at two epochs, near maximum
and minimum phase, and the flux densities at all four wavelengths
are systematically higher for the first set of observations than the
second.  The mean flux densities for these two epochs are listed in Table A2;
the differences are $\Delta ~ = ~ (F_{\rm max} - F_{\rm min})/F_{\rm max}$
= 32\% at \f, 26\% at \ff, 25\% at \fff\ and
31\% at 100~$\mu$m.

Can these variations be attributed to the stellar variability?  The 
huge variations in Mira variables at visible wavelengths are due largely
to the changing photospheric temperature; the variation in the
bolometric magnitude is much smaller, about a factor of 2 (e.g. Petit
1982; Hoffmeister et al. 1985).  Because of the variation in stellar
effective temperature, the star reaches maximum light later at longer 
wavelengths in the visible and near infrared (e.g. Le Bertre 1992).
In particular, the variation at visible wavelengths leads the total
light variation by about 0.1 of a period.  The two groups of observations in
Table A1 were thus made close to maximum and minimum luminosity.

We modeled the object as a star with a circumstellar envelope and varied the
luminosity of the model star.  The envelope contains silicate grains,
has a dust loss rate of $8\; 10^{-10}$~\Msun\ y$^{-1}$
and is assumed to have the same outflow speed as the gas (9.5 $\rm
km~s^{-1}$).  The star is assumed to be a black body of temperature
2400 K and luminosity 3000~$\rm L_{\odot}$ at minimum, and 2800 K and
6000~$\rm L_{\odot}$ at maximum.  The resulting model flux densities
are listed in Table A2 and are reasonably close to the observed values
(note that the discrepancy at 12~$\mu$m may be caused in part by the
saturation of the detectors at these flux densities exceeding 1000~Jy).
The variation in the IRAS flux densities can thus be fully explained by the
stellar variability.

These results show that caution is required in the interpretation of IRAS data
for variable stars; colors must be calculated from data taken at the
same epoch, as should models of the circumstellar envelope.
Variability introduces a significant amount of scatter into the IRAS
colors, especially $[12] - [25]$, which as a result does not provide the
clean measure of the stellar mass loss rate which the models predict
(see, for example, the discussion by Habing 1996).

\begin{table}
\caption[]{Observed and model IRAS flux densities for $\chi$ Cyg. The
PSC fluxes are also listed}
\begin{tabular}{lllllllll}
\multicolumn{5}{c}{Observed (averages)}
\smallskip\cr
Phase& $F12$ & $F25$ & $F60$ & $F100$\cr
     & (Jy)  & (Jy)  & (Jy)  & (Jy) \cr
\hline\cr
max& 1804& 568& 96& 20\cr
min& 1223& 420& 72& 14\smallskip\cr
PSC& 1688& 459& 81& 18\cr
\hline
\medskip\cr
\multicolumn{5}{c}{Calculated}
\smallskip\cr
Phase& $F12$ & $F25$ & $F60$ & $F100$\cr
     & (Jy)  & (Jy)  & (Jy)  & (Jy) \cr
\hline\cr
max& 1486& 576& 86& 20\cr
min& 1108& 486& 79& 17\cr
\hline
\end{tabular}
\end{table}

\renewcommand{\thetable}{\arabic{table}}
\setcounter{table}{0}
\renewcommand{\baselinestretch}{1}

\tabcolsep 4pt
\begin{table*}[t]
\caption[]{\label{Tab:IRAS}
IRAS co-added flux densities for S stars, grouped according to their location 
in the ($K - [12]$, \ccc) diagram}

\begin{tabular}{rrrrrrrrllllllllllllllllll}
\noalign{Region A: Stellar photospheres}
\medskip\cr
\multicolumn{2}{c}{GCGSS \hfil IRAS}& F2.2   &Ref& F12 &   F25  &   F60 &   F100& 
 Tc
&LRS&VC& Sp & Var & $P$ & $\Delta V$ & Name \cr
& & (Jy) & & (Jy) & (Jy) & (Jy) & (Jy) & & & & & & (d) & (mag)\cr
\hline\cr
  26 &   01113+2815 &  139.4 & 2& 11.95&    3.02&   0.63&       &    no &    & F&
S3/2 &    &    &     & HR 363\cr
  79 &   03377+6303 &  502.0 & 2& 42.23&   10.86&   1.82&       &    no &  18& 
& S4/2 & Lb &    & 0.1 & BD Cam, HR 1105\cr
 133 & 05199$-$0842 &   92.1 & 2&  7.94&    2.06&       &       &    no &  16& S&
S4,1 &    &    &     & HD 35155\cr
 382 &   07392+1419 &  331.4 & 2& 25.91&    6.51&   1.08&   0.37&       &  18& 
& M3S  & SR &    & 0.2 & NZ Gem, HD 61913\cr
 729 & 11098$-$3209 &  112.8 & 2&  9.21&    2.45&   0.39&       &       &    & S&
Swk  &    &    & 0.1 & NSV 5129, HR 4346\cr
 796 & 12272$-$4127 &  207.2 & 1& 17.77&    4.47&   0.69&       &       &  18&
S&M3-IIIa&   &    &     & NSV 5655, HR 4755\cr
 804 & 13079$-$8931 &  167.6 & 1& 15.23&    3.88&   0.67&       &       &  17& S&
S5,1 & Lb:&    &     & BQ Oct, HD 110994\cr
 826 & 13372$-$7136 &  251.4 & 1& 24.13&    6.43&   1.08&       &       &  18& S&
S6,2 &    &    &     & HD 118685\cr
 938 & 16425$-$1902 &  150.1 & 2& 12.24&    3.06&  $^a$ &       &    no &  31& S&
Swk  &    &    &     & HD 151011      \cr
1315 &   22521+1640 &  283.4 & 2& 27.58&    7.33&   1.23&   0.67&    yes&    & S&
S4/1 & SRb& 50 & 0.3 & HR Peg, HR 8714\cr
1322 &   23070+0824 &  938.4 & 2& 84.67&   20.31&   3.20&   0.96&    no &  18& 
& M4S  & SRa& 93 & 0.3 & GZ Peg, 57 Peg\cr
\hline
\end{tabular}
\medskip\\
a: Strong cirrus contamination
\medskip\\

\begin{tabular}{rrrrrrrrlllllllllllllllll}
\noalign{Region B: Small ZrO index and no \ccc\ excess}
\medskip\cr 
\multicolumn{2}{c}{GCGSS \hfil IRAS} & F2.2   &Ref& F12 &   F25  &   F60 &   F100&
 Tc &LRS&VC&  Sp & Var & $P$ & $\Delta V$ & Name \cr
& & (Jy) & & (Jy) & (Jy) & (Jy) & (Jy) & & & & & & (d) & (mag)\cr
\hline\cr
   8 & 00192$-$2020 & 1356.0& 2& 199.1&   81.66& 14.57+&   6.08+&yes&  16 &   &
M5-6Se&SRc& 159 & 1.9 & T Cet\cr
 134 & 05208$-$0436 &   80.2& 8& 20.56&   10.85&   1.65&        &   &  43 &  U&
M4Swk  &  &     &     & V535 Ori\cr
 168 &   05495+1547 &   26.1& 1& 11.35&    3.52&   0.62&        &   &     &  F&
S7.5,1e&M & 494 & 4.9 & Z Tau    \cr
 221 & 06266$-$1148&   27.6& 7&       &        &       &        &   &     &  F&
S-*2e  &  &     &     &\cr
&\multicolumn{3}{r}{JD 2\ts445\ts420:}
                              &   8.97&    4.10&   0.94&     \cr
&\multicolumn{3}{r}{JD 2\ts445\ts620:}
                              &  14.57&    7.64&   1.14&\smallskip\cr
 265 & 06466$-$2022&    87.2& 2&  8.34&    2.41&   0.47&        &   &   16&   &
M4S    &  &     &     & HD 49683      \cr
 323 & 07117$-$1430&    40.6& 2&  4.17&    1.52&   0.30&        &   &     &   & 
      &  &     &     & \cr
 408 & 07461$-$3705 &       &  & 14.11&    6.25&   0.87&        &   &   14&  F&
S6*1   &  &     &     & \cr
 436 & 07545$-$4400 &   51.6& 1&      &        &       &        &   &   22&  E&
S4,2   & M&  340& 4.7 & SU Pup\cr
&\multicolumn{3}{r}{JD 2\ts445\ts470:}
                              &  15.44&    6.88&   1.29&    \cr
&\multicolumn{3}{r}{JD 2\ts445\ts645:}  
                              &  25.03&   11.95&   1.96&\smallskip\cr
 446 & 07573$-$6509 &   27.6& 1&  5.45&    1.72&   0.27&        &   &     &   &
S7,2   &M & 280 & 2.7 & X Vol    \cr
 474 & 08098$-$2809 &   55.5& 2&  7.07&    2.23&   0.42&        &   &     &  S&
S4,2   &  &     &     & $-$27:5131  \cr
 533 & 08348$-$3617 &   55.5& 1& 11.25&    3.44&   0.53&        &   &   18&  S&
S5,2   &  &     &     & $-$36:4827  \cr
 589 &   09076+3110 & 2808.0& 2&489.88&  210.29& 34.57+&  11.71+&yes&   22&   &
M6S    &SRc& 120& 1.5 & RS Cnc(OH?)\cr
 626 & 09411$-$1820 &       &  &  10.0&    4.10&   0.71&        &   &     &  F&
M0S    &M &  300& 5.0 & FM Hya       \cr
 704 & 10436$-$3459 &  133.2& 1& 31.32&   13.98&   1.93&    1.09&   &   42&  E&
S5,4   &SR&  104& 2.0 & Z Ant\cr
 903 &   15492+4837 & 1295.0& 2&205.41&   98.75& 17.13+&   7.14+&yes&   41&   &
M6.5S  &SRb& 148& 1.5 & ST Her\cr
 914 & 16097$-$6158 &       &  & 12.24&    4.51&   0.70&        &   &     &  F&
S4,1   & M&  323& 5.0 & Y TrA      \cr
 948 & 16552$-$5335 &   39.1& 1&  8.80&    2.90&       &        &    &    &   &
S4,4   &  &     &     & \cr
1099 &   19008+1210 &  105.8& 2& 11.54&    3.62&       &        & yes&    & S &
S5/2   &Lb&     & 0.6 & V915 Aql   \cr
1131 & 19226$-$2012 &   20.5& 7&   7.32&   2.94&   0.50&        &    &    & F&
M8Swk   &M & 332 & 7.0 & TT Sgr\cr
1165 &   19486+3247 & 5813. &2 &       &       &       &        & yes&    & E &
S7/1.5e&M & 408 & 6.  & $\chi$ Cyg (SiO)\cr
&\multicolumn{3}{r}{JD 2\ts445\ts450:}
                           &1781.65$^a$& 552.84& 95.47+& 21.32  \cr
&\multicolumn{3}{r}{JD 2\ts445\ts640:}
                           &1215.31$^a$& 414.40& 71.64+& 14.50  
\smallskip\cr
1211 &   20213+0047 &152.9  &2 & 22.54&    7.27&   1.29&        &    &  17& S &
S7,2   &M & 364 &  4.5& V865 Aql  \cr
1346 & 23595$-$1457 &120.3  &2 & 13.11&    3.88&   0.73&        & yes&  16& F &
S5-7/1.5-3e&M&351 &  7.7& W Cet    \cr
\hline
\end{tabular}
\medskip\\
a: detector probably saturated\\
+: the `zero-crossing' flux density \Fz\ has been adopted instead of the template 
flux density $F_{\rm t}$, indicating a possibly
resolved shell (see Sect.~2.2)

\end{table*}
\clearpage

\addtocounter{table}{-1}
\begin{table*}
\caption[]{Continued.}
\begin{tabular}{rrrrrrrrlllllllllllllllllllllll}
\noalign{Region C: \cc\ and \ccc\ excesses, silicate emission common}
\medskip\cr
\multicolumn{2}{c}{GCGSS \hfil IRAS}
& F2.2   &Ref& F12 &   F25  &   F60 &   F100&   Tc
&LRS&VC& Sp & Var & $P$ & $\Delta V$ & Name \cr
& & (Jy) & & (Jy) & (Jy) & (Jy) & (Jy) & & & & & & (d) & (mag)\cr
\hline\cr
   6 &   00135+4644&   37.4& 2& 15.96&    6.24&   1.57&   0.85&        &  16& F&
S4/7e &  M  & 346& 6.9 & X And    \cr
   9 &   00213+3817&  453.3& 2& 335.74& 175.66& 26.41+&   9.45&     yes&    & E&
S5-7/4-5e&M  & 409& 9.1 & R And (SiO)\cr
  14 &   00445+3224&   96.5& 2& 37.75&   18.58&   3.97&   2.10&        &  22& F&
S6/2e &  M  & 430& 7.8 & RW And   \cr
  28 &   01159+7220&  147.4& 2& 343.76& 192.52& 30.14+&  10.88&        &  22& E&
S4,6e &  M  & 612& 8.2 & S Cas (SiO)\cr
  36 &   01266+5035&   34.1& 6&  6.59&    1.93&   0.45&       &        &    &  &
S6/8e &  M  & 355& 5.3 & RZ Per    \cr
  49 &   02143+4404&  275.7& 2& 166.42&  70.94&  14.01&   5.43&     yes&  22& E&
S7/1e &  M  & 396& 7.9 & W And (SiO)\cr
 149 &   05374+3153&  280.8& 2&  45.33&  23.32&   5.00&       &     yes&  43&  &
M2S   &  Lc &    & 0.2 & NO Aur    \cr
 231 &   06331+1415&  136.9& 2&  21.95&  10.35&2.61$^a$&   3.90&     no &  42& F&
S8,5  &  SRa&1145& 1.4 & DY Gem\cr
 283 &   06571+5524&  118.1& 2&       &       &       &       &        & 16& F&
S5/5e &  M  & 378& 7.1&R Lyn (SiO)\cr
&\multicolumn{3}{r}{JD 2\ts445\ts420:}
                              &  17.55&   5.61&   1.25&       \cr
&\multicolumn{3}{r}{JD 2\ts445\ts620:}
                              &  28.46&   9.46&   2.01&       \smallskip\cr
 307 &   07043+2246&   88.0& 2&       &       &       &       &     yes& 16& F&
S3,9e &  M  & 370& 8.0&R Gem\cr
&\multicolumn{3}{r}{JD 2\ts445\ts435:}
                              &  22.23&   7.53&   2.21&   1.40\cr
&\multicolumn{3}{r}{JD 2\ts445\ts620:}
                              &  42.33&  16.81&   3.73&   1.48\smallskip\cr
 316 &   07092+0735&       &  &  12.23&   5.63&   1.06&   0.36&        &    & E&
Se    &  M  & 420& 4.8 & WX CMi     \cr
 326 &   07149+0111&   70.5& 2&       &       &       &       &        &  16&  &
M7se: &  M  & 395& 7.4 & RR Mon\cr
&\multicolumn{3}{r}{JD 2\ts445\ts440:}
                              &  27.76&  11.75&   2.31&   0.85\cr
&\multicolumn{3}{r}{JD 2\ts445\ts625:}
                              &  15.77&   7.21&   1.69&   1.03\smallskip\cr
 341 & 07197$-$1451&       &  &  20.24&  10.47&   1.86&       &        &  27& E& 
     &  M  & 314& 4.0 & TT CMa     \cr
 347 &   07245+4605&  964.7& 2& 133.50&  65.92&13.20+$^d$&5.58+&    yes&  23&  &
M6S   &  SRc& 110& 2.5 & Y Lyn (SiO)\cr
 614 & 09338$-$5349&   68.0& 1&       &       &       &       &        &  01& E&
S7,8e &  M  & 408& 3.0 & UU Vel\cr
&\multicolumn{3}{r}{JD 2\ts445\ts510:}  
                              &  11.53&   5.98&   1.77&       \cr
&\multicolumn{3}{r}{JD 2\ts445\ts580:}  
                              &  17.79&   9.37&       \smallskip\cr
 649 & 09564$-$5837&       &  & 156.15&  71.37&  14.31&       &        &  15&
E&S6.5/1-&  SRb& 109& 1.3 & RR Car \cr
 656 & 10017$-$7012&   43.7& 1&   6.02&   1.72&   0.37&       &        &   &  &
S5,6   & M   &    & 3.5 & KN Car       \cr
 816 & 13136$-$4426&  356.8& 1&  56.55&  20.77&   4.04&   1.78&        &  43&
C&S6/8=CS&  SR & 114& 2.0 & UY Cen\cr
 821 & 13240$-$5742&       &  &  13.40&   6.27&  $^b$ &       &        &  14& F&
S6*3  &  SR & 198& 1.9 & EE Cen\cr
 861 & 14372$-$6106&   43.3& 9&  21.38&   9.13&       &       &        &  15& E& 
     &     &    &     & CSV2170\cr
 872 & 15030$-$4116&  111.8& 1&  21.81&   9.23&   2.18&       &        &    & F&
S7,8e &  M  & 326& 5.6 & GI Lup        \cr
 931 & 16334$-$3107&  377.0& 2&  52.49&  21.88&4.22$^c$&  $^b$&        &  16&
E&S8/4var&  SRa& 194& 3.5 & ST Sco\cr
 954 & 17001$-$3651&  453.3& 1&       &       &       &       &        &  22& E&
S7,2  &   M & 449& 8.2 & RT Sco\cr
&\multicolumn{3}{r}{JD 2\ts445\ts410:}
                              & 170.65&  69.64&  15.37&         \cr
&\multicolumn{3}{r}{JD 2\ts445\ts600:}  
                              & 269.51& 118.53&  25.01&       \smallskip\cr
$-$  &   17081+6422&  377.0& 2&  61.96&  25.52&   4.87&   2.27&        &    &  & 
     &   Lb&    & 0.4 & TV Dra \cr 
1093 & 18575$-$0139&   37.7& 1&  17.18&   5.92&   1.83&       &        &    &  &
SC    &  M: &    & 3.8 & VX Aql      \cr
1096 & 18586$-$1249&  242.3& 2&  51.78&  19.19&   4.03&       &        &  21& E&
S6/3e &   M & 395& 8.8 & ST Sgr       \cr
1112 &   19111+2555&   18.7& 6&       &       &       &       &        &  41&  &
SC    &   M & 438& 5.8 & S Lyr\cr
&\multicolumn{3}{r}{JD 2\ts445\ts440:}
                              &  43.81&  21.43&   4.39&        \cr
&\multicolumn{3}{r}{JD 2\ts445\ts630:}  
                              &  84.17&  38.23&   7.53&   2.73\smallskip\cr
1115 & 19126$-$0708&  286.0& 2&        &       &       &      &        &  22& E&
S6/6e &   M & 490& 7.0 & W Aql (SiO)\cr
&\multicolumn{3}{r}{JD 2\ts445\ts440:}
                          & 1397.52$^e$& 719.87&132.67+& 36.87\cr
&\multicolumn{3}{r}{JD 2\ts445\ts630:}
                          & 1057.16$^e$& 481.93&100.58+& 24.90\smallskip\cr
1117 & 19133$-$1703&  166.1& 2&  42.27&  14.62&  4.76+&  4.76+&    yes &    & F&
S5/6e &  M  & 394& 5.8 & T Sgr\cr
1150 &   19354+5005&  107.7& 2& 107.44&  52.86& 12.22+&   5.52&     yes&  22& E&
S6/6e &   M & 426& 8.3 & R Cyg (SiO)\cr
1159 &   19451+0827&   15.3& 7&   7.69&   3.53&   0.84&       &        &    & I&
Se    &   M & 607& 2.0 & QU Aql   \cr
1175 & 19545$-$1122&       &  &   7.59&   3.80&       &       &        &  29& E&
M6S   &     &    &     & V1407 Aql  \cr
\hline
\end{tabular}
\medskip\\
a: close weak source\\
b: profile badly distorted by close source\\
c: close weak source apparent on some scans\\
d: moves to Region B after removing the extended shell contribution\\
e: Detector possibly saturated\\
+: the `zero-crossing' flux density \Fz\ has been adopted instead of the template 
flux density $F_{\rm t}$, indicating a possibly
resolved shell (see Sect.~2.2)
\end{table*}
\clearpage
\newpage

\addtocounter{table}{-1}
\begin{table*}[t]
\caption[]{Region C (continued)}
\begin{tabular}{rrrrrrrrlllllllllllllllllllllll}
\multicolumn{2}{c}{GCGSS \hfil IRAS}
& F2.2   &\multicolumn{2}{c}{Ref \hfil F12} &   F25  &   F60 &   F100&   Tc
&LRS&VC& Sp & Var & $P$ & $\Delta V$ & Name \cr
& & (Jy) & & (Jy) & (Jy) & (Jy) & (Jy) & & & & & & (d) & (mag)\cr
\hline\cr
1200 &   20114+7702&       &  &       &       &       &       &        &    & I&
S5/6e &   M & 326& 6.9 & SZ Cep\cr
&\multicolumn{3}{r}{JD 2\ts445\ts394:}
                         &   7.27&   2.71&   0.68&       \cr
&\multicolumn{3}{r}{JD 2\ts445\ts580:}
                         &   9.67&   3.45&   0.72&       \smallskip\cr
1268 & 21172$-$4819&   14.9& 1&   7.15&   2.34&   0.63&       &        &    & I&
S2,5  &     &    &     & $-$48:13866    \cr
C3107$^a$&22036+3315&  44.1& 2&  15.92&   7.28&   1.60&       &        &    & C&
SC    &   M & 439& 6   & RZ Peg\cr
1294 & 22196$-$4612& 4211. & 1&       &       &       &       &     yes&  42& E&
S5,7  &  SRb& 150& 1.3 & $\pi^1$ Gru\cr
&\multicolumn{3}{r}{JD 2\ts445\ts470:}
                              & 969.23& 471.63& 88.71+&  23.88\cr
&\multicolumn{3}{r}{JD 2\ts445\ts645:}
                          &1421.10$^b$& 419.22& 92.25+&  25.80\smallskip\cr
1345 &   23554+5612&   66.7& 2&  52.07&  28.10&   8.80&   5.28&        &  42& E&
S6/6e &   M & 476& 6.9 & WY Cas     \cr
1347 &   00001+4826&   18.2& 6&  50.54&  24.54&   4.15&       &        &  21& E&
S5/6e &   M & 396& 3.3 & IW Cas\cr
\hline
\end{tabular}
\medskip\\
a: Star number from the {\it General Catalogue of Carbon
Stars} (Stephenson 1989)\\ 
b: Detector possibly saturated\\
+: the `zero-crossing' flux density \Fz\ has been adopted instead of the template 
flux density $F_{\rm t}$, indicating a possibly
resolved shell (see Sect.~2.2)
\bigskip\\
\end{table*}
\clearpage
\newpage

\addtocounter{table}{-1}
\begin{table*}[t]
\caption[]{Continued.}
\begin{tabular}{rrrrrrrrlllllllllllllllllllllll}
\noalign{Region D: No silicate emission, weak $K - [12]$ excess, some \ccc\ 
excess}
\medskip\cr
\multicolumn{2}{c}{GCGSS \hfil IRAS}
& F2.2   &\multicolumn{2}{c}{Ref \hfil F12} &   F25  &   F60 &   F100&   Tc
&LRS&VC& Sp & Var & $P$ & $\Delta V$ & Name \cr
& & (Jy) & & (Jy) & (Jy) & (Jy) & (Jy) & & & & & & (d) & (mag)\cr
\hline\cr
  12 &   00435+4758&   64.9& 2&   8.47&   2.49&   $^a$& $^a$  &     yes& 01& S&
S5/3e &  M  & 277& 7.7& U Cas\cr
  17 &   00486+3406&   36.3& 5&  4.56&    1.44&   0.35&       &        &   &  &
S6/3e &  M  & 328& 7.2& RR And    \cr
  20 &   00578+5620&  169.2& 2&  21.87&   7.81&   2.43&       &        & 16& F&
S6/3  &  SRb& 136& 1.8& V365 Cas\cr
  29 &   01186+6634&   46.6& 2&   5.10&   1.90&   0.42&       &        &   &  & 
     &     &    &    & \cr
  57 &   02228+3753&   95.6& 2&   9.89&   2.80&  0.66+&   0.70&        & 01& S&
S8,8  &  SR & 159& 1.1&BI And    \cr
 103 &   04352+6602&  398.5& 2&  42.56&  11.64&   3.70&   2.21&     yes& 17& S&
S4,7e &  M  & 373& 7.1&T Cam     \cr
 116 &   04543+4829&  175.5& 2&  12.81&   4.27&   1.38&       &     yes&   & F&
S5,8  &  SRb& 182& 0.7&TV Aur    \cr
 160 &   05440+1753&       &  &   4.72&   1.55&   0.50&       &        &   &  & 
     &  SRa& 364& 2.5&EI Tau    \cr
 237 &   06347+0057&   51.1& 2&   6.18&   2.21&   0.85&       &        &   &  & 
     &  Lb &    & 0.9&CX Mon     \cr
 312 &   07095+6853&  138.2& 2&  14.57&   5.98&  1.94+&   1.87&     yes&   &  &
M5S   & Lb  &    & 0.6&AA Cam     \cr
 387 & 07399$-$1045&  177.2& 2&  19.48&   7.25&   2.08&       &        & 18& S&
S3,6  & SRb &    & 2.8&SU Mon     \cr
 524 & 08308$-$1748&   65.5& 2&   5.85&   2.00& $^a$  &       &        &   &  & 
     & M   & 250& 1.5&SZ Pyx\cr
 556 & 08461$-$7051&   53.0& 1&   5.70&   1.49&   0.46&   0.33&        &   &  &
SC    & Lb: &    & 0.4&UX Vol       \cr
 617 & 09358$-$6536&   39.5& 1&   4.00&   1.22&   0.27&       &        &   &  &
S7,8  &     &    &    &      \cr
 696 & 10389$-$5149&  109.7& 1&  10.61&   3.21&   1.11&       &        &18 & S&
S5,2  & Lb  &    & 0.4&HP Vel       \cr
 787 & 12135$-$5600&       &  &  19.78&   6.55&   1.65&       &        &18 &  &
SC6/8e& M   &    & 4.3& BH Cru\cr
 803 &   12417+6121&   39.5& 3&   4.39&   1.38&   0.35&   0.39&   yes  &   &  &
S3/6e & M   & 226& 5.6&S UMa     \cr
 830 & 13440$-$5306&  129.5& 1&  15.60&   5.66&   2.08&       &        &18 &  &
SC    & Lb  &    & 1.2&AM Cen     \cr
 834 & 13477$-$6009&  387.6& 1&  49.16&  14.57&   3.42&       &        &17 & S&
S8,5  &  SR & 307& 3.3&VX Cen     \cr
 904 & 15548$-$7420&   58.1& 1&   6.12&   1.93&   0.65&   0.59&        &   &  &
SC    & SRa & 152& 3.5&VY Aps       \cr
 923 & 16209$-$2808&   71.2& 2&   7.71&   4.10&   1.27&       &        &   &  & 
     &     &    &    & \cr
 940 & 16472$-$6056&   49.2& 1&   5.71&   2.06&   0.72&       &        &   &  &
SC    & M:  &    & 2.3&LQ Ara      \cr
 978 & 17206$-$2826&  319.4& 2&  39.93&  11.48&   2.72&       &        &17 & S&
S5,4  & SRb & 320& 1.4&V521 Oph    \cr
     &   17553+4521&  545.0& 2&  55.41&  17.50&4.14+$^d$&3.20+&    yes &17 & 
&MII-III& SRb & 120& 0.9&OP Her\cr
1011 & 17562$-$1133&       &  &   8.45&   3.30&  $^c$ &       &        &01 &  & 
     &     &    &    &\cr
1056 & 18310$-$3541&   53.0& 7&   5.99&   2.05&   0.62&       &        &   &  &
S5,4  &  L  &    & 0.5&V3574 Sgr    \cr
1141 &   19311+2332&  258.2& 2&  32.48&  13.87&   3.06&       &        &16 & F&
S6/5  &   Lb&    & 3.7&EP Vul (SiO)\cr
1188 &   20026+3640&  356.8& 2&  40.22&  15.45&   5.37&       &    yes &31 & S&
S7,5  & SRb & 212& 3.0&AA Cyg    \cr
1310 &   22479+5923&   91.3& 2&  10.03&   3.25&  $^a$ &       &        &   & F&
S6/2  & SRb & 60 & 1.4&CV Cep\cr
1308 &   22476+4047&  766.3& 2& 100.51&  32.66& 10.25+& 10.07+&        &16 & S&
S7.5/1& SRb & 174& 1.7&RX Lac\cr
1339 &   23380+7009&   60.9& 2&   7.95&   2.89&   0.86&       &        &   & F&
M6S   &     &    &    &S1339       \cr
\hline
\end{tabular}
\medskip\\
a: profile badly distorted by close source\\
c: strong cirrus contamination\\
d: moves to Region B after removing the extended shell contribution\\
+: the `zero-crossing' flux density \Fz\ has been adopted instead of the template 
flux density $F_{\rm t}$, indicating a possibly
resolved shell (see Sect.~2.2)
\end{table*}

\clearpage
\newpage

\addtocounter{table}{-1}
\begin{table*}[]
\caption[]{Continued.}
\begin{tabular}{rrrrrrlrlllllllllllllllllllllll}
\noalign{Region E: Strong \ccc\ excess, many resolved shells}
\medskip\cr
\multicolumn{2}{c}{GCGSS \hfil IRAS}
& F2.2   &\multicolumn{2}{c}{Ref \hfil F12} &   F25  &   F60 &   F100&   Tc
&LRS&VC& Sp & Var & $P$ & $\Delta V$ & Name \cr
& & (Jy) & & (Jy) & (Jy) & (Jy) & (Jy) & & & & & & (d) & (mag)\cr
\hline\cr
  82 &   03452+5301&   66.7& 2&   7.66&   3.36&2.40$^b$&      &        & 01& F&
S5,8  &  Lb &    & 0.8&WX Cam\cr
  89 &   04123+2357&  258.5& 2&  25.13&   7.63&  3.30+&   2.56&     yes&   & S& 
     &     &    &    &+23:654    \cr
 114 &   04497+1410&  1010.& 2&  87.25&  22.19&  7.52+&   3.57&     yes& 18&  &
S3,1  &  SRb& 30 & 0.2&$o^1$ Ori\cr
 117 &   04599+1514&   88.8& 2&  11.11&   3.45&   1.44&       &        & 17&  &
SC    &  SRb& 370& 1.1&GP Ori    \cr
 212 &   06197+0327&  144.7& 2&  17.16&   4.63&  2.66+&  5.88+&        & 18& S&
SC    &  SR & 310& 1.1&FU Mon    \cr
 422 & 07507$-$1129&   79.5& 2&   7.21&   1.82&   0.48&       &     yes&   & I&
S5/2  & Lb  &    & 0.1&NQ Pup\cr
 540 & 08388$-$5116&    3.9& 1&   5.39&   2.02&   0.97&       &        &   &  &
S6,8  &     &    &    &S325\cr
 652 & 10015$-$4634&   58.7& 1&   6.31&   1.84&   1.23&       &        &   &  &
S5,8  &     &    &    &       \cr
 817 & 13163$-$6031&   94.7& 1&  21.52&  12.91&   7.01&       &        &   &  &
SC5:/8&     &    &    & TT Cen\cr
 929 & 16316$-$5026&   71.2& 8&       &       &       &       &        &42 &  &
Se    &     &    &    &\cr
&\multicolumn{3}{r}{JD 2\ts445\ts380:}
                              &  65.34&  31.24& 16.70+& 30.18+\cr
&\multicolumn{3}{r}{JD 2\ts445\ts600:}
                              &  98.46&  45.58& 20.72+& 38.74+\smallskip\cr
1189 &   20044+2417&  129.5& 2&  19.16&   7.55&  4.95+&  7.30+&        &16 & F&
S4,2  & SRa & 370& 2.0&DK Vul    \cr
1195 & 20100$-$6225&  133.2& 1&  11.03&   2.82&  1.77+&  3.27+&    yes &18 & S&
S4,4  &     &    &    &HD 191630\cr
1196 & 20120$-$4433&  187.2& 1&  39.13&  25.58& 13.35+& 25.88+&        & 16& F&
S4,4  & SRb & 223& 2.6&RZ Sgr\cr
\hline
\end{tabular}
\medskip\\
b: close weak source apparent on some scans\\
+: the `zero-crossing' flux density \Fz\ has been adopted instead of the template 
flux density $F_{\rm t}$, indicating a possibly
resolved shell (see Sect.~2.2)

\end{table*}
\clearpage
\newpage

\begin{table*}
\caption[]{\label{Tab:extended}
S stars whose circumstellar envelopes are resolved at 60~$\mu$m, listed
according to the region to which they belong in the ($K - [12]$, \ccc)
diagram (see Fig.~\protect\ref{3}). In column Rem, YPK+ and YPK$-$
denote stars found to be extended or not, respectively, by Young et al. (1993a).
Stars with no YPK label were not examined by Young et al. (1993a). The
mass loss rates and wind velocities listed in columns $\mathaccent 95 M$ and
$V_{\rm e}$, respectively, are taken from Table~6
}
\begin{tabular}{lllrrrllrcl}
IRAS       & Name & Var & $F60$ &
$F_{\rm z}/F_{\rm p}$ & $W25$ & $W50$ & $QF$ &\multicolumn{1}{c}{$\mathaccent 95 M$}
&\multicolumn{1}{c}{$V_{\rm e}$}
& Rem.\cr
&          &      & (Jy)& & ($'$)  &($'$)& &\multicolumn{1}{c}{ (\Msun\ y$^{-1}$)}
& \multicolumn{1}{c}{(\kms)}\cr 
\hline\cr
\noalign{\hfil Region B\hfil}
\smallskip\cr
00192$-$2020&T Cet     &  SRc & 14.6 & 1.05 & 1.96 & 1.44& 14.5& 4.0(-8)$^a$& 6.9&
YPK$-$,b\cr
19486+3247 & $\chi$ Cyg&  M   &      &      &      &     &     & 3.5(-7)   & 9.5&
YPK+,f\cr
\multicolumn{3}{r}{JD 2\ts445\ts450:}
                              & 95.5 & 1.01 & 1.96 & 1.45&  9.5\cr
\multicolumn{3}{r}{JD 2\ts445\ts640:}
                              & 71.6 & 1.002& 1.93 & 1.42&  0.4\cr
09076+3110 & RS Cnc    &  SRc & 34.6 & 1.04 & 1.97 & 1.44& 29.2& 1.7(-7)   & 7.4&
YPK+,b,f\cr
           & RS Cnc    &      &      &      &      &     &     & 4.5(-8)   & 2.6
& e\cr
15492+4837 & ST Her    &  SRb & 17.1 & 1.02 & 1.95 & 1.43&  6.7& 1.7(-7)   & 9.5&
YPK+,b\cr

\medskip\cr
\noalign{\hfil Region C \hfil}
\smallskip\cr
00213+3817 & R And     &  M   & 26.4 & 1.05 & 2.01 & 1.47& 30.6& 7.5(-7)   & 9.3&
YPK$-$\cr
01159+7220 & S Cas     &  M   & 30.1 & 1.01 & 1.94 & 1.42& 10.4& 3.5(-6)   &20.0&
YPK$-$\cr
07245+4605 & Y Lyn     &  SRc & 13.2 & 1.15 & 2.00 & 1.45& 30.9&
2.0(-7)   & 8.6& YPK+,d\cr
19126-0708 & W Aql     &  M   &      &      &      &     &     & 1.1(-5)   &18.7&
YPK$-$\cr
\multicolumn{3}{r}{JD 2\ts445\ts440:}
                              &132.7 & 1.006& 1.93 & 1.44&  5.8\cr
\multicolumn{3}{r}{JD 2\ts445\ts630:}
                              &100.6 & 1.02 & 1.95 & 1.44&  7.5\cr
19133$-$1703&T Sgr     & M    &  4.8 & 1.11 & 2.05 & 1.48&  8.0& 3.5(-7)   &14.1&
b\cr 
19354+5005 & R Cyg     &  M   & 12.2 & 1.03 & 1.96 & 1.44& 10.0& 1.5(-6)   &10.5&
YPK+,f\cr
22196$-$4612&$\pi^1$ Gru& SRb &      &      &      &     &     & 4.0(-7)   &11.0&
YPK+,f\cr
\multicolumn{3}{r}{JD 2\ts445\ts470:}
                              & 88.7 & 1.04 & 1.98 & 1.44& 37.9\cr
\multicolumn{3}{r}{JD 2\ts445\ts645:}
                              & 92.2 & 1.07 & 2.02 & 1.49& 38.3
\medskip\cr
\noalign{\hfil Region D \hfil}
\smallskip\cr
02228+3753 & BI And    & SR   &  0.7 & --   & 2.17 & 1.58&  2.6& \cr
07095+6853 & AA Cam    & Lb   &  1.9 & --   & 2.08 & 1.57&  4.1& 2.5(-7)   &17.9&
\cr
17553+4521 & OP Her    & SRb  &  4.1 & 1.15 & 2.16 & 1.53& 42.4&           &   
&b\cr
22476+4047 & RX Lac    & SRb  & 10.2 & 1.16 & 2.11 & 1.52& 23.6& 1.5(-8)   & 3.4&b
\medskip\cr
\noalign{\hfil Region E \hfil}
\smallskip\cr
04123+2357 & BD+23$^\circ$654&                              
                         --   &  3.3 & 1.20 & 2.18 & 1.53&  7.2& \cr
04497+1410 & $o^1$ Ori & SRb  &  7.5 & 1.48 & 2.89 & 1.71& 34.8& $<4.0$(-8)&   
& c\cr  
06197+0327 & FU Mon    & SR   &  2.7 & 1.40 & 2.42 & 1.62& 12.6& 9.0(-8)   & 2.8&
b\cr
16316$-$5026&S929      & --\cr   
\multicolumn{3}{r}{JD 2\ts445\ts380:}
                              & 16.7 & 1.50 & 3.08 & 1.81& 24.8&           &   
&b\cr      
\multicolumn{3}{r}{JD 2\ts445\ts600:}
                              & 20.7 & 1.89 & --   & 1.87& 24.1&           &   
&b\cr   
20044+2417 & DK Vul    & SRa  &  4.9 & 1.22 & 2.32 & 1.60& 25.3& 2.5(-7)   &
5.0&b\cr
20100$-$6225&HD 191630 & --   &  1.8 & 1.35 & 2.57 & 1.67& 24.0& $<9.0$(-7)&   
&b,c\cr 
20120$-$4433&RZ Sgr    & SRb  & 13.3 & 1.26 & 2.28 & 1.58& 62.1& 1.8(-6)   &
8.8&YPK+,b,d\cr
\hline
\end{tabular}
\bigskip\\
Notes:\\
a: numbers between parentheses refer to power of ten\\
b: 100~$\mu$m profile wider than point-source template\\
c: listed as extended at \fff\ in the IRAS {\it Small Scale Structure
Catalogue}\\  
d: listed as extended at 100 $\mu$m in the IRAS {\it Small Scale Structure
Catalogue}\\
e: two-components wind\\
f: circumstellar shell resolved in CO (Sahai 1992; Stanek et al. 1995)
\end{table*}
\clearpage
\newpage

\begin{table}
\caption[]{\label{Tab:maser}
Masers in S stars }
\begin{tabular}{llllllll}
Star& IRAS& SiO& H$_2$O& OH& Refs\cr
\hline\cr
\noalign{\bf Region A}\cr
BD~Cam& 03377+6303& -& N& -& B96\cr
NZ~Gem& 07392+1419& -& N& -& B96\cr
HR~Peg& 22521+1640& N& N& -& B96, ~Z89\cr
GZ~Peg& 23070+0824& N& N& -& D73, ~H89\cr
& & & & & \cr
\noalign{\bf Region~B}\cr
T~Cet& 00192-2020& N& N& N& A89a, ~A90, ~D76,\cr
& & & & & H94, ~L78, ~P71\cr
V535~Ori& 05208-0436& -& N& N& D89, ~H91\cr
Z~Tau& 05495+1549& N& -& N& B78, ~J91\cr
RS~Cnc& 09076+3110& N& N& Y?& D73, ~D76, ~A92,\cr
& & & & & D82, ~K77, ~L78,\cr
& & & & & L90, ~P92, ~R76\cr
Z~Ant& 10436-3459& -& -& N& H91\cr
ST~Her& 15492+4837& -& N& N& D73, ~D76, ~D82,\cr
& & & & & P79\cr
S~Her & 16496+1501& N& N& -& B77a, ~B84, ~B87,\cr
& & & & & B90, ~E88, ~S81\cr 
$\chi$ Cyg& 19486+3249& Y& N& N& A89b, ~A92,\cr
& & & & & B87, ~B94, ~C83a,\cr
& & & & & C96, ~D76, ~J84,\cr
& & & & & J85, ~J87, ~J91,\cr
& & & & & N85, ~O85, ~P92\cr
W~Cet& 23595-1457& N& N& -& B77a, ~B84, ~L76,\cr
& & & & & L78\cr
& & & & & \cr
\noalign {\bf Region~C}\cr
R~And& 00213+3817& Y& N& N& B77a, ~B78, ~B87,\cr
& & & & & B94, ~B96, ~C96,\cr
& & & & & D78, ~F75, ~J91,\cr
& & & & & T94\cr
RW~And& 00445+3224& N& N& N& B77a, ~B78, ~C83b,\cr
& & & & & C96, ~F75, ~F78,\cr
& & & & & K77, ~W72\cr
S~Cas& 01159+7220& Y& N& N& B77a, ~B94, ~F73,\cr 
& & & & &  T94\cr
RZ~Per& 01266+5035& N& -& -& B77a\cr
W~And& 02143+4404& Y& N& N& A89b, ~B77a, ~B87,\cr
& & & & &  B94, ~C96, ~D96,\cr
& & & & &  D78, ~K77, ~O80,\cr
& & & & &  T94, ~W72\cr
NO~Aur& 05374+3153& -& N& N& B75, ~B81, ~B96\cr
R~Lyn& 06571+5524&  Y& N& -& B77a, ~B94, ~K84,\cr
& & & & &  S81\cr
R~Gem& 07043+2246&  N& N& N& B77a, ~B78, ~B96,\cr
& & & & &  J91, ~L78, ~W72\cr
RR~Mon& 07149+0111& N& -& N& K77, ~L78\cr
TT~CMa& 07197-1451& N& N& N& B90, ~B96, ~H91\cr
Y~Lyn& 07245+4605& Y& N& N& A90, ~B75, ~B81,\cr
& & & & & D73, ~D82\cr
RR~Car& 09564-5837& N& N& -& H94, ~L77\cr
ST~Sco& 16334-3107& N& N& N& B77a, ~B94, ~J88,\cr
& & & & &  L77, ~W72\cr
RT~Sco& 17001-3651& N& N& N& B77b, ~C71, ~L76,\cr
& & & & &  L77, ~L78, ~S88\cr
\hline
\end{tabular}
\end{table}

\addtocounter{table}{-1}
\begin{table}
\caption[]{
Masers in S stars (Continued)}
\begin{tabular}{llllllll}
Star& IRAS& SiO& H$_2$O& OH& Refs\cr
\hline\cr
\noalign{\bf Region ~C (continued)}\cr
TV~Dra& 17081+6422& N& -& N& B94, ~W72\cr
ST~Sgr& 18586-1249& N& N& -& B77a\cr
W~Aql& 19126-0708& Y& N& N& B77a, ~B84, ~B87,\cr
& & & & &  B94, ~D89, ~L78,\cr
& & & & &  W72, ~Z79\cr
T~Sgr& 19133-1703& N& N& -& B77a, ~B90\cr
R~Cyg& 19354+5005& Y& N& N& B94, ~B96, ~C96,\cr
& & & & & P92, ~W72, ~Z79\cr
RZ~Peg& 22036+3315& -& N& -& B96\cr
$\pi^1$ Gru& 22196-4612& N& N& -& D89, ~H90\cr
WY Cas& 23554+5612& -& N& -& B77a, ~C83b\cr
& & & & & \cr
\noalign{\bf Region ~D}\cr
U~Cas& 00435+4758& N& N& -& B77a, ~C83b, ~J91,\cr
& & & & & S81\cr
RR~And& 00486+3406& N& N& N& B77a, ~F78\cr
S29   & 01186+6634& -& N& -& C83b\cr
BI~And& 02228+3753& -& N& -& C83b\cr
T~Cam& 04352+6602& N& N& -& B77a, ~B94, ~D76,\cr
& & & & & J91\cr
TV~Aur& 04543+4829& N& -& -& B77a, ~B94\cr
EI~Tau& 05440+1753& N& -& -& B77a\cr
AA~Cam& 07095+6853& -& N& -& B96\cr
SU~Mon& 07399-1045& N& -& -& B94\cr
SZ~Pyx& 08308-1748& -& N& -& C83b\cr
S~UMa& 12417+6121& N& N& -& B77a, ~B96, ~D78\cr
VX~Cen& 13477-6009& N& -& -& P92\cr
OP~Her& 17553+4521& -& N& -& B96\cr
EP~Vul& 19311+2332& Y& -& N& B94, ~R76\cr
AA~Cyg& 20026+3640& N& N& -& B77a, ~B94, ~B96,\cr
& & & & & D82\cr
RX~Lac& 22476+4047& -& N& N& D73, ~D82, ~W72\cr
& & & & & \cr
\noalign{\bf Region ~E}\cr
$o^1$ Ori& 04497+1410& -& N& N& B96, ~D73, ~P71\cr
GP~Ori& 04599+1514& N& -& -& L78\cr
FU~Mon& 06197+0327& N& -& -& B77a\cr
RZ~Sgr& 20120-4433& -& -& N& C71\cr
\hline
\end{tabular}

\vspace*{\fill}

\clearpage
\newpage

\medskip
\end{table}

\begin{table}
\def\ref{\hangindent 2pc \hangafter 1}
{\small
\noindent{References to Table~3.
The parentheses at the end of each 
reference indicate which maser is probed by the paper}\\

\hrulefill\\

\ref
A89a: Allen D.A., Hall P.J., Norris R.P., Troup E.R., Wark R.M., Wright A. E.,
1989, MNRAS 236, 363 (SiO)

\ref
A89b: Alcolea J., Bujarrabal V., Gallego J. D., 1989, A\&A 211, 187
(SiO)

\ref
A90: Alcolea J., Bujarrabal V., Gomez-Gonzalez J.,  1990, A\&A 231, 431
(SiO)

\ref
A92: Alcolea J., Bujarrabal V., 1992, A\&A 253, 475 (SiO)

\ref
B75: Bowers P.F., 1975, AJ 80, 512 (OH)

\ref
B77a: Blair G.N., Dickinson D.F., 1977, ApJ 215, 552 (SiO, $\rm H_2O$)

\ref
B77b: Bowers P.F., Kerr F.J., 1977, A\&A 57, 115 (OH)

\ref
B78: Bowers P.F., Sinha R.P., 1978, AJ 83, 955 (OH)

\ref
B81: Bowers P.F., 1981, AJ 86, 1930 (OH)

\ref
B84: Bowers P.F., Hagen W., 1984, ApJ 285, 637 ($\rm H_2O$)

\ref
B87: Bujarrabal V., Planeses P., del Romero A., 1987, A\&A 175, 164
(SiO)

\ref
B90: Benson P.J., Little-Marenin I.R., Woods T.C., Attridge J.M.,
Blais K.A., Rudolph D.B., Rubiera M.E., Keefe H.L., 1990, ApJS 74, 911
(SiO, $\rm H_2O$, OH)

\ref
B94: Bieging J.H., Latter W.B., 1994, A\&A 422, 765 (SiO)

\ref
B96: Benson P.J., Little-Marenin I.R., 1996, ApJS 106, 579 ($\rm H_2O$)

\ref
C71: Caswell J.L., Robinson B.J., Dickel H.R., 1971, Ap. Letters 9, 61
(OH)

\ref
C83a: Clemens D.P., Lane A.P., 1983, ApJ 266, L117
(SiO)

\ref
C83b: Crocker D.A., Hagen W., 1983, A\&AS 54, 405 ($\rm H_2O$)

\ref
C96: Cho S.H., Kaifu N., Ukita N., 1996, A\&AS 115, 117 (SiO)

\ref
D73: Dickinson D.F., Bechis K.P., Barrett, A.H., 1973, ApJ 180, 831
($\rm H_2O$)

\ref
D76: Dickinson D.F., 1976, ApJS 30, 259 ($\rm H_2O$)

\ref
D78: Dickinson D.F., Snyder L.E., Brown L.W., Buhl D., 1978, AJ
  83, 36 (SiO)

\ref
D82: Dickinson D.F., Dinger A.S., 1982, ApJ 254, 136 ($\rm H_2O$)

\ref
D89: Deguchi S., Nakada Y., Forster J.R., 1989, MNRAS 239, 825
($\rm H_2O$)

\ref
E88: Engels D., Schmid-Burgk J., Walmsley C.M., 1988, A\&A 191, 283

\ref
F73: Fillit R., Foy R., Gheudin M., 1973, Ap. Letters 14, 135 (OH)

\ref
F75: Foy R., Heck A., Mennessier M.O., 1975, A\&A 43, 175 (OH)

\ref
F78: Fix J.D., Weisberg J.M., 1978, ApJ 220, 836 (OH)

\ref
H89: Heske A., 1989, A\&A 208, 77 (SiO)

\ref
H90: Haikala L.K., 1990, A\&AS 85, 875  (SiO)

\ref
H91: te Lintel Hekkert P., Caswell J.S., Habing H.J., Haynes R.F.,
Norris R.P., 1991, A\&AS 90, 327 (OH)

\ref
H94: Haikala L.K., Nyman L.A., Forsstrom V., 1994, A\&AS 103, 107
(SiO)

\ref
J84: Jewell P.R., Batrla W., Walmsley C.M., Wilson T.L., 1984,
  A\&A 130, L1  (SiO)

\ref
J85: Jewell P.R., Walmsley C.M., Wilson T.L., Snyder L.E., 1985,
  ApJ 298, L55  (SiO)

\ref
J87: Jewell P.R., Dickinson D.F., Snyder L.E., Clemens D.P., 
  1987, ApJ 323, 749  (SiO)

\ref
J91: Jewell P.R., Snyder L.E., Walmsley C.M., Wilson T.L., 
  Gensheimer P.D., 1991, A\&A 242, 211 (SiO)

\ref
K77: Kolena J., Pataki L., 1977, AJ 82, 150 (OH)

\hrulefill\\
}
\end{table}

\begin{table}
\def\ref{\hangindent 2pc \hangafter 1}
References to Table 3 (continued).\\
\small{
\hrulefill\\

\ref
K84: Kuiper T.B.H., Swanson P.N., Dickinson D.F., Rodr\'\i guez Kuiper
E.N., Zimmermann P., 1984, ApJ 286, 310 ($\rm H_2O$)

\ref
L76: L\'epine J.R.D., Paes de Barros M.H., Gammon R.H., 1976,
  A\&A 48, 269  ($\rm H_2O$)

\ref
L77: L\'epine J.R.D., Paes de Barros M.H., 1977, A\&A 56, 219
($\rm H_2O$)

\ref
L78: L\'epine J.R.D., Le Squeren A.M., Scalise E., 1978, ApJ 225, 869
(SiO)

\ref
L90: Lewis B.M., Eder J., Terzian Y., 1990, ApJ 362, 634  (OH)

\ref
N85: Nyman L.A., Olofsson H., 1985, A\&A 147, 309 (SiO)

\ref
O80: Olnon F.M., Winnberg A., Matthews H. E., Schultz G.V., 1980,
  A\&AS 42, 119 (OH, $\rm H_2O$)

\ref
O95: Olofsson H., Rydbeck O.E.H., Nyman L.A., 1985, A\&A 150, 169
(SiO)

\ref
P71: Pashchenko M.I., Slysh V., Strukov I., Fillit R., Gheudin M.,
Nguyen-Quang-Rieu, 1971, A\&A 11, 482 (OH)

\ref
P79: Pashchenko M.I., Rudnitskij G.M., 1979, Astron. Tsirk. 1040, 4

\ref
P92: Patel N.A., Joseph A., Ganesan R., 1992, J. Astron. Astrophys.
  13, 241 (SiO)

\ref
R76: Rudnitskij G.M., 1976, Soviet AJ 20, 693 (OH)

\ref
S81: Spencer J.H., Winnberg A., Olnon F.M., Schwartz P.R., Matthews
H.E., Downes D., 1981, AJ 86, 392 (SiO)
 
\ref
S88: Sivagnanam P., Le Squeren A.M., Foy R., 1988, A\&A 206, 285 (OH)

\ref
T94: Takaba H., Ukita N., Miyaji T., Miyoshi M., 1994, PASJ 46, 629
($\rm H_2O$)

\ref
W72: Wilson W.J., Barrett A.H., 1972, A\&A 17, 385 (OH)

\ref
Z79: Zuckerman B., 1979, ApJ 230, 442 (SiO)

\hrulefill\\

}

\end{table}
\clearpage
\newpage

\begin{table*}
\caption[]{\label{Tab:COstars}
S stars observed in CO at the Caltech Submillimeter Observatory}
\bigskip
$$\vbox{
\tabskip 1em plus 2em
\halign to \hsize{\hfil$\rm #$\hfil& & \hfil$\rm #$\hfil \cr
GCGSS& Other& \alpha(1950)& \delta(1950)& \it{l}& \it{b}& SpT& Var& \it{P} 
& \it{V}_{LSR}&
r.m.s.\cr
& & & & & & & & (d)& (km~s^{-1})& (mK)\cr
\hline\cr
& & & & & & & & & & \cr
89& HD~26816 & 04~12~23.7& +23~57~15& 171.6& -19.0& S & Lb: & - & - & 17\cr
117& GP~Ori& 04~59~56.7& +15~14~58& 186.0& -15.7 & SC7/8 & SRb &
370:& +86,~+92 & 37\cr
149& NO~Aur& 05~37~26.8& +31~53~43& 176.9& +0.7  & M2IIIS& Lc&     -  &  -7,~-3&
13\cr                                
212& FU~Mon& 06~19~46.1& +03~27~00& 206.6& -4.9  & SC  & SR&       310&      -46,
 ~-39& 10\cr                               
283& R~Lyn& 06~57~10.8& +55~24~06& 161.2& +23.4  &S5/5e& Mira&     378&  +9,~+28&
18\cr                                
422& NQ~Pup& 07~50~43.6& -11~29~41& 230.5& +8.04& S5/2 & Lb & - & - & 42\cr
589& RS~Cnc& 09~07~37.8& +31~10~03& 194.5& +42.1 & M6IIIaSe& SRc?& 120&    
+7,~+11&23\cr                             
626& FM~ Hya& 09~41~10.3& -18~20~43& 252.8& +25.5& M0S & Mira&     300&  -&  
17\cr
704& Z~Ant& 10~43~40.2& -34~59~17& 275.9& +20.9  & S5,4& SR&       104&      -& 
   10\cr
796& HR~4755& 12~27~16.8& -41~27~35& 298.6& +21.0& M2II-III&       -  &  -&     
-2& 10\cr
803& S~UMa& 12~41~45.5& +61~22~00& 124.6& +56.0  & S3/6e& Mira&    226&     
+5,~+18& 10\cr                            
816& UY~Cen& 13~13~36.8& -44~26~27& 307.6& +17.9 &S6/8,CS& SR&     114&  -23&39\cr
\hline\cr
}}$$                                       
\bigskip                                   
\bigskip                                   
\parindent=0pc                                      
                                           
{Notes to Table~\thetable:}\\
1. The spectral type in column 8 is from the GCGSS\\
2. Column 10 gives the observed range of optical velocities, taken from the
following references: Brown et al. 1990 (GCGSS 149 and 589),
Udry et al. (1997) and Jorissen et al. (1997) (GCGSS 117 and 212), 
Wilson 1953 (GCGSS 283, 803 and 816), 
Hoffleit (1982) (GCGSS 796)\\
3. All stars observed in the CO(2--1) line except GP Ori, NQ Pup and UY Cen,
which were observed in the CO(3--2) line

\end{table*}
\clearpage
\newpage

\begin{table*}
\caption[]{\label{Tab:COresults}
CO(2--1) data for the four S stars detected by
the CSO observations}
\bigskip
$$\vbox{
\tabskip 1em plus 2em
\halign to \hsize{\hfil$\rm #$\hfil& & \hfil$\rm #$\hfil \cr
GCGSS& & I_{CO}& T(peak)& V_c& V_e\cr 
& & (K \times km~s^{-1})& (K)& (km~s^{-1})& (km~s^{-1})\cr
\hline\cr 
& & & & & & \cr
212& FU~Mon& 0.63 \pm 0.05& 0.15 \pm 0.02& -41.7 \pm 0.2& 2.8 \pm 0.3\cr
283& R~Lyn& 0.89 \pm 0.19& 0.08 \pm 0.02& +15.7 \pm 1.4& 9.3 \pm 1.7\cr
589& RS~Cnc& 13.3~ \pm 0.9~& 0.78~~& +7.7~~& 7.4~~\cr
& & & 1.17~~& +6.8~~& 2.6~~\cr
704& Z~Ant& 0.73 \pm 0.09& 0.07 \pm 0.02& -15.6 \pm 1.0& 7.4 \pm 1.3\cr
& & & & & & & & &\cr
\hline\cr
}}$$
\bigskip
\bigskip
\end{table*}

\clearpage
\newpage

\newlength{\widthsav}
\newlength{\oddsidemarginsave}
\newlength{\evensidemarginsave}
\setlength{\widthsav}{\textwidth}
\setlength{\textheight}{16cm}
\setlength{\textwidth}{24cm}
\setlength{\oddsidemarginsave}{\oddsidemargin}
\setlength{\evensidemarginsave}{\evensidemargin}
\setlength{\oddsidemargin}{2pt}
\setlength{\evensidemargin}{2pt}

\tabcolsep 3pt

\begin{landscape}
\begin{table*}
\caption[]{\label{Tab:COlit}
A compilation of CO data for S stars (columns 3--10; for data prior to 1990, 
see Loup et al. 1993) and mass loss rates 
fitting these data (columns 11--19), assuming $T_{\star}$ = 2500 K, $R_{\star}  
= ~ 2.5 \times 10^{13}$ cm, and $f =  n(\mbox{\rm CO})/n(\mbox{H}_2) = 
6.5 \times 10^{-4}$}
\begin{tabular}{rc|cccccccc|ccccccccccccc}
& & \multicolumn{7}{c}{Observations}& & &\multicolumn{8}{c}{Model}\cr
\cline{3-10}\cline{11-19}
IRAS& Star& Line& HPBW& rms& $I_{\rm CO}$& $T_{\rm MB}$ & $V_{\rm c}$& $V_{\rm e}$
& Ref
& $D$ & $V_{\rm e}$ & ${\rm d}M/{\rm d}t$ &\multicolumn{2}{c}{radius}& Line & HPBW 
& $T_{\rm MB}$
& $I_{\rm CO}$\cr
\cline{14-15}  
& & & $''$& (K)& (K~km/s)& (K)& (km/s)& (km/s)&
& (pc)& (km/s)& (M$_{\odot}$~yr$^{-1}$)& (cm) & $('')$ & &
($''$) & (K)& (K km/s)\cr
\hline\cr
\noalign{\bf{Region~A:}}
03377+6303  & BD~Cam& 1$-$0& 50&   0.02&  $-$&    $-$&     $-$&      $-$&   BL94
& 92& (8.5)&$<1.6\;10^{-8}$& & & & & \cr
& &  2$-$1& 13&   0.05&    $-$&    $-$&     $-$&      $-$&   SL95 \cr
& & & & & & & & & & & & & & & & & \cr
12272$-$4127& HR~4755& 2$-$1& 30& 0.01& $-$& $-$& $-$&  $-$& this~work & 145& (8.5)
&$<2.3
\;10^{-8}$& & & & & \cr
& & & & & & & & & & & & & & & & & \cr
13372$-$7136& GCGSS~826& 1$-$0& 45& 0.01&  $-$& $-$&  $-$& $-$&  SL95 & 130& (8.5)
&$<5.8 \;10^{-8}$& & & & & \cr
& & & & & & & & & & & & & & & & & \cr
22521+1640  & HR~Peg& 1$-$0& 50& 0.01 &  $-$&  $-$&  $-$& $-$&   BL94 & 125& (8.5)
&$<6.3 \;10^{-8}$& & & & & \cr
\hline \cr
\noalign{\bf{Region~B:}}
00192$-$2020& T~Cet& 1$-$0& 100& 0.14& $-$& $-$& $-$& $-$& K97 & & & & & & & & & \cr
& & 2$-$1& 30& 0.05& 2.6& 0.28& +23.9& 6.9& K97& 130& 6.9&4.6
10$^{-8}$& 1.26  $10^{16}$& 6.5&  2$-$1 & 30 & 0.24& 3.32\cr
& & & & & & & & & & & & & & & & & \cr
*09076+3110& RS~Cnc& 1$-$0& 33& 0.07& 7.3&  0.77& +7.5&  7.0& N92\cr
& & 2$-$1& 30& 0.02& 13.3&  0.78& +7.7&  7.4& this~work& 200& 7.4& 1.9
$10^{-7}$& 2.61 $10^{16}$& 8.7 & 2$-$1 & 30 & 0.79& 10.63\cr
& & 2$-$1& 30& &  &        1.17& +6.8&  2.6& this~work &    & 2.6& 5.2
$10^{-8}$& 1.81 $10^{16}$& 6.0 & 2$-$1 & 30 & 1.18&  5.61\cr
& & 3$-$2& 20& 0.07& 29.8&  4.3&  +6.8&  4.8& S95\cr
& & 3$-$2& 20& 0.07& 21.9&  2.63& +6.8&  5.8& K97\cr
& & & & & & & & & & & & & & & & & \cr
09410$-$1820& FM~Hya& 2$-$1& 30& 0.02&  $-$&  $-$&  $-$&   $-$&
this~work & 1800&  (8.5)& $< 1.0\;  10^{-6}$& & & & & \cr
& & & & & & & & & & & & & & & & & \cr
10436$-$3459& Z~Ant& 2$-$1& 30& 0.01& 0.73& 0.07& $-$15.6&  7.4&
this~work& 900&  7.4& 2.9 $10^{-7}$ & 3.26 $10^{16}$ & 2.4 & 2$-$1 &
30 & 0.07&  0.84\cr
& & & & & & & & & & & & & & & & & \cr
15492+4837& ST~Her& 1$-$0& 22&  $-$&   4.9&   0.24& $-$5.3& 8.6& KJ94
& 290& 9.5& 2.0 $10^{-7}$& 2.43  $10^{16}$ & 5.6 & 1$-$0 & 22 & 0.17& 3.47\cr
& & 2$-$1& 13&  $-$&  17.4&  1.05& $-$4.6& 8.8& KJ94
&    &    &              &                 &     & 2$-$1 & 13 & 1.12& 20.50\cr
& & 2$-$1& 13& 0.10&  $-$&  0.45& $-$4.4& 9.1& SL95\cr
& & 3$-$2& 20& 0.07& 5.1& 0.34& $-$4.4& 12.0&  K97
&    &    &              &                 &     & 3$-$2 & 20 & 0.41 & 9.1 \cr
& & & & & & & & & & & & & & & & & \cr
19486+3247& $\chi$~Cyg& 1$-$0& 23&  0.10&  $-$&  2.20&  +9.6& 10.2&
H90 & 106& 9.5& 2.6  $10^{-7}$& 2.85  $10^{16}$& 18 & 1$-$0& 23 & 1.68& 37.79\cr
& & 1$-$0&  45&  $-$&  10.6&   $-$&  +10&   10&   M90 & & & &&& 1$-$0&
45 & 0.68& 13.45\cr
& & 1$-$0& 50& 0.02&11.8&  0.77& +9.7&  8.9& BL94	& & & &&&
1$-$0& 50 & 0.57& 11.14\cr
& & 1$-$0&100& 0.07& 3.6&  0.27& +9.7& 10.1& K97	& & & &&&
1$-$0& 100 &0.16& 3.00\cr
& & 2$-$1& 13& 0.24&   $-$&  6.2 &+10.0&  8.8& SL95	& & & &&&
2$-$1& 13 & 5.47& 105.2\cr
& & 2$-$1& 25& 0.03&25.4&  1.77& +9.9&  9.0& BL94	& & & &&&
2$-$1& 25 & 2.45& 43.76\cr
& & 2$-$1& 30& 0.15&22.7&  1.63& +9.9&  9.9& K97	& & & &&&
2$-$1& 30 & 1.85& 32.59\cr
& & 3$-$2& 20& 0.19&63.0&  4.7 & +7.8&  9.9& Y95	& & & &&&
3$-$2& 20 & 2.81& 49.28\cr
& & 3$-$2& 20& 0.19&41.5&  3.0 &+10.6&  9.2& S95\cr	
& & 3$-$2& 20& 0.31&44.3&  3.37&+10.4&  9.8& K97\cr
& & 4$-$3& 15& 0.24&52.7&  4.0 & +9.9&  9.0& Y95        & & & &&&
4$-$3& 15 & 3.55& 63.02\cr
\hline
\end{tabular}
\end{table*}
\clearpage
\newpage

\addtocounter{table}{-1}
\renewcommand{\thetable}{\arabic{table} (continued)}

\begin{table*}
\caption[]{}
\begin{tabular}{rc|cccccccc|ccccccccccccc}
& & \multicolumn{7}{c}{Observations}& & &\multicolumn{8}{c}{Model}\cr
\cline{3-10}\cline{11-19}
IRAS& Star& Line& HPBW& rms& $I_{\rm CO}$& $T_{\rm MB}$ & $V_{\rm c}$& $V_{\rm e}$
& Ref
& $D$ & $V_{\rm e}$ & ${\rm d}M/{\rm d}t$ &\multicolumn{2}{c}{radius}& Line & HPBW 
& $T_{\rm MB}$
& $I_{\rm CO}$\cr
\cline{14-15}  
& & & $''$& (K)& (K~km/s)& (K)& (km/s)& (km/s)&
& (pc)& (km/s)& (M$_{\odot}$~yr$^{-1}$)& (cm) & $('')$ & &
($''$) & (K)& (K km/s)\cr
\hline\cr
\noalign{\bf{Region~B (continued):}}
\smallskip\cr
20213+0047&  V865~Aql& 2$-$1& 25& 0.03& $-$& $-$&  $-$& $-$&   BL94& 840& (8.5)
& $<3.0\;  10^{-7}$& & & & & \cr
& & & & & & & & & & & & & & & & & \cr
23595$-$1457& W~Cet& 1$-$0& 23& 0.19&  $-$& $-$& $-$& $-$& H90&  950& (8.5)
& $< 3.4\;  10^{-7}$ & & & & & \cr
& & 1$-$0& 50& 0.02& $-$& $-$& $-$& $-$&   BL94\cr
& & 1$-$0& 45& 0.04& $-$& $-$& $-$& $-$&   SL95\cr
& & 3$-$2& 20& 0.03& $-$& $-$& $-$& $-$&   Y95\cr
\hline\cr
\noalign{\bf{Region C:}}
\smallskip\cr
00213+3817& R~And& 1$-$0&  45& $-$& 3.71&  $-$& $-$16& 11& M90 & 490&
9.3& $8.6\;  10^{-7}$& $5.66\;  10^{16}$& 7.7 & 1$-$0& 45 &  0.22& 3.71\cr
& & 1$-$0& 50& 0.04& 5.73& 0.44& $-$15.8& 9.3& BL94 & & & & & & 1$-$0&
50 & 0.18& 3.02\cr
& & 2$-$1& 25& 0.05& 6.5&  0.73& $-$15.6& 8.5& BL94 & & & & & & 2$-$1&
25 & 0.85& 11.46\cr
& & & & & & & & & & & & & & & & & \cr
01159+7220& S~Cas&  1$-$0& 23& 0.36&  $-$&    0.44& $-$30.8& 22.0&
H90& 860& 20.0& $4.0\; 10^{-6}$& $1.0\;  10^{17}$& 8.0 & 1$-$0& 23 & 0.43& 15.6\cr
& &  1$-$0& 45&  $-$& 3.43&  $-$& $-$29& 18&  M90   & & & & & & 1$-$0
& 45 & 0.12& 4.29\cr
& &  1$-$0& 33& 0.11& 6.0& 0.23& $-$26.0& 19.0&  N92& & & & & & 1$-$0
& 33 & 0.22& 7.8\cr
& &  1$-$0& 50& 0.03& 4.63&  0.14& $-$31.0&  22.0& BL94& & & & & &
1$-$0& 50 & 0.10& 3.49\cr
& & 2$-$1& 25& 0.03& 8.1& 0.29& $-$30.0& 19.7& BL94    & & & & & &
2$-$1& 25 & 0.35& 11.62\cr
& & & & & & & & & & & & & & & & & \cr
02143+4404& W~And& 1$-$0& 33& 0.09& 4.5&  0.43& $-$35.0& 10.4&  N92&
630& 8.5& $8.0\;  10^{-7}$& $5.65\;  10^{16}$& 5.9 & 1$-$0& 33 & 0.27& 4.15\cr
& & 1$-$0& 50& 0.02& 1.37& 0.12& $-$34.2&  7.8& BL94& & & & & & 1$-$0&
50 & 0.12& 1.83\cr
& & 2$-$1& 25& 0.01& 3.0&  0.29& $-$34.4&  6.5& BL94& & & & & & 2$-$1&
25 & 0.52& 7.04\cr
& & 3$-$2& 20& 0.08& 7.0&  0.59& $-$36.7&  8.9& Y95 & & & & & & 3$-$2&
20 & 0.55& 7.43\cr
& & & & & & & & & & & & & & & & & \cr
05374+3153& NO~Aur& 2$-$1& 13& 0.05&  $-$&    $-$&     $-$&    $-$&  SL95& 620
& (8.5)& $< 1.1\;  10^{-7}$& & & & & \cr
& & 2$-$1& 30& 0.01&  $-$&    $-$&     $-$&    $-$&  this~work\cr
& & & & & & & & & & & & & & & & & \cr
06331+1415& DY~Gem& 1$-$0& 50& 0.04&  $-$&    $-$&     $-$&    $-$&
BL94& 890& 8.0& $2.4\; 10^{-7}$& $2.87\;  10^{16}$& 2.1 & 1$-$0& 50 & 0.01& 0.18\cr
& & 2$-$1& 13& 0.05&  $-$&   0.25& $-$16.0&  8.0& SL95& & & & & &
2$-$1& 13 & 0.25& 3.50\cr
& & & & & & & & & & & & & & & & & \cr
06571+5524& R~Lyn& 1$-$0& 50& 0.01& 0.32& 0.03& +15.3&  7.5&  BL94&
950& 8.0& $3.5\;  10^{-7}$& $3.51\;  10^{16}$& 2.5 &1$-$0 & 50 & 0.017& 0.27\cr
& &  2$-$1&  25&  0.01&  0.54&  0.04& +16.6&  7.7&  BL94& & & & & &
2$-$1& 25 & 0.092& 1.25\cr
& &  2$-$1&  30&  0.02&  0.94&  0.08& +15.7&  9.3& this~work & & & & &
& 2$-$1& 30 & 0.064& 0.87\cr  
\hline
\end{tabular}
\end{table*}
\vfill
\clearpage
\newpage

\addtocounter{table}{-1}
\renewcommand{\thetable}{\arabic{table} (continued)}
\begin{table*}
\caption[]{}
\begin{tabular}{rc|cccccccc|ccccccccccccc}
& & \multicolumn{7}{c}{Observations}& & &\multicolumn{8}{c}{Model}\cr
\cline{3-10}\cline{11-19}
IRAS& Star& Line& HPBW& rms& $I_{\rm CO}$& $T_{\rm MB}$ & $V_{\rm c}$
& $V_{\rm e}$& Ref
& $D$ & $V_{\rm e}$ & ${\rm d}M/{\rm d}t$ &\multicolumn{2}{c}{radius}
& Line & HPBW & $T_{\rm MB}$
& $I_{\rm CO}$\cr
\cline{14-15}  
& & & $''$& (K)& (K~km/s)& (K)& (km/s)& (km/s)&
& (pc)& (km/s)& (M$_{\odot}$~yr$^{-1}$)& (cm) & $('')$ & &
($''$) & (K)& (K km/s)\cr
\hline\cr
\noalign{\bf{Region~C (continued):}}
\smallskip\cr
07043+2246& R~Gem& 1$-$0& 23 &0.10&  $-$&  0.65& $-$59.2&  6.0&  H90&
850& 5.4& $2.3\;  10^{-7}$& $7.63\;  10^{17}$&60.0 & 1$-$0& 23 & 0.10& 1.07\cr
& & 2$-$1& 13& 0.12&  $-$&  1.40& $-$59.1&  4.8& SL95& & & & & &
2$-$1& 13 & 0.55& 4.99\cr
& & 3$-$2& 20& 0.07& 1.7&  0.28& $-$59.0&  5.5& K97  & & & & & &
3$-$2& 20 & 0.18& 1.57\cr
& & & & & & & & & & & & & & & & & \cr
07149+0111& RR~Mon& 1$-$0& 45& 0.01&  $-$&    $-$&     $-$&    $-$&  SL95 
& 1200& (8.5)& $ < 7.0\;  10^{-7}$& & & & & \cr
& & & & & & & & & & & & & & & & & \cr
07245+4605& Y~Lyn& 1$-$0& 33& 0.08& 2.2&  0.20& $-$0.7&  8.6&  N92 &
330& 8.6& $2.3\;  10^{-7}$ & $2.73\;  10^{16}$& 5.5 &  1$-$0& 33 & 0.14& 2.50\cr
& & & & & & & & & & & & & & & & & \cr
09338$-$5349& UU~Vel& 1$-$0& 45& 0.11&  $-$&    $-$&     $-$&    $-$&  SL95
& 1260& (8.5)& $< 4.0\;  10^{-6}$& & & & & \cr
& & & & & & & & & & & & & & & & & \cr
13136$-$4426& UY~Cen& 2$-$1& 23& 0.02&  $-$&   0.07& $-$28.6& 13.1&
SL95 & 550& 13.1& $2.1\;  10^{-7}$& $2.36\;  10^{16}$& 2.9 & 2$-$1& 23
& 0.07& 1.67\cr
& & 3$-$2& 20& 0.04&  $-$&  $-$& $-$&  $-$&  this~work\cr
& & & & & & & & & & & & & & & & & \cr
16334$-$3107& ST~Sco& 1$-$0& 50& 0.02& 0.61& 0.06& $-$4.5&  6.5& BL94
& 540& 7.1& $1.2\;  10^{-7}$& $1.98\;  10^{16}$& 2.5 & 1$-$0 & 50 & 0.01 &  0.15 \cr
& & 1$-$0& 45& 0.02&  $-$&   0.14& $-$1.2&  7.3& SL95 & & & & & &
1$-$0& 45 & 0.01& 0.19\cr
& & 2$-$1& 25& 0.02& 1.17&   0.14& $-$5.8&  7.6& BL94 & & & & & &
2$-$1& 25 & 0.09& 1.24\cr
& & & & & & & & & & & & & & & & & \cr
17001$-$3651& RT~Sco& 1$-$0& 45& 0.02&  $-$&   0.20& $-$44.5&  13.1&
SL95& 490& 11.0& $4.6\;  10^{-7}$& $3.7\;  10^{16}$& 5.0 & 1$-$0& 45 &
0.06& 1.36\cr
& & 2$-$1& 23& 0.05&  $-$&   0.18& $-$44.4&  9.6& SL95 & & & & & &
2$-$1& 23 & 0.31& 5.92\cr
& & & & & & & & & & & & & & & & & \cr
17081+6422& TV~Dra& 1$-$0& 50& 0.02& 0.13& 0.03& +22.4&  3.7& BL94&
540& 5.0& $4.6\;  10^{-8}$& $1.34\;  10^{16}$& 1.6 & 1$-$0& 50 & 0.003& 0.034\cr
& & 2$-$1& 13& 0.04& 1.0&  0.15& +22& 6&  O93 & & & & & & 2$-$1& 13 & 0.17& 1.67\cr
& & 2$-$1& 25& 0.01& 0.40& 0.05& +21.1&  5.0& BL94 & & & & & & 2$-$1&
25 & 0.05& 0.46\cr
& & 2$-$1& 13& 0.09&  $-$&   0.30& +21.2&  5.1& SL95\cr
& & & & & & & & & & & & & & & & & \cr
18575$-$0139& VX~Aql& 1$-$0& 45& 0.12&  $-$&  $-$&  $-$& $-$&  SL95 &
1400& 7.8& $3.2\;  10^{-7}$& $3.41\;  10^{16}$& 1.6 & 1$-$0& 45 & 0.01& 0.15\cr
& &  2$-$1& 23& 0.01&  $-$& 0.05& +6.7&  7.8& SL95& & & & & & 2$-$1&
23 & 0.05& 0.67\cr
& & & & & & & & & & & & & & & & & \cr
18586$-$1249& ST~Sgr& 1$-$0& 50& 0.01& 0.42& 0.03& +54.6&  9.3& BL94 &
670& 10.0& $3.5\;  10^{-7}$& $3.25\;  10^{16}$ & 3.2 & 1$-$0& 50 & 0.022 & 0.04 \cr
& & 1$-$0& 45& 0.01&  $-$&  0.06& +55.1&  9.6& SL95 & & & & & & 1$-$0
& 45 & 0.027& 0.54\cr
& & 2$-$1& 25& 0.04&  $-$&    $-$&     $-$&    $-$&  BL94 & & & & & &
2$-$1& 25 & 0.12& 2.06\cr
& & 2$-$1& 23& 0.03&  $-$&  0.11& +49.8& 11.2& SL95& & & & & & 2$-$1&
23 & 0.14& 2.53\cr
& & & & & & & & & & & & & & & & & \cr
19111+2555& S~Lyr& 2$-$1& 13& 0.04&  $-$&   0.40& +51.2& 13.9& SL95 &
1500& 13.9& $2.1\;  10^{-6}$& $8.10\;  10^{16}$& 3.6 & 2$-$1& 13 & 0.40& 9.0\cr
\hline
\end{tabular}
\end{table*}
\vfill
\clearpage
\newpage

\addtocounter{table}{-1}
\renewcommand{\thetable}{\arabic{table} (continued)}
\begin{table*}
\caption[]{}
\begin{tabular}{rc|cccccccc|ccccccccccccc}
& & \multicolumn{7}{c}{Observations}& & &\multicolumn{8}{c}{Model}\cr
\cline{3-10}\cline{11-19}
IRAS& Star& Line& HPBW& rms& $I_{\rm CO}$& $T_{\rm MB}$ & $V_{\rm c}$
& $V_{\rm e}$& Ref
& $D$ & $V_{\rm e}$ & ${\rm d}M/{\rm d}t$ &\multicolumn{2}{c}{radius}
& Line & HPBW & $T_{\rm MB}$
& $I_{\rm CO}$\cr
\cline{14-15}  
& & & $''$& (K)& (K~km/s)& (K)& (km/s)& (km/s)&
& (pc)& (km/s)& (M$_{\odot}$~yr$^{-1}$)& (cm) & $('')$ & &
($''$) & (K)& (K km/s)\cr
\hline\cr
\noalign{\bf{Region~C (continued):}}
\smallskip\cr
19126$-$0708& W~Aql& 1$-$0& 23& 0.09&  $-$&  2.31& $-$24.6& 19.3& H90&
610& 18.7& $1.3\;  10^{-5}$& $2.12\;  10^{17}$& 23.2 & 1$-$0& 23 & 2.98& 102.4\cr
& & 1$-$0& 45&  $-$&  26.2&   $-$&  $-$25&   20&   M90\cr
& & 1$-$0& 45& 0.05& 28.4&  0.96& $-$24.0& 18.1& N92& & & & & &1$-$0&
45 & 1.14& 35.54\cr
& & 1$-$0& 50& 0.03& 27.7& 0.98& $-$24.0& 17.7& BL94& & & & & &1$-$0& 50& 0.95
& 29.5\cr
& & 1$-$0& 45& 0.22&  $-$&  1.30& $-$24.1&  18.3&  SL95\cr
& & 2$-$1& 25& 0.04& 43.1&  1.60& $-$24.1& 17.4& BL94& & & & & &2$-$1&
25 & 26.7& 78.2\cr
& & 2$-$1& 30& 0.03& 54.8&  2.10& $-$25.2& 19.3& K97& & & & & &2$-$1&
30 & 1.96& 56.98\cr
& & 2$-$1& 13& 0.16& $-$& 6.05& $-$24.0& 18.8& SL95& & & & & & 2$-$1&
13 & 6.98&  217.3\cr
& & 2$-$1& 23& 0.25& $-$& 2.04& $-$24.3& 18.0& SL95& & & & & & 2$-$1&
23& 3.05& 90.0\cr
& & 3$-$2& 20& 0.10& 93.2&  3.51& $-$23.8& 19.6& K97& & & & & &3$-$2&
20 & 2.70& 76.95\cr
& & & & & & & & & & & & & & & & & \cr
19133$-$1703& T~Sgr& 1$-$0& 50 &0.02 & $-$ &   $-$ &   $-$&  $-$&  BL94\cr
& & 1$-$0& 45& 0.02&  $-$&   $-$&   $-$&  $-$&  SL95\cr
& & 2$-$1& 25& 0.04&  $-$&   $-$&   $-$&  $-$&  BL94\cr
& & 2$-$1& 23& 0.02& $-$& 0.06& +10.8& 14.1& SL95& 810& 14.1&$ 4.0\;  10^{-7}$
&$ 3.22
\;  10^{16}$& 2.7 & 2$-$1& 23 & 0.06& 1.53\cr
& & & & & & & & & & & & & & & & & \cr
19354+5005& R~Cyg& 1$-$0& 50& 0.01& 2.52& 0.15 &$-$17.5&  9.9& BL94&
900& 10.5&$1.7\;10^{-6}$& $8.16\;  10^{16}$& 6.1& 1$-$0& 50 & 0.11& 2.0\cr
& & 2$-$1& 25& 0.02& 5.04& 0.33& $-$17.2& 11.4& BL94& & & & & &2$-$1&
25 & 0.43& 7.06\cr
& & 3$-$2& 20& 0.17& 14.6&  1.3& $-$18.2& 10.4& S95 & & & & & &3$-$2&
20 & 0.44& 7.18\cr
& & & & & & & & & & & & & & & & & \cr
22036+3315& RZ~Peg& 2$-$1& 13& 0.04&  $-$&  0.21& $-$23.4& 12.6& SL95&
1560& 12.6&$ 1.1\;  10^{-6}$& $5.93\;  10^{16}$& 2.5 & 2$-$1& 13 & 0.21& 4.57\cr
& & & & & & & & & & & & & & & & & \cr
*22196$-$4612& $\pi^1$~Gru& 1$-$0& 45& 0.05& 6.9&  0.33& $-$13.2&
19.4& N92& 160& 11.0&$ 4.6\;  10^{-7}$& $3.70\;  10^{16}$& 15.5 &
1$-$0& 45 & 0.53& 11.80\cr
& & 1$-$0& 45& 0.08&  $-$&   0.60& & & SL95\cr
& & 2$-$1& 23&  $-$&  $-$& 2.20& $-$12.5& 11.0& S92& & & & & & 2$-$1&
23 & 2.20& 45.67\cr
& & 2$-$1& 23& 0.06&  $-$&   2.00&  & &     SL95\cr
& & & & & & & & & & & & & & & & & \cr
23554+5612& WY~Cas& 2$-$1& 13& 0.04&  $-$&  0.50& +7.5& 13.3& SL95&
1300& 13.3& $1.8 \; 10^{-6}$& $7.65\;  10^{16}$& 3.9 & 2$-$1& 13 & 0.50& 1.08\cr
\hline
\end{tabular}
\end{table*}
\vfill
\clearpage
\newpage

\addtocounter{table}{-1}
\renewcommand{\thetable}{\arabic{table} (continued)}
\begin{table*}
\caption[]{}
\begin{tabular}{rc|cccccccc|ccccccccccccc}
& & \multicolumn{7}{c}{Observations}& & &\multicolumn{8}{c}{Model}\cr
\cline{3-10}\cline{11-19}
IRAS& Star& Line& HPBW& rms& $I_{\rm CO}$& $T_{\rm MB}$ & $V_{\rm c}$
& $V_{\rm e}$& Ref
& $D$ & $V_{\rm e}$ & ${\rm d}M/{\rm d}t$ &\multicolumn{2}{c}{radius}
& Line & HPBW & $T_{\rm MB}$
& $I_{\rm CO}$\cr
\cline{14-15}  
& & & $''$& (K)& (K~km/s)& (K)& (km/s)& (km/s)&
& (pc)& (km/s)& (M$_{\odot}$~yr$^{-1}$)& (cm) & $('')$ & &
($''$) & (K)& (K km/s)\cr
\hline\cr
\noalign{\bf{Region~D:}}
\smallskip\cr
00578+5620& V365~Cas& 1$-$0& 50& 0.01&  $-$&  $-$&  $-$&  $-$&  BL94\cr
& &  2$-$1& 25& 0.03&  $-$&  $-$& $-$& $-$&   BL94\cr
& &  2$-$1& 13& 0.02&  $-$& 0.07& $-$2.1&  7.2&  SL95& 800& 7.2&
$7.0\;  10^{-8}$&$ 1.53\;  10^{16}$&1.3 & 2$-$1& 13 & 0.07& 0.98\cr
& & & & & & & & & & & & & & & & & \cr
04352+6602& T~Cam& 1$-$0& 50& 0.02&  $-$&  $-$&  $-$& $-$&   BL94\cr
& & 2$-$1& 25& 0.05&  $-$&  $-$&  $-$&  $-$&  BL94& 520& (8.5)& $< 2.2\;  10^{-7}$& 
& & & & \cr
& & & & & & & & & & & & & & & & & \cr
04543+4829& TV~Aur& 1$-$0& 50& 0.02&  $-$&  $-$&  $-$& $-$&   BL94\cr
& & 2$-$1& 25& 0.03&  $-$&  $-$& $-$& $-$&   BL94& 780& (8.5)
& $< 2.6\;  10^{-7}$& & & & & \cr
& & & & & & & & & & & & & & & & & \cr
07095+6853& AA~Cam& 2$-$1& 13& 0.02&  $-$&   0.06& $-$43.5& 17.9&
SL95& 880& 17.9&$ 2.9\;  10^{-7}$& $2.67\;  10^{16}$& 2.0 & 2$-$1& 13
& 0.06& 2.14\cr
& & & & & & & & & & & & & & & & & \cr
07399$-$1045& SU~Mon& 1$-$0& 50& 0.01&  $-$& $-$&  $-$& $-$&  BL94& 780
& (8.5)& $< 2.6\;  10^{-7}$& & & & & \cr
& &  1$-$0& 45& 0.01&  $-$&  $-$&  $-$&  $-$&  SL95\cr
& &  2$-$1& 25& 0.03&  $-$&  $-$&  $-$& $-$&  BL94\cr
& & & & & & & & & & & & & & & & & \cr
12417+6121& S~UMa& 2$-$1&  30&  0.01&  $-$&  $-$&  $-$& $-$& this~work&1650
& (8.5)&$ <4.8\;  10^{-7}$& & & & & \cr
& & & & & & & & & & & & & & & & & \cr
*13440$-$5306& AM~Cen& 1$-$0& 45& 0.02&  $-$&   0.08& $-$27.4&  5.4&
SL95& 910& 5.4&$ 3.5\;  10^{-7}$& $4.13\;  10^{16}$& 3.0 & 1$-$0& 45 &
0.054& 0.55\cr
& &  2$-$1& 23& 0.04&  $-$& 0.20& & & SL95& & & & & & 2$-$1& 23 & 0.27& 2.33\cr
& & & & & & & & & & & & & & & & & \cr
13477$-$6009& VX~Cen& 1$-$0&  45&  0.03&  $-$& $-$& $-$& $-$&   SL95& 530
& (8.5)&$ <4.6\;  10^{-7}$& & & & & \cr
& & & & & & & & & & & & & & & & & \cr
17206$-$2826& V521~Oph& 1$-$0&  50&  0.01& $-$& $-$&  $-$&  $-$& BL94& 580
& (8.5)&$ < 2.5\;  10^{-7}$& & & & & \cr
& & 1$-$0&  45&  0.01& $-$& $-$& $-$& $-$& SL95\cr
& & & & & & & & & & & & & & & & & \cr
19311+2332& EP~Vul& 1$-$0& 50& 0.02& 0.94& 0.12& $-$0.2&  5.2& BL94&
650& 5.4&$ 2.9\;  10^{-7}$&$ 3.70\;  10^{16}$& 3.8&1$-$0& 50 & 0.07 & 0.68 \cr
& & 1$-$0& 45& 0.01&  $-$&   0.08& $-$0.7&  5.6& SL95& & & & & &
1$-$0& 45 & 0.08& 0.84\cr
& & 2$-$1& 25& 0.02& 1.56& 0.20& +0.1&  5.5& BL94& & & & & & 2$-$1& 25
& 0.34& 2.99\cr
& & 2$-$1& 13& 0.14& $-$& 0.52& $-$0.1& 4.7& SL95\cr
& & & & & & & & & & & & & & & & & \cr
20026+3640& AA~Cyg& 1$-$0& 50& 0.03& 1.51& 0.11& +22.7&  9.4&  BL94&
550& 7.1&$ 3.1\;  10^{-7}$& $3.46\;  10^{16}$& 4.2 &1$-$0& 50 & 0.06& 0.80\cr
& &  2$-$1& 13& 0.13&  $-$&  0.95& +27.9&  4.8& SL95& & & & & & 2$-$1&
13 & 1.05& 12.82\cr
& & & & & & & & & & & & & & & & & \cr
22476+4047& RX~Lac& 1$-$0& 45& 0.13& $-$& $-$&  $-$& $-$&  SL95\cr
& & 2$-$1& 13& 0.12& $-$& 0.68& $-$15.4& 3.4& SL95& 380& 3.4&$ 4.0\;
10^{-8}$ & $1.4\;  10^{16}$& 4.1 & 2$-$1& 13 & 0.67& 4.42\cr
\hline
\end{tabular}
\end{table*}
\vfill
\clearpage
\newpage

\addtocounter{table}{-1}
\renewcommand{\thetable}{\arabic{table} (continued)}
\begin{table*}
\caption[]{}
\begin{tabular}{rc|cccccccc|ccccccccccccc}
& & \multicolumn{7}{c}{Observations}& & &\multicolumn{8}{c}{Model}\cr
\cline{3-10}\cline{11-19}
IRAS& Star& Line& HPBW& rms& $I_{\rm CO}$& $T_{\rm MB}$ & $V_{\rm c}$
& $V_{\rm e}$& Ref
& $D$ & $V_{\rm e}$ & ${\rm d}M /{\rm d}t$ &\multicolumn{2}{c}{radius}
& Line & HPBW & $T_{\rm MB}$
& $I_{\rm CO}$\cr
\cline{14-15}  
& & & $''$& (K)& (K~km/s)& (K)& (km/s)& (km/s)&
& (pc)& (km/s)& (M$_{\odot}$~yr$^{-1}$)& (cm) & $('')$ & &
($''$) & (K)& (K km/s)\cr
\hline\cr
\noalign{\bf{Region~E:}}
\smallskip\cr
04123+2357& GCGSS89& 2$-$1& 30& 0.02& $-$& $-$& $-$& $-$& this~work& 650
& (8.5)&$ < 1.7\;  10^{-7}$& & & & & \cr
& & & & & & & & & & & & & & & & & \cr
04497+1410& $o^1$~Ori& 1$-$0& 45& 0.02&   $-$&  $-$&  $-$&   $-$&  SL95&330
& (8.5)&$ < 2.0\;  10^{-7}$& & & & & \cr
& & & & & & & & & & & & & & & & & \cr
04599+1514& GP~Ori& 3$-$2& 20& 0.04& $-$& $-$& $-$& $-$& this~work& 1100& (8.5)
&$ < 5.2\;  10^{-7}$& & & & & \cr
& & & & & & & & & & & & & & & & & \cr
*06197+0327& FU~Mon& 2$-$1& 30& 0.01& 0.66& 0.16& $-$41.7&  2.8&
this~work& 860& 2.8&$ 1.0\;  10^{-7}$& $7.70\;  10^{17}$&59.8& 2$-$1&
30 & 0.14& 0.63\cr
& & & & & & & & & & & & & & & & & \cr
07507$-$1129& NQ~Pup& 3$-$2& 20& 0.04& $-$& $-$& $-$& $-$& this~work
& 1160& (8.5)&$ < 6.3\;  10^{-7}$& & & & & \cr
& & & & & & & & & & & & & & & & & \cr
13163$-$6031& TT~Cen& 1$-$0& 45& 0.04&  $-$& 0.05& +5.2& 24.0& SL95&
1070& 24.7& $4.6\;  10^{-6}$& $1.0\;  10^{17}$& 6.2&1$-$0& 45 & 0.06& 2.67\cr
& &  2$-$1& 23& 0.05&  $-$& 0.24& +5.5& 25.4& SL95& & & & & &2$-$1& 23
& 0.21& 8.92\cr
& & & & & & & & & & & & & & & & & \cr
20044+2417& DK~Vul& 1$-$0& 50& 0.02& 1.33& 0.04& $-$13.9&  8.9& BL94&
910& 5.0&$ 2.9\;  10^{-7}$& $3.83\;  10^{16}$& 2.8 & 1$-$0& 50 & 0.04& 0.38\cr
& &  2$-$1& 25& 0.02& 1.41& 0.21& $-$14.2&  5.3&  BL94& & & & & &
2$-$1& 25 & 0.20& 1.64\cr
& &  2$-$1& 13& 0.05&  $-$&  1.10& $-$14.2&  4.7&  SL95& & & & & &
2$-$1& 13 & 0.74& 5.93\cr
& & & & & & & & & & & & & & & & & \cr
20100$-$6225& GCGSS1195& 1$-$0& 45& 0.04&  $-$&  $-$&  $-$&  $-$&   SL95
& 900& (8.5)& $< 1.0\;  10^{-6}$& & & & & \cr
& &  2$-$1& 23& 0.14&  $-$&   $-$& $-$& $-$&  SL95\cr
& & & & & & & & & & & & & & & & & \cr
20120$-$4433& RZ~Sgr& 1$-$0& 45& 0.03& $-$& 0.36& $-$31.6& 14& SL95&
820& 8.8&$ 2.1\;  10^{-6}$& $9.90\;  10^{16}$& 8.1&1$-$0& 45 & 0.30& 4.36\cr
& & 2$-$1& 23& 0.04& $-$& 0.98& $-$31.2& 8.8& SL95& & & & & & 2$-$1&
23 & 1.08& 14.6\cr
\hline\cr
\end{tabular}

{References:}\\
H90: Heske 1990;
M90: Margulis et al 1990;
S92: Sahai 1992;
N92: Nyman et al. 1992;
N93: Nyman et al. 1993;
O93: Omont et al. 1993; 
KJ94: Kahane \& Jura 1994;
BL94: Bieging \& Latter 1994;
SL95: Sahai \& Liechti 1995;
Y95: Young 1995;
S95: Stanek et al. 1995;
K97:  Knapp et al. 1997b
\medskip\\
{Notes:}\\
1. 06197+0327, FU Mon: Data from Sahai \& Liechti (1995) not used --
see text;\\
2. 09076+3110, RS Cnc: CO(2-1) line profile has two components -- see
text;\\
3. 13440-5306, AM Cen: Values for central velocity $\rm V_c$ and outflow 
velocity $V_{\rm e}$ given by SL95 are averages of
the values for the CO(1-0) and CO(2-1) lines;\\
4. 22196-4612, $\pi^1$ Gru: SL95 do not list values for $\rm V_c$
and $V_{\rm e}$ because of the line wings\\
\end{table*}

\clearpage
\newpage

\end{landscape}

\setlength{\textheight}{\textwidth}
\setlength{\textwidth}{\widthsav}
\setlength{\oddsidemargin}{\oddsidemarginsave}
\setlength{\evensidemargin}{\evensidemarginsave}

\end{document}